\documentclass[a4paper,english,nofootinbib,pra,superscriptaddress,twocolumn]{revtex4-2}

\usepackage{amsmath,amsfonts}
\usepackage{graphicx}
\usepackage{amssymb}
\usepackage{enumerate}
\usepackage{epsfig}
\usepackage{epstopdf}
\usepackage{mathrsfs}
\usepackage[colorlinks,citecolor=blue,linkcolor=blue]{hyperref}
\usepackage{textcomp}
\usepackage{subcaption}
\usepackage{dcolumn}
\usepackage{amsthm}
\usepackage[latin9]{inputenc}
\usepackage{babel}
\usepackage{xcolor}

\newcommand{\be}{\begin{equation}}
\newcommand{\ee}{\end{equation}}
\newcommand{\bey}{\begin{eqnarray}}
\newcommand{\eey}{\end{eqnarray}}
\newcommand{\ba}{\begin{array}}
\newcommand{\ea}{\end{array}}
\newcommand{\bi}{\begin{itemize}}
\newcommand{\ei}{\end{itemize}}
\newcommand{\bem}{\begin{enumerate}}
\newcommand{\eem}{\end{enumerate}}
\newcommand{\bw}{\begin{widetext}}
\newcommand{\ew}{\end{widetext}}

\newcommand{\ra}{\rangle}
\newcommand{\la}{\langle}

\newcommand{\ov}{\overline}

\newcommand{\wh}{\widehat}
\newcommand{\ww}{\widetilde}

\newcommand{\bk}{{\bf k}}

\newcommand{\bJ}{{\bf J}}
\newcommand{\bp}{{\bf p}}
\newcommand{\bP}{{\bf P}}
\newcommand{\bq}{{\bf q}}
\newcommand{\bQ}{{\bf Q}}

\newcommand{\bs}{{\bf s}}
\newcommand{\bS}{{\bf S}}

\newcommand{\bx}{{\bf x}}

\newcommand{\C}{{\mathcal{C}}}
\newcommand{\cs}{\mathcal{S}}

\newcommand{\E}{{\cal E}}

\newcommand{\HH}{{\mathscr{H}}}

\newcommand{\LL}{{\mathcal{L}}}
\newcommand{\M}{\mathcal{M}}

\newcommand{\R}{{\mathcal{R}}}
\newcommand{\SP}{{\mathscr{S}}}

\newcommand{\WW}{{\mathscr{W}}}

\newcommand{\VV}{{\mathscr{V}}}

\newcommand{\bbs}{\boldsymbol}

\begin{document}

 \title{A framework for quantum theory of elementary physical entities
 }

\author{Wen-ge Wang}\email{wgwang@ustc.edu.cn}
\affiliation{ Department of Modern Physics, University of Science and Technology of China,
	Hefei 230026, China}
\affiliation{CAS Key Laboratory of Microscale Magnetic Resonance, USTC, 
Hefei 230026, China}

 \date{\today}

\begin{abstract}
 A unified framework, which is directly established on the quantum ground,
 is proposed for elementary physical entities, called \emph{modes} in this paper.
 The framework is mainly built upon five basic assumptions, which loosely speaking have the following contents. 
 (i) The state space of each mode is given by the direct product of a momentum-state space and a spinor-state space,
 the latter of which is certain representation space  of the $SL(2,C)$ group (a covering group of the Lorentz group);
 (ii) spinor states of modes have a layer-type structure
 and modes are either fermionic or bosonic, depending on their helicity properties;
 (iii) there are three fundamental processes --- free evolution,
 vacuum fluctuation (emergence or vanishing of a pair of fermionic modes that possess exactly opposite physical properties), 
 and two fundamental interaction processes
 (change of two fermionic modes into one bosonic mode and the reverse);
 (iv) vacuum fluctuation happens instantly;
 and (v) the time evolution operator is constructed from operators, which map state spaces
 of incoming modes of fundamental processes to those of outgoing modes. 
 The time evolution operator turns out to be a function of 
 quantum fields that are constructed from creation and annihilation operators for free-mode states,
 whose interaction part has a local feature. 
 As an example, a simple model of modes is studied
 and is compared with the first-generation part of the standard model (SM). 
 Concerning electroweak interactions, the studied model has a time evolution operator, 
 whose main body is formally similar to that of the SM.
 Besides, it predicts $\frac 13$ and $\frac 23$ electronic changes for quark-type modes,
 gives an interpretation to the color degree of freedom, and contains certain modes that behave like dark matters. 
\end{abstract}

 \maketitle


\section{Introduction}\label{sect-intro}


 One of the ultimate goals of physics is to establish a theory, in which elementary physical entities and 
 their interactions may be described in a unified way.
 This goal is partially achieved in the standard model (SM), the most successful quantum field theory (QFT)
 that has ever been built (see textbooks, e.g., Refs.\cite{Weinberg-book,Peskin,Itzy}),
 in which the electromagnetic and weak interactions are described in a unified way
 based on the local gauge group of $U(1)\otimes SU(2)$, known as the Glashow-Weinberg-Salam (GWS) electroweak theory,
 meanwhile, the strong interaction is described based on the group of $SU(3)$ in an independent way.
 The SM still has some unsatisfactory features and the topic of going beyond the SM
 has attracted lots of attention (see, e.g., Refs.\cite{GGS99,KO01,Polch-book,Ross84,Raby93,book-fundam-QFT});
 for example, a satisfactory explanation to neutrino masses
 is still lacking \cite{Fukuda98,Ahmad02,Araki05,Adamson08,An12,Ahn12,Abe12,Abe14,GN03,SV06}.

 In this paper, a framework is proposed for developing theories in which elementary physical entities
 may be described in a unified way.
 It is directly built on a quantum ground, different from QFTs which need to start from classical fields. 
 To distinguish from the terminology used in the SM, 
 we use the name of \emph{``mode''} to refer to an elementary physical entity that lies at the fundamental level.

 The framework is based on five basic assumptions (BAs).
 One BA states that mode states consist of only two parts --- a momentum-state part and a spinor-state part.
 Here, spinor-state spaces are given by representation spaces of the so-called $SL(2,C)$ group
 (a covering group of the proper and orthochronous Lorentz group
 \cite{Penrose-book,Kim-group,CM-book,Corson}).
 This BA  implies that properties of the spinor spaces used
 should be sufficient for characterizing differences among mode species.
 Another BA states that vacuum fluctuation (VF) is a fundamental process, 
 which is described in the most direct way, as emergence/vanishing of two modes 
 that possess exactly opposite physical properties from/into the vacuum. 
 Ideas behind the rest assumptions are as follows:
 Modes are either fermionic or bosonic, depending on their helicity properties;
 there are two fundamental interaction processes, as the change of
 two fermionic modes into one bosonic mode and the reverse;
 VF happens instantly;
 and the time evolution operator is derived from operators that map
 state spaces of incoming modes to those of outgoing modes of fundamental processes.

 Upon the above-discussed five BAs and with the help of a few additional assumptions, we are to 
 show that many basic properties of state spaces of modes and of the time evolution operator may be determined.
 In particular, the time evolution operator is found to have a concise form, written as a function of 
 quantum fields that are constructed from creation and annihilation operators for free-mode states.
 A framework is thus established for developing theories for elementary physical entities.
 An interesting question is whether this framework may allow the construction of a specific model,
 which does not contain so many assumptions as  the SM, meanwhile,
 whose time evolution operator has a form similar to that of the SM.
 We are to show that such a model is constructable.

 The paper is organized  as follows. 
 For the sake of easy reading,  in Sec.\ref{sect-notation},
 we give a short list for notations and conventions to be used. 
 Detailed contents of the five BAs are given in Sec.\ref{sect-framework}, 
 and state spaces of modes are discussed in Secs.\ref{sect-basic-modes} - \ref{sect-compound-modes}. 
 The time evolution operator is discussed  in Sec.\ref{sect-$H$-operator}. 
 The above-mentioned specific mode is given in Sec.\ref{sect-physical-model}, 
 and its similarities to and differences from the first-generation part of the SM
 are discussed in Sec.\ref{sect-compare-SM}.
 Finally, summary and discussions are given in Sec.\ref{sect-conclusion}.

\section{Notations and conventions}\label{sect-notation}

 It proves convenient to use Dirac's ket-bra notation for spinor states,
 particularly in derivations related to the time evolution operator. 
 Such an abstract notation system for Weyl spinors has been discussed in detail  in Ref.\cite{pra16-commu}
 and its main points are given in Appendix \ref{sect-recall-Weyl-spinor} of this paper.
 The abstract notation for four-component vectors is discussed in Appendix \ref{sect-vector-abstract},
 after a brief discussion of their basic properties in the ordinary notation in Sec.\ref{sect-recall-vector}.
 A brief discussion of $SL(2,C)$ transformations are given in Appendix \ref{sect-SL2C-transf}. 

 In this section, we list major conventions, notations, and abbreviations to be used.
 We give here brief explanations to them and leave detailed discussions (if needed) to later sections.

\vspace{0.3cm}
\noindent \emph{1. Conventions}.
 \\ (i) An \emph{overline} above a spinor indicates its {complex conjugate}
 and similar for symbols related to spinors.
 \footnote{This is a convention usually used in the mathematical theory of spinors.
 Under this convention, e.g., $\ov U$  for a Dirac spinor $U$ refers to
 its complex conjugate, but not  to $U^\dag \gamma^0$.}
 \\ (ii)  Repeated index implies a summation over it,
 unless otherwise stated.  
 \\ (iii)   $|\psi \phi\ra \equiv |\psi\ra | \phi\ra $ and its bra is written as $\la \phi \psi | \equiv  \la \phi| \la \psi|$.

\vspace{0.3cm}
\noindent \emph{2. Frequently used notations for spinors}.
 \\ (a) 
 $\WW$ and $\ov \WW$: the two smallest nontrivial representation spaces of the $SL(2,C)$ group,
 spanned by two-component Weyl spinors, which are the complex conjugate space of each other.
 \\ (b) 
 $|S^A\ra$ of $A=0,1$: a basis in the space $\WW$.
 Its complex conjugate, as a basis in $\ov\WW$, is written as $|\ov S^{A'}\ra$ with a primed index $A' = 0', 1'$.
 \\ (c) 
 $|\kappa\ra = \kappa_A|S^A\ra$: an expansion of an arbitrary Weyl spinor $|\kappa \ra \in \WW$, 
 with expansion coefficients $\kappa_A$.
 The corresponding spinor $\ov{|\kappa \ra} $ in $\ov\WW$
 is expanded as $\ov{|\kappa \ra} = \ov\kappa_{A'}|\ov S^{A'}\ra$.
 \\ (d) 
 $\epsilon^{AB}$ and $\epsilon_{AB}$: symbols for raising and lowering indices of spinors in $\WW$, respectively, 
 both having the matrix expression of $ \left( \begin{array}{cc} 0 & 1 \\ -1 & 0 \end{array} \right)$.
 For example, $\kappa^A = \epsilon^{AB} \kappa_B$ and $\kappa_A = \kappa^B \epsilon_{BA}$. 
 The corresponding symbols for $\ov\WW$ are written as $\epsilon^{A'B'}$ and $\epsilon_{A'B'}$, respectively,
 described by the same matrix. 
\\ (e) $\VV$: a four-component vector space, which is isomorphic to $\WW \otimes \ov\WW$. 
 A basis in $\VV$ is written as $|T_\mu\ra$ with $\mu =0,1,2,3$. 
 \\ (f) $|S_{AB'}\ra$: a shorthand writing of $|S_A\ra |\ov S_{B'}\ra$.

\vspace{0.3cm}
\noindent \emph{3. Spinor bra for scalar product.}
 
 We introduce spinor bras for writing $SL(2,C)$-scalar products of spinors in an abstract notation,
 and call them \emph{spinor bra for scalar product}. 
 As seen in the explicit expression of the scalar product of Weyl spinors (as elementary spinors),
 which is given in Eq.(\ref{ww-chi-kappa=chi-kappa}) of Appendix \ref{sect-SL2C-transf}, 
 the product involves no complex-conjugated components. 
 \footnote{This is different from the case of inner product for vectors in a Hilbert space.
 In fact, to introduce $c$-number products in a theory in which the $SL(2,C)$ group lies at the fundamental level, 
 $SL(2,C)$-scalar product is the one that should be introduced first, while, inner product (with Hilbert space)
 may be introduced at a later stage. 
 }
 For this reason, no complex conjugate component should be involved in the definition of 
 spinor bra for scalar product, in contrast to bra used in inner product. 
 To indicate this difference explicitly, we write the former as $\la\la \cdot |$ with double ``$\la $'', 
 while using the ordinary notation for the latter. 
 Below are explicit notations for bras. 
 \\ (a)
 $\la\la S^A|$: the spinor bra for scalar product, which corresponds to the ket  spinor $|S^A\ra$,
 giving a basis in a space dual to $\WW$.
 The scalar product of $|S^A\ra$ and $|S^B\ra$ is written as
\begin{align}\label{}
 \la\la S^{A}|S^{B}\ra = \epsilon^{A B}. 
\end{align}
 \\ (b) 
 $\la\la \kappa|$: the spinor bra for scalar product, which corresponds to the spinor $|\kappa\ra$. 
 The scalar product of two Weyl spinors $|\kappa\ra$ and $|\chi\ra$, written as $\la\la \kappa | \chi \ra$, 
 has the expression in Eq.(\ref{ww-chi-kappa=chi-kappa}), i.e., 
\begin{align}\label{<k|c>}
  \la\la \kappa | \chi \ra \equiv \kappa_A \chi^A.
\end{align}
 Equation (\ref{<k|c>}) requires the following  expansion of $\la\la \kappa|$,
\begin{gather}\label{<kappa|-expan}
 \la\la \kappa | = \la\la S^{A}| \kappa_A,
\end{gather}
 which has the same expansion coefficients as the ket $|\kappa\ra$, not involving complex conjugation.
 \\ (c) $\la\la \Psi|$: the spinor bra for scalar product, which corresponds to
 a generic ket spinor $|\Psi\ra $.
 If $|\Psi\ra$ is expanded as
 $|\Psi\ra = c_1|\Psi_1\ra + c_2 |\Psi_2\ra$, then, the bra  $\la\la \Psi|$ is written as
\begin{align}\label{<Psi|}
 \la\la \Psi| = c_1\la\la \Psi_1| + c_2 \la\la \Psi_2 |,
\end{align}
 consistent with Eq.(\ref{<kappa|-expan}).
 \\ (d) $\la \wh \psi|$: the bra of a ket spinor $|\psi\ra$, 
 which is obtained by certain 
 operation that involves complex conjugation (ACC operation to be discussed below).
 It may be used in the construction of inner product.

\vspace{0.3cm}
\noindent \emph{4. Notions for modes and their states.}
 \\ (i) $M$: a generic mode.
 \\ (ii) $F$: a generic fermionic mode.
 \\ (iii) $B$: a generic bosonic mode.
 \\ (iv) Basic mode: a mode whose spinor space is either $\WW$ (called $b$-mode) or $\ov\WW$ (called $\ov b$-mode). 
 \\ (v) $|M^{\alpha}_{\bp}\ra $: a state of a mode $M$ with a three-momentum $\bp$ and an index $\alpha=(r,\varrho)$,
 where $r=0,1$ is an index for helicity
 \footnote{The helicity index $r$ is not influenced by the operation of complex conjugation. }
 and $\varrho$ indicates the sign of $p^0$ of a four-momentum $p^\mu$.
 \\ (vi) $|\ov\Psi\ra $:
 a shorthand writing for the complex conjugate of  ${|\Psi} \ra$, i.e., $|\ov\Psi\ra = \ov{|\Psi \ra}$,
 except for $\Psi =  M, f, B$, and $ \bp$. 
 \\ (vii) $|w^r(\bp)\ra, |u^\alpha(\bp)\ra $, and $|v^\alpha(\bp)\ra$: helicity states of $b$-mode. 
 \\ (viii) Layer structure: a structure of spinor states of modes, expressed as a one-column matrix
 with one row corresponding to one layer (under certain restriction to be 
 given in BA2 of the next section). 
 \\ (ix) $f$: a fermionic mode, whose first layer lies in a space $\WW$ and is described by $|u^\alpha(\bp)\ra$.
 \\ (x) $|\bp\ra$ and $\la \bp|$: ket and bra of momentum state, as in the ordinarily-used notation. 

\vspace{0.3cm}
\noindent \emph{5. Overlines  not indicating complex conjugate}.
 \\ (a) $\ov M$: the antimode of $M$.
 Two modes are the antimode of each other, if their layers have a one-to-one correspondence
 and spaces of corresponding layers have the relationship of complex conjugation. 
 \\ (b)  $|\ov M^{\alpha}_{ \bp  }\ra $: a state of a mode $\ov M$. 
 \\ (c) $\ov f$: the antimode of $f$, with the first layer state as $|\ov v^\alpha(\bp)\ra$.
\begin{subequations}\label{ov-def-nonspinor}
\begin{align}\label{ov-p}
  \text{(d)} & \qquad  \ov \bp  := -\bp \ \ \qquad \text{for three-momentum}, \qquad 
 \\  \text{(e)} & \qquad  \ov \alpha  := (r, -\varrho) \ \ \qquad \text{for  $\alpha = (r,\varrho)$}. \qquad   \label{ov-alpha}
\end{align}
\end{subequations}

\vspace{0.3cm}
\noindent \emph{6. Some  abbreviations}.
 \\ (i) BA$i$: the $i$-th basic assumption (BA).
\\ (ii) FIP: fundamental interaction process.
\\ (iii) VF: vacuum fluctuation.
\\ (iv) BPIP: basic positive-$p^0$ interaction process. 
\\ (v) MA: assumption used in a specific model.
\\ (vi) $d\ww p$: a symbol used for an integration over $\bp$, 
\begin{align}\label{}
 d\ww p := \frac{1}{|\bp|} d^3p.
\end{align}
\\ (vii) ACC operation: a combined operation of taking antimode-counterpart state and complex conjugation.

 
\section{Basic assumptions}\label{sect-framework}

 In this section, we give five BAs, as the foundation of the framework to be proposed. 
 We first discuss two BAs about mode states in Sec.\ref{sect-BA1-2},
 then, in Sec.\ref{sect-dual-state-space} discuss a strategy to be used for constructing dual state spaces,
 and, finally, give the rest three BAs  about fundamental processes in Sec.\ref{sect-BA3-5}. 
 
 \subsection{Basic assumptions for states of modes}\label{sect-BA1-2}

 In this section, we introduce two BAs concerning basic properties of states of modes.

 The first BA, indicated as BA1, is about the global structure of mode states. 
 According to experiences that have been accumulated by experimental and theoretical 
 studies of the physical world during the past several hundreds of years, 
 states of physical entities should possess at least two types of degree of freedom --- 
 one related to position and the other to spin --- 
 and obey the well known {Lorentz symmetry}.

 At the fundamental level of theory, the simplest method of respecting the Lorentz symmetry
 is to assume that state spaces of modes may be given by representation spaces of the $SL(2,C)$ group,
 a covering group of the proper and orthochronous Lorentz group.
 Elements in these representation spaces are called \emph{spinors}. 
 In fact, the  theory of spinors supplies
 the most powerful mathematical tool for dealing with the Lorentz symmetry  \cite{Penrose-book,Kim-group,CM-book,Corson}.

 As is  well known in quantum mechanics, 
 the position representation and the momentum representation 
 are mathematically equivalent, linked by a Fourier transformation.
 But, when establishing the foundation of a theory that emphasizes spin --- more exactly helicity states,
 it proves convenient to use the momentum representation, 
 because the helicity of a particle should  be computed with respect to a given three-momentum.

 One basic idea of this paper is that 
 modes possess two types of degree of freedom only, 
 one related to momentum and the other related to spinor. 
 Since there is no essential difference among momenta of difference physical entities, 
 the above idea requires that spinor properties should be sufficient for distinguishing among mode species.

 Summarizing the above discussions, we state BA1 as follows.
\begin{itemize}
  \item \textbf{BA1}. The state space of a mode consists of only two parts: a momentum-state part and a spinor-state part,
  the latter of which is a representation space of the $SL(2,C)$ group. 
\end{itemize}
 We use $|\bp\ra$ to indicate a momentum state with a three-momentum $\bp$,
 and use $\E_{\rm mom}$ to denote the space spanned by all $|\bp\ra$.
 The spinor space of a mode $M$ is denoted by $\E^S_M$.
 The state space of $M$, indicated as $\E_M$, is 
 then given by their direct product, i.e., $\E_M = \E_{\rm mom} \otimes \E^S_M$. 
 The space dual to $\E_{\rm mom}$ is spanned by the ordinary bras $\la \bp|$,
 for which the well-known Lorentz-invariant inner product is written as
\begin{gather}\label{IP-pb}
 \la \bp'|\bp\ra = |\bp| \delta^3(\bp-\bp').
\end{gather}

 The second BA, indicated as BA2, is about the spinor structure of modes.
 As is known in the theory of spinors, there are two smallest nontrivial representation spaces of $SL(2,C)$, 
 which we indicate by $\WW$ and $\ov \WW$.
 Other representation spaces are constructed from $\WW$ and $\ov \WW$.
 We call modes, whose spinor spaces are $\WW$ or $\ov\WW$, \emph{basic modes}.
 More exactly, we call a basic mode with $\WW$ a \emph{$b$-mode} and
 that with $\ov \WW$ an \emph{$\ov b$-mode} (pronunc anti $b$-mode). 
 Moreover, a mode is called a \emph{unit mode},
 if its spinor space is $\WW$, or $\ov \WW$, or a direct product of some of $\WW$ and $\ov \WW$.

 BA2 basically states a {layer structure} of spinor states, 
 which refers to a one-column matrix, whose rows are given by spinor states of unit modes.
 We call a row in such a matrix a \emph{layer}.  
 Physically, it is natural to require that all the layers within a same mode should 
 possess the same helicity. 
 Besides, like the spin-statistics relationship established in QFT \cite{Pauli40,SW64}, 
 we relate half-integer spin to fermionic mode and integer spin to bosonic mode, satisfying
 \begin{subequations}\label{commut-FB}
\begin{align}\label{fermi-BM}
 & |F\ra |F'\ra = -|F'\ra |F\ra \quad & \text{for fermionic modes},
 \\ & |B\ra |B'\ra = |B'\ra |B\ra \quad & \text{for bosonic modes}. 
\end{align}
\end{subequations}
 To summarize, we state BA2 as follows. 
\begin{itemize}
  \item \textbf{BA2}. Each mode's spinor state has a layer structure, 
 with all its layers possessing a same helicity;
 the mode is fermionic if its layers have a half-integer helicity, otherwise, is bosonic. 
\end{itemize}
 Note that, under the requirement of identical helicity of all the layers within one mode,
 a layer structure does not exactly correspond to a direct sum.
 (Different rows in the matrix form of a direct sum are independent.)

\subsection{About a space dual to $\E_M$}\label{sect-dual-state-space}

 A problem is met in the construction of inner product, or a like, for vectors in 
 the state space $\E_M$ of a generic mode $M$.
 The problem is that no definite direct product may be formed from 
 a space, which is spanned by spinor bras for scalar product
 [e.g., spanned by $\la\la S^A|$  in Eq.(\ref{<kappa|-expan})],
 and the space spanned by the ordinary momentum-state bras $\la \bp|$.
 This is because expansion coefficients in the former do not involve complex conjugation 
 [see Eq.(\ref{<Psi|})],  while, those in the latter do. 
 \footnote{ 
 As an illustration, one may consider $|\Psi\ra = |\psi_1\ra |\psi_2\ra$, where 
 $|\psi_1\ra = a ( c_0|S^0\ra + c_1|S^1\ra)$ as the spinor part and $|\psi_2 \ra = \frac 1a |\bp\ra $ as the momentum part, 
 with $c$-number parameters $a\ne 0, c_1, c_2$. 
 One finds that the bra in the space spanned by $\la\la S^A|$ is written as
 $\la\la \psi_1| = a ( c_0 \la\la S^0| + c_1 \la\la S^1|)$, 
 while, the momentum-state bra is $\la \psi_2 | = \frac 1{a^*} \la \bp|$.
 Thus, although $|\Psi\ra$ is independent of the number $a$, 
 the product of $\la\la \psi_1|$ and $ \la \psi_2 |$ is proportional to $a/a^*$.
 }

 To solve the above-discussed problem, one method is to make use of some operation
 that maps $\E^S_M$ to another space of ket, which involves complex conjugation in certain way.
 Then, one may map the new space to a space of spinor bras for scalar product,
 \footnote{In fact, a related problem is met in the relativistic quantum mechanics
 when constructing inner product for Dirac spinors, 
 though usually formulated in a different mathematical language.
 The solution given there  makes use of a transformation like $U^\dag \gamma^0$ in the ordinary terminology.
 Unfortunately, this solution is specific to Dirac spinors and is not applicable to Weyl spinors.
}

 More exactly, a solution to this problem may be obtained by taking the following steps:
 \\  (i)  Introduce an operation, which maps $\E^S_M$ to a new space of kets such that complex conjugation is involved. 
 \\ (ii) Map the above-obtained space to a space of spinor bras for scalar product,
 such that scalar products exist between the obtained bras and those kets in $\E^S_M$. 
 \\ (iii) Construct a space dual to $\E_M$, by taking the direct product of the space obtained in ``(ii)''
 and the ordinary dual momentum-state space. 
 \\ 
 Exact ways of taking the steps will be discussed in detail in later sections, 
 e.g., in Sec.\ref{sect-ACC} for basic modes.
\footnote{ The solution to be given is related to the so-called chiral representation of $\gamma^\mu$ matrices
 [cf.~Eq.(\ref{gamma-mu})].
}

\subsection{Basic  assumptions for fundamental processes}\label{sect-BA3-5}
 
 In this section, we give the rest three BAs, which concern fundamental processes.
 These processes determine the time evolution of modes. 
 We assume that a continuous {time parameter}, indicated by $t$, may be used when discussing changes of mode states.

 Before discussing the BAs, we need to introduce several concepts. 
 One concept is \emph{antimode}.
 By definition, two modes are said to be the antimode of each other, 
 if they have the same number of layers and spinor spaces of their corresponding layers, from top to bottom, 
 have the relationship of complex conjugation. 
 For example, $b$-mode and $\ov b$-mode are the antimode of each other. 
 Another concept is \emph{VF pair}, refering to a pair of fermionic modes,
 which are the antimode of each other
 and possess opposite four-momentum $p^\mu = (p^0,\bp)$ and opposite angular momentum.

 The third BA (BA3) is about the types of fundamental process. 
 Besides free process, we assume two other fundamental processes.
 One is \emph{VF process}, in which a VF pair emerges from or vanishes into the vacuum.
 The other is \emph{fundamental interaction process}  (FIP), which has two types indicated as FIP-$1$ and FIP-$2$. 
 In an FIP-$1$, two fermionic modes, which have a same number of layers
 and whose top-layer spinor spaces are the complex conjugate of each other, change to one bosonic mode
 subject to conservation of three-momentum.
 The reverse of an FIP-$1$ is an  FIP-$2$.
 Now, we state BA3 as follows.
\begin{itemize}
  \item \textbf{BA3}. Modes may undergo three types of \emph{fundamental processes}: free process, 
  VF process, and FIP. 
\end{itemize}

 For each fundamental process, an operator to be called \emph{$H$-operator} is of importance.
 It maps the state space of the  incoming mode(s) of this process to  that of the outgoing mode(s). 
 We use $H_{\rm FP}$, with FP=free,  VF, and FIP, to indicate $H$-operators of
 free process, VF process, and FIP, respectively.
 And, we define $p^0$ as the expectation value of $H_{\rm free}$, i.e., 
\begin{align}\label{p0-def}
 p^0 := \text{expectation value of $H_{\rm free}$ for  $|\bp\ra$}.
\end{align}

 The fourth BA (BA4) is about a time feature of VF. 
 In fact, experimentally, no finite time scale has ever been observed for VF.
 This suggests that a VF pair may exist instantly only;
 in other words, its existence is associated with one instant $t$ only.
 We assume that this is related to the negative-$p^0$ fermionic mode in the VF pair.
\begin{itemize}
 \item \textbf{BA4}. A fermionic mode with a negative $p^0$ may exist instantly only.
\end{itemize}
 Since an FIP may contain a negative-$p^0$ fermionic mode, too,
 BA4 implies that both VF process and FIP should happen instantly.

 Finally, we discuss the fifth BA (BA5), which is for the equation of time evolution. 
 Due to BA4, such an equation should be for states
 that do not contain any negative-$p^0$ fermionic mode. 
 We call a process  a \emph{basic positive-$p^0$ interaction process} (BPIP), 
 if it is a combination of one FIP and some VF pair (if needed), 
 such that all its incoming and outgoing fermionic modes have positive $p^0$.
 The corresponding combination of $H_{\rm FIP}$ and $H_{\rm VF}$
 is called the $H$-operator of the BPIP,  indicated as $H_{\rm BPIP}$. 
 Now, we are ready to introduce BA5. 
\begin{itemize}
  \item \textbf{BA5}. The time evolution of a state vector $|\Psi\ra$,
  which  do not contain any negative-$p^0$ fermionic mode, obeys the following Schr\"{o}dinger-type equation, 
\begin{align}\label{dPsi-dt}
 i\frac{d|\Psi\ra}{dt}  = \left( H_0 + H_{\rm int} \right)|\Psi\ra.
\end{align}
\end{itemize}
 On the right-hand side (rhs) of Eq.(\ref{dPsi-dt}), 
 the time evolution operator, namely $\left( H_0 + H_{\rm int} \right)$, is given by
\begin{subequations}\label{H0-Hint}
\begin{align}
 & H_0 = {\sum_{M}}' H_{\rm free}, \label{H0}
 \\ & H_{\rm int} = \sum_{\text{BPIP}} H_{\rm BPIP}, \label{H-int}
\end{align}
\end{subequations}
 where the prime on the rhs of (\ref{H0}) means that $p^0>0$ for fermionic modes $M$. 

\section{Basic modes}\label{sect-basic-modes}

 In this section, we discuss properties of basic modes.
 Specifically, we discuss their helicity states in Sec.\ref{sect-basic-property-b-mode},
 a property of antimodes in Sec.\ref{sect-antimode-counterpart-state-bmode}, and
 inner products in Sec.\ref{sect-ACC}.

\subsection{Helicity states}\label{sect-basic-property-b-mode}

 For a given momentum $\bp$,  the helicity operators for $b$-mode and $\ov b$-mode, 
 indicated as $H^{\rm cy}_{b}$ and $H^{\rm cy}_{\ov b}$, respectively, are written as
\begin{gather}\label{hcy}
 H^{\rm cy}_b = \frac{\bs_b \cdot \bp}{|\bp|} \quad \& \quad
 H^{\rm cy}_{\ov b} = \frac{\bs_{\ov b} \cdot \bp}{|\bp|}.
\end{gather}
 Here, $\bs_b$ and $\bs_{\ov b}$ represent the (spinor) angular momentum operators 
 of the two modes, respectively, with $\bs_b \equiv (s_b^1, s_b^2, s_b^3)$ and similar for $\bs_{\ov b}$. 
 (See Appendix \ref{app-am-basic-mode} for detailed discussions about $\bs_b$ and $\bs_{\ov b}$.)
 These two operators have the relationship of negative complex conjugate [see Eq.(\ref{sk=-sk'-com})], i.e., 
\begin{gather}\label{sk=-sk'}
 \bs_{\ov b} = - {\ov \bs}_b.
\end{gather}
 We recall that $\ov \bs_b \equiv ( \bs_b)^*$  by the convention of overline implying 
 complex conjugation (Sec.\ref{sect-notation}).

 We use $|w^r(\bp)\ra$ to indicate helicity states of a $b$-mode, with helicities $h^r_b$
 labelled by an index $r$, 
\begin{gather}\label{Hcy-b}
 H^{\rm cy}_b |w^r(\bp)\ra = h^r_b|w^r(\bp)\ra.
\end{gather}
 Equation (\ref{sk=-sk'}) implies that helicity states of an $\ov b$-mode are given by
 complex conjugates of $|w^r(\bp)\ra$, namely, by $|\ov w^r(\bp)\ra$,
\begin{gather}\label{Hcy-ovb}
 H^{\rm cy}_{\ov b} |\ov w^r(\bp)\ra = h^r_{\ov b}|\ov w^r(\bp)\ra.
\end{gather}
 Detailed discussions on properties of these helicity states are given in Appendix \ref{app-helicity};
 in particular,
\begin{align}\label{hr-b}
  h^r_b = \frac{(-1)^r}{2}, \qquad
  h^r_{\ov b} =  \frac{ (-1)^{r+1}}{2},
\end{align}
 where $r=0,1$ (mod $2$), and 
\begin{gather}\label{ww-eps}
 \la\la w^r(\bp) |w^{s}(\bp)\ra = \epsilon^{r s},
\end{gather}
 where $\epsilon^{r s}$ has the same matrix expression as $\epsilon^{AB}$.
 Further introducing $\epsilon_{r s}$ with the same matrix form, 
 one may use these two symbols to raise and lower the helicity index $r$, e.g., 
\begin{align}\label{w^r_s}
 |w_r (\bp) \ra =  |w^s (\bp) \ra \epsilon_{sr}, \quad
 |w^r (\bp) \ra = \epsilon^{rs} |w_s (\bp) \ra .
\end{align}
 From Eq.(\ref{w^r_s}), it is seen that
\begin{align}\label{w^r-w_s}
 |w_0 (\bp)\ra = -|w^1 (\bp)\ra , \quad |w_1 (\bp)\ra = |w^0 (\bp)\ra,
\end{align}
 and, hence, the two sets of $\{|w^r (\bp) \ra\}$ and $\{|w_r (\bp) \ra\}$ give equivalent bases in
 the space $\WW$.

\subsection{Antimode-counterpart states}\label{sect-antimode-counterpart-state-bmode}

 As discussed in Sec.\ref{sect-dual-state-space}, for the purpose of getting a space
 which is dual to $\E_M$ and useful in the construction of inner product, 
 one needs a type of spinor bra, whose expansion coefficients are the
 complex conjugates of those of the corresponding  ket. 
 The operation of complex conjugation, though giving the wanted change of expansion coefficients, 
 is not capable of completing this task by itself, because for example it relates 
 a ket basis of $\{|S_A\ra \}\in \WW$
 to a bra basis of $\{ \la \ov S_{A'}|\}$, between which no scalar product may be formed.

 To complete the above-mentioned task,  an additional operation is needed, 
 which transfers ket bases to their complex-conjugate space, meanwhile, leaves the expansion coefficients unchanged.  
 Such an operation is  basis-dependent and, hence, 
 its introduction should be based on physical consideration.
 It seems that the most natural way is to make use of helicity states. 
 More exactly, we observe that a $b$-mode and its antimode ($\ov b$-mode) may share a same value of helicity.
 Based on this, we introduce a concept of {antimode-counterpart state}: 
 For a state of an arbitrary mode with a given momentum and a given helicity, 
 its \emph{antimode-counterpart state} refers to a state of its antimode,
 which possesses the same momentum and the same helicity.

 Let us discuss the antimode-counterpart state of a $b$-mode state $|w^r (\bp) \ra$,
 indicated as $|{w^r}(\bp)\ra_{\rm ac}$ with the subscript ``ac'' standing for ``antimode-counterpart state''. 
 From Eqs.(\ref{sk=-sk'}) and (\ref{hr-b}), one sees that $|{w^r}(\bp)\ra_{\rm ac}$ should be 
 given by $|\ov w^{r+1} (\bp) \ra$, apart from a phase factor. 
 Since a major goal is to construct inner product,  the phase factor is of importance. 
 To determine it, we note that the helicity of $|\ov w^{r+1} (\bp) \ra$ is equal to
 that of $|\ov w_{r} (\bp) \ra$ according to  Eq.(\ref{w^r-w_s});
 meanwhile, from $|{w^r}(\bp)\ra$ to $|\ov w_{r} (\bp) \ra$,
 the position of $r$ is directly moved by
 the most important symbol in the theory of spinors, i.e., by the $\epsilon$-symbol.

 Thus, within the framework supplied by the theory of spinors, the simplest 
 and also most natural choice of $|{w^r}(\bp)\ra_{\rm ac}$ is $|\ov w_{r} (\bp) \ra$.
 Similarly, from $|w_r (\bp) \ra$, one gets an antimode-counterpart state as $|\ov w^r (\bp) \ra$.
 It turns out that these two ways of getting antimode-counterpart state are not equivalent. 
 In fact, they lead to different inner products [see Eq.(\ref{IP-ww-main}) to be given below]. 
 For this reason, we introduce a label $\varrho = \pm$ to make a distinction and write, by definition,
\begin{subequations}\label{bm-ac}
\begin{align}\label{bm-ac+}
 & |{w^r_\varrho}(\bp)\ra_{\rm ac} := |\ov w_{r\varrho} (\bp)\ra \qquad \text{for $\varrho =+$, }
 \\  & |w_{r \varrho}(\bp)\ra_{\rm ac} := |\ov w^r_\varrho (\bp) \ra \qquad \text{for $\varrho =-$. } \label{bm-ac-}
\end{align}
\end{subequations}
 Direct derivation shows that [see Eq.(\ref{bm-ac2-app})]
\begin{subequations}\label{bm-ac2}
\begin{align}\label{}
 \label{bm-ac+2} & |w_{r\varrho}(\bp)\ra_{\rm ac} = -|\ov w^r_\varrho (\bp)\ra \qquad \text{for $\varrho =+$, }
 \\ \label{bm-ac-2} & |w^r_{\varrho}(\bp)\ra_{\rm ac} = -|\ov w_{r\varrho} (\bp)\ra \qquad \text{for $\varrho =-$. }
\end{align}
\end{subequations}
 The label $\varrho$ does not obey the summation convention
 and there is no difference between $\varrho$ in the upper and lower positions. 

\subsection{Dual state space}\label{sect-ACC}

 In this section, we construct spaces dual to $\E_{b}$ and $\E_{\ov b}$, and use them to 
 get products like inner product. 
 To this end, we make use of an operation to be called ACC operation as defined below. 
 \begin{itemize}
  \item An \emph{ACC operation} on a state first changes the state to
  its antimode-counterpart state, then, takes complex conjugation.
\end{itemize}
 Note that the operation of taking antimode-counterpart state is applied to helicity states only
 and does not influence expansion coefficients on helicity states.

 We use $|{w^r}(\bp)\ra_{\rm ACC}$ to indicate the result of an ACC operation on a state $|{w^r}(\bp)\ra$,
 and similar for other spinors. 
 Making use of Eq.(\ref{bm-ac}), one finds the following results of ACC operation on helicity states 
 (Appendix \ref{app-antimode-counterpart}), 
\begin{subequations}\label{hat-w-w-main}
\begin{align}\label{hat-w-w+1-main}
 | {w^r_\varrho}(\bp)\ra_{\rm ACC} & = \varrho | w_{r\varrho}(\bp)\ra,
 \\ | {w_{r\varrho}}(\bp)\ra_{\rm ACC} & = - \varrho |w^{r}_\varrho(\bp)\ra. \label{hat-w-w-1-main}
\end{align}
\end{subequations}
 It is of interest to note that, within an ACC operation, 
 taking antimode-counterpart state and taking complex conjugation are in fact commutable 
 (Appendix \ref{app-antimode-counterpart}).

 Expanding the above-obtained ACC kets on the basis of $|S^A\ra$ and making use of Eq.(\ref{<kappa|-expan})
 to change kets to spinor bras for scalar product, one gets bras which we call \emph{ACC-bras}.
 The ACC-bras involve complex conjugation and, hence, we use to $\la \cdot |$ to indicate them;
 in addition, we use a hat to indicate that ACC operation has been used to get the bras. 
 Thus, according to Eq.(\ref{hat-w-w-main}), we get the following ACC-bras of helicity states,
\begin{subequations}\label{hat-w-2}
\begin{align}\label{hat-w-2-u}
 \la \wh{w^r_\varrho}(\bp)| & = \varrho \la w_{r\varrho}(\bp)|,
 \\ \la  \wh{w_{r\varrho}}(\bp)| & = - \varrho \la w^{r}_\varrho(\bp)|. \label{hat-w-2-d}
\end{align}
\end{subequations}
 Then, making use of Eqs.(\ref{ww-eps}) and Eq.(\ref{hat-w-w-main}) and noting Eq.(\ref{eps-delta}), 
 it is easy to check that
\begin{subequations}\label{IP-ww-main}
\begin{align}
 & \la \wh {w^r_\varrho}(\bp)| {w^s_\varrho}(\bp)\ra = \varrho\ \delta^{rs}, 
 \\ & \la \wh {w_{r\varrho}}(\bp)| {w_{s\varrho}}(\bp)\ra = \varrho\ \delta_{rs}.
\end{align}
\end{subequations}
 For an arbitrary spinor $|\phi\ra$ in the space $\WW$, written as $|\phi\ra = c_{r} |{w^r_\varrho}(\bp)\ra$, 
 the ACC operation gives
\begin{align}\label{wh-kappa}
 |{\phi}\ra_{\rm ACC} & :=  \ov{|\phi\ra_{\rm ac}}  \equiv |\ov \phi\ra_{\rm ac} = \ov c_{ r} |{w^r}(\bp)\ra_{\rm ACC} ,
\end{align}
 with $  \ov c_{ r} \equiv c^*_{ r} $. 
 Then, one has
\begin{align}\label{wh-kappa-bra}
 \la \wh{\phi}| &  = \la \wh{w^r}(\bp)|  \ov c_{ r}. 
\end{align}
 We call the scalar product of an ACC-bra and a ket, say, $\la \wh{\phi}|\phi'\ra$, an \emph{ACC-scalar product}.
 Making use of Eq.(\ref{IP-ww-main}), one finds that the
 ACC-scalar product is an inner product for $\varrho=+$,  
 while, it is a negative inner product for $\varrho=-$.
 (See Appendix \ref{app-inner-product} for a proof in a generic case.)

 Thus, with respect to helicity states, we have constructed a space dual to $\E^S_b$,
 which is spanned by ACC-bras obtained from kets in $\E^S_b$ by the ACC operation. 
 We call such a dual space an \emph{ACC-dual space}. 
 This space satisfies requirements given at the first and second steps discussed in Sec.\ref{sect-dual-state-space}.

 It is easy to see that the direct product of the ACC-dual space and the ordinary dual momentum-state space
 gives a space, which is dual to the state space $\E_b$ of $b$-mode.
 For the sake of brevity in presentation, we introduce the notation of $\wh{\la \bp|}$ and set it as
\begin{align}\label{p-ACC}
   \wh{\la \bp|} : = \la \bp |.
\end{align}
 Then, for an arbitrary vector 
 $|\psi\ra = \int d\ww p \ c_{\bp r} |\bp\ra |{w^r_\varrho}(\bp)\ra$ with coefficients $c_{\bp r} $,  one gets that
 \begin{align}\label{bra-wh-kappa}
 \la \wh{\psi}| & := \int d\ww p \ \ov c_{\bp r} \la \bp | \la \wh{w^r}(\bp) |.
\end{align}
 When momentum states are included, we speak of ACC-bra in the sense of Eq.(\ref{bra-wh-kappa}).

 It is easy to check that the above discussions are also valid for $\ov b$-mode.
 In particular, Eqs.(\ref{bm-ac})-(\ref{IP-ww-main}) are valid under  complex conjugation.

 Finally, since different signs of $\varrho $ correspond to different types of inner product, 
 we assume that modes with different signs of $\varrho$ behave as \emph{different species} of mode.
 For this reason, states with different signs of $\varrho$ are always orthogonal.

\section{Unit modes besides basic modes}\label{sect-other-unit-modes}

 In this section, we discuss unit modes besides basic modes, 
 whose spinor spaces are obtained from direct products of some of $\WW$ and $\ov \WW$. 
 From the angle of spinor, such a unit mode may be regarded as being constructed from some basic modes.
 We assume that, in this understanding, the ``ingredient'' basic modes share a common momentum. 
 Then, for ``ingredient'' basic modes within a unit mode, the fermionic relationship in Eq.(\ref{fermi-BM}) implies that
 \begin{subequations}\label{SAB-ov-commu-two}
\begin{align}
 & |S_A\ra | \ov S_{B'}\ra  = - | \ov S_{B'}\ra |S_A\ra, \label{SA-ovSB-commu}
 \\ \label{SAB-commu}  & |S_A\ra |S_{B}\ra  = - |S_{B}\ra |S_A\ra,
 \\  & |\ov S_{A'}\ra |\ov S_{B'}\ra  = - |\ov S_{B'}\ra |\ov S_{A'}\ra.
\end{align}
\end{subequations}

\subsection{$z$-mode}\label{sect-z-mode}

 In this section, we discuss a unit mode, whose spinor space is obtained from
 the direct product space of $\WW \otimes \ov\WW$.
 As is known in the spinor theory, $\WW \otimes \ov\WW$ is isomorphic to a four-component-vector space,
 which we denote by $\VV$. 
 We call this mode  \emph{$z$-mode}.
 Clearly, the space $\VV$ coincides with its complex conjugate, i.e., $\ov \VV = \VV$,
 and hence $z$-mode is itself's antimode.

 For brevity, sometimes we write $|S_A\ra |\ov S_{B'}\ra$ as $|S_{AB'}\ra$
 and $\la\la \ov S_{B'}| \la\la S_{A}|$ as $\la\la S_{B'A}|$.
 The space $\VV$ is spanned by four vectors $|T^\mu\ra$ of $\mu = 0,1,2,3$, which
 are linked to spinors in $\WW \otimes \ov\WW$ through 
 an \emph{EW-symbol operator} $\sigma$ defined by  
\begin{equation}\label{sigma}
 \sigma := \sigma^{\mu AB'} |T_\mu\ra \la\la S_{B'A}|,
\end{equation}
 where $\sigma^{\mu AB'}$ indicate the so-called Enfeld-van der Waerden symbols,  in short \emph{EW-symbols}
 (see also Appendix \ref{sect-vector-abstract}).
 Scalar products of these vectors are written as $\la\la T^\mu|T^\nu\ra = g^{\mu\nu}$ [see Eq.(\ref{Tmu-Tnu})],
 where $g^{\mu\nu}$ is the Minkovski's metric in Eq.(\ref{gmunu}).

 We use $\bJ_\VV$ to denote the angular-momentum operator that acts on the space $\VV$.
 It should come from the sum of $\bs_b$ and $\bs_{\ov b}$,
 namely, $(\bs_b I_{\ov \WW} + \bs_{\ov b} I_{\WW})$, where $I_\WW$ and $I_{\ov \WW}$
 are identity operators in $\WW$ and $\ov \WW$ [see Eqs.(\ref{I-W}) and (\ref{I-ovW})], respectively. 
 Since $\sigma$ maps $\WW\otimes \ov\WW$ to $\VV$
 and the reverse map is written as $\sigma^{-1} = \sigma^T$ [Eq.(\ref{s-T-s})], $\bJ_\VV$ has the following form, 
\begin{align}\label{J-VV}
 \bJ_\VV \equiv (\bJ_\VV)^{\mu\nu}  |T_\mu\ra  \la\la T_\nu| 
 = \sigma \left( \bs_b I_{\ov \WW} + \bs_{\ov b} I_{\WW} \right) \sigma^T.
\end{align}
 Making use of the explicit expressions of the EW-symbols in Eq.(\ref{sigma^AB}), one finds that (Appendix \ref{app-Jz})
\begin{align}\label{J3-matrix}
 (J^3_\VV)^{\mu\nu} = \left( \begin{array}{cccc} 0 & 0 & 0 & 0 \\ 0 & 0 & i & 0
  \\ 0 & -i & 0 & 0 \\ 0 & 0 & 0 & 0  \end{array} \right).
\end{align}

 We use $|\varepsilon^\lambda(\bk)\ra$ of $\lambda =0,1,2,3$ 
 to denote helicity states of a $z$-mode with a momentum $\bk$ and helicities $h^\lambda_z$, 
\begin{align}\label{}
 H^{\rm cy}_z |\varepsilon^\lambda(\bk)\ra = h^\lambda_z |\varepsilon^\lambda(\bk)\ra,
\end{align}
 where $H^{\rm cy}_z = \bJ_\VV \cdot \bk / |\bk|$.
 For $\bk_0 = (0,0,|\bk|)$, one has
\begin{align}\label{J3-h-lambda-z}
 \left[ \left( J^3_\VV \right)^{\mu\nu}  |T_\mu\ra  \la\la T_\nu | \right] |\varepsilon^\lambda(\bk_0)\ra
 = h^\lambda_z |\varepsilon^\lambda(\bk_0)\ra.
 \end{align}
 Using Eq.(\ref{J3-matrix}), one finds the following eigenstates, 
\begin{subequations}\label{varepsilon-bk_0}
\begin{align}\label{}
 |\varepsilon^0(\bk_0)\ra &= |T^0\ra ,
 \\ |\varepsilon^1(\bk_0)\ra &= \frac{1}{\sqrt 2} (|T^1\ra + i|T^2\ra ),
 \\ |\varepsilon^2(\bk_0)\ra &= \frac{1}{\sqrt 2} (|T^1\ra - i|T^2\ra ),
  \\ |\varepsilon^3(\bk_0)\ra &= |T^3\ra ,
\end{align}
\end{subequations}
 with helicities
\begin{align}\label{varepsilon-helicity}
  h^1_z = 1, \quad h^2_z =-1, \quad h^0_z = h^3_z =0.
\end{align}
 The helicity states $|\varepsilon^\lambda(\bk)\ra$ are then obtained by
 the $SO(3)$ rotation that transforms the direction of $\bk_0$ to that of $\bk$.

 Generically, one may consider a vector $|\psi\ra$,
 which is an eigenstate of both $s_b^k I_{\ov \WW}$ and $s_{\ov b}^k I_{\WW}$, 
 with eigenvalues of $a$ and $b$, respectively, i.e., 
\begin{align}\label{}
 s^k_b I_{\ov \WW}|\psi\ra = a|\psi\ra, \quad  s^k_{\ov b} I_{\WW}|\psi\ra = b|\psi\ra. 
\end{align}
 One easily checks that $|\Psi\ra = \sigma |\psi\ra$ is an eigenstate of $J^k_\VV$ with an eigenvalue $(a+b)$, i.e., 
\begin{align}\label{Jk-Psi}
 J^k_\VV |\Psi\ra = (a+b) |\Psi\ra.
\end{align}
 The helicity states $|\varepsilon^\lambda(\bk_0)\ra$ in Eq.(\ref{varepsilon-bk_0}) may also be written in this form.
 Indeed, from Eqs.(\ref{varepsilon-bk_0}) and (\ref{|T>-|SS'>}), one finds that
\begin{subequations}\label{varepsilon-S}
\begin{align}\label{}
 |\varepsilon^0(\bk_0)\ra &= \frac{1}{\sqrt 2} \sigma ( |S^{11'}\ra + |S^{00'}\ra ) ,
 \\ |\varepsilon^1(\bk_0)\ra &= - {\sqrt 2} \sigma  |S^{10'}\ra,
 \\ |\varepsilon^2(\bk_0)\ra &= -{\sqrt 2} \sigma  |S^{01'}\ra,
  \\ |\varepsilon^3(\bk_0)\ra &= \frac{1}{\sqrt 2} \sigma (|S^{11'}\ra - |S^{00'}\ra);
\end{align}
\end{subequations}
 then, making use of Eqs.(\ref{sk=-sk'}) and (\ref{wr-SA}), one checks that 
 $|\varepsilon^\lambda(\bk_0)\ra$ may also be written in the form of $|\Psi\ra = \sigma |\psi\ra$.

 Finally, we discuss ACC-bras. 
 Since  the antimode of a $z$-mode is itself,
 the antimode-counterpart state of $|\varepsilon^{\lambda}(\bk)\ra$ is just itself, i.e.,
 $|\varepsilon^{\lambda}(\bk)\ra_{\rm ac} = |\varepsilon^{\lambda}(\bk)\ra$.
 Then, one finds that
\begin{align}\label{whvep=ovvep}
 \la \wh{\varepsilon^{\lambda}}(\bk)| = \la\la \ov\varepsilon^{\lambda}(\bk)|.
\end{align}
 Due to the relation of $|\ov T^\mu\ra = -|T^\mu\ra$ in Eq.(\ref{ovT=T}), Eq.(\ref{varepsilon-bk_0}) implies that
\begin{subequations}\label{ov-varepsilon}
\begin{align}\label{}
 |\ov{\varepsilon}^0(\bk)\ra  =-|\varepsilon^0(\bk)\ra,
 \quad |\ov{\varepsilon}^3(\bk)\ra  =-|\varepsilon^3(\bk)\ra,
 \\  |\ov{\varepsilon}^1(\bk)\ra  =  -|\varepsilon^2(\bk)\ra,
 \quad |\ov{\varepsilon}^2(\bk)\ra  = -|\varepsilon^1(\bk)\ra.
\end{align}
\end{subequations}
 We use $\wh g^{\lambda \lambda'}$ to indicate ACC-scalar products of the $|\varepsilon\ra$-vectors, i.e.,
\begin{align}\label{<wh-vep|vep>}
 \wh g^{\lambda \lambda'} := \la \wh{ \varepsilon^\lambda}(\bk)|\varepsilon^{\lambda'}(\bk)\ra.
\end{align}
 Making use of Eq.(\ref{ovT=T}), it is easy to check that
\begin{align}\label{wh-g-g}
 \wh g^{\lambda \lambda'}  = - g^{\mu \mu'} \quad \text{with $\lambda =\mu, \lambda'=\mu'$}.
\end{align}
 The direct product of the ACC-bra space spanned by $\la \wh{\varepsilon^{\lambda}}(\bk)|$ 
 and the space spanned by $\la \bk|$ gives a space dual to $\E_z$.

\subsection{s-mode, $\ov s$-mode, and others }\label{sect-s-mode}

 In this section, we first discuss unit modes, whose spinor spaces, denoted by  $\SP$ and $\ov \SP$,
 are related to the two product spaces of  $\WW \otimes \WW$ and  $\ov\WW \otimes \ov\WW$, respectively.
 Then, we argue that there is no further unit mode of interest physically.

 We call modes, whose spinor spaces are  $\SP$ and $\ov \SP$, \emph{$s$-mode} and \emph{$\ov s$-mode}, respectively. 
 As a consequence of Eq.(\ref{SAB-commu}), 
 $|S_0\ra |S_0\ra = |S_1\ra |S_1\ra =0$, $|S_1\ra |S_0\ra = - |S_0\ra |S_1\ra $,
 and similar for their complex conjugates. 
 This implies that, say, $\SP$ may not be isomorphic to $\WW \otimes \WW$. 
 In fact, both $\SP$ and $\ov \SP$ are one-dimensional $SL(2,C)$-scalar spaces (see Appendix \ref{app-scalar-mode}).
 We use $|s\ra$ and $|\ov s \ra $ to denote basis scalars in $\SP$ and $\ov \SP$, respectively.

 In fact, the physical requirement of fermionicness, from which Eq.(\ref{SAB-ov-commu-two}) is gotten, 
 lies outside the mathematical theory of spinors. 
 It may render unusual properties to $s(\ov s)$-mode.
 In particular, a scalar space like $\SP$ may not accommodate any nontrivial angular momentum operator,
 while, the space $\WW^{(1)} \otimes \WW^{(2)}$ has a nontrivial angular momentum operator 
 written as $ \left( \bs^{(1)}_b I^{(2)}_{\WW} + \bs^{(1)}_b I^{(2)}_{\WW} \right)$.
 For this reason, at the present stage, it is unclear what type of angular-momentum property should be 
 assigned to $s(\ov s)$-mode.
 Anyway, both $s$-mode and $\ov s$-mode should be bosonic.

 In discussions to be given below, 
 we assume that physical impact of the above-discussed indefiniteness of $s(\ov s)$-mode
 should be reduced as much as possible. 
 More exactly, we assume that
 (i) $s$-mode and $\ov s$-mode may not exist independently, 
 (ii) they do not need to satisfy the requirement of identical layer-helicity stated in BA2, 
 and (iii) they have vanishing ACC-scalar products, i.e., 
\begin{align}\label{<s|s>=0}
 \la \wh s|s\ra =\la \wh {\ov s}|\ov s\ra =0. 
\end{align}
 Due to the vanishing ACC-scalar products in Eq.(\ref{<s|s>=0}), 
 $s(\ov s)$-mode may not be detected directly by experiments.

 Finally, we argue that there is no further unit mode of interest physically. 
 Clearly, due to the relation in Eq.(\ref{SAB-commu}), 
 no unit mode may be constructed from either three $b$-modes, or from three $\ov b$-modes.
 Moreover, a product of $|S_A\ra |S_{B}\ra |\ov S_{C'}\ra$ is not of interest physically,
 because it must be proportional to $|s\ra |\ov S_{C'}\ra$,
 which transforms likes a Weyl spinor $ |\ov S_{C'}\ra$, but, 
 always has a vanishing ACC-scalar product due to Eq.(\ref{<s|s>=0}). 
 By similar arguments, products of more than three of $\WW$ and $\ov \WW$  are not of interest, either. 
 
\section{Multilayer modes}\label{sect-compound-modes}

 In this section, we discuss multi-layer modes. 
 Specifically, we give some preliminary discussions in Sec.\ref{sect-angular-m-r}, 
 discuss fermionic and bosonic modes in Sec.\ref{sect-BM-layer-modes} and Sec.\ref{sect-bosonic-modes},
 respectively, and give identity operators for states of modes in Sec.\ref{sect-identity}. 
 According to  previous discussions, there are only five species of unit mode,
 whose spinor states may be used as mode layers,
 i.e., $b$-mode, $\ov b$-mode, $z$-mode, $s$-mode, and $\ov s$-mode;
 we indicate the corresponding layers as
 $b$-layer, $\ov b$-layer, $z$-layer, $s$-layer, and $\ov s$-layer, respectively. 
 
\subsection{Preliminary discussions}\label{sect-angular-m-r}

 In this section, we first discuss a notation to be used,
 then,  give a formal discussion for layer states,
 and, finally,  generalize Eq.(\ref{sk=-sk'}) to a generic mode.

\noindent 1. \emph{Notation of $\alpha$}.
\\  
 When writing helicity states of multilayer modes, it is inconvenient to use the index $r$.
 For example, for a fermionic mode consisting of one $b$-layer and one $\ov b$-layer, 
 if the $b$-layer lies in a state $|w^r_\varrho(\bp)\ra$, 
 then, the $\ov b$-layer should lie in a state $|\ov w_{r,\varrho}(\bp)\ra$ with the label $r$ in the lower position,
 such that the two layers possess the same helicity obeying BA2 [cf.~Eqs.(\ref{hr-b}) and (\ref{w^r_s})]. 
 The inconvenience lies in that
 the helicity label $r$ for the whole mode does not have a definite position.
 In addition, the label $\varrho$ should also be indicated when writing a state of a fermionic mode.

 To solve the above problems, 
 we introduce an index $\alpha$, $\alpha := (r,\varrho)$, and use it in the following way
 of indexing helicity states, 
\begin{subequations}\label{uv-alpha}
\begin{gather}\label{}
 |u^\alpha(\bp)\ra := |w^r_\varrho(\bp)\ra ,
 \\ |v^\alpha(\bp)\ra := |w_{r\varrho}(\bp)\ra .
\end{gather}
\end{subequations}
 Clearly,  for the mode in the above-discussed example, the label $\alpha$ has a definite (upper) position. 
 
 Two remarks: 
 (i) $|u^\alpha(\bp) \ra$ and $|\ov v^\alpha(\bp)\ra$ \emph{have the same helicity},
 in contrast to that $|w^r(\bp)\ra$ and $|\ov w^r(\bp)\ra$ have opposite helicities. 
 And, (ii) since the label $\varrho$ does not obey the summation convention, 
 repeated $\alpha$ implies a summation only over $r\in \alpha$, not including $\varrho$.

 From Eqs.(\ref{hat-w-2}) and (\ref{uv-alpha}), one sees that
\begin{subequations}\label{whu-v}
\begin{gather}\label{hat-u-v}
 \la \wh{u}^{\alpha}(\bp)| = \varrho(\alpha) \la v^{\alpha}(\bp)|,
 \\  \la \wh{v}^{\alpha}(\bp)| = - \varrho(\alpha) \la u^{\alpha}(\bp)| , \label{hat-u-v-2}
\end{gather}
\end{subequations}
 where $\varrho(\alpha)$ indicates the sign $\varrho$ in $\alpha$.
 Making use of Eqs.(\ref{uv-alpha}), (\ref{ww-eps}), and (\ref{eps-delta}), it is direct to check that
\begin{align}\label{<v|u>=del}
 \la v^\alpha(\bp) | u^\beta(\bp)\ra = -\la u^\beta(\bp) | v^\alpha(\bp)\ra = \delta^{\alpha \beta},
\end{align}
 where $\beta = (s,\varrho')$
 and $\delta^{\alpha \beta} := \delta^{rs} \delta^{\varrho \varrho'}$, as well as that
\begin{gather}\label{IP-uv}
 \la \wh{u}^{\alpha}(\bp)|{u^{\beta}(\bp)}\ra = \la \wh{v}^{\alpha}(\bp)|{v^{\beta}}(\bp)\ra =
 \varrho(\alpha) \delta^{\alpha \beta}.
\end{gather}

\vspace{0.2cm}
\noindent  2. \emph{Layer states}.
\\  
 W use $e^{i\vartheta_m^M} |\LL_m^M\ra$ to indicate the spinor state of the $m$-th layer of a mode $M$.
 We call $\vartheta_m^M$ the \emph{layer phase} and call $|\LL_m^M\ra$ the \emph{layer spinor}. 
 For a given mode, BA2 requires that $|\LL_m^M\ra$ should possess the same helicity for all $m$. 

\vspace{0.2cm}
\noindent 3. \emph{An angular-momentum relation for generic antimodes}.
 \\  
 For a generic mode $M$, 
 we use $\E_M^{S -\text{ext}}$ to indicate the space, which is spanned by
 basis spinors like those of $\E^S_M$, but, not obeying the requirement of identical layer helicity imposed by BA2. 
 Clearly, this space $\E_M^{S -\text{ext}}$ is just the direct sum of all those spaces, 
 each corresponding to one layer of $M$. 
 We use $\bS_M \equiv  (S^1_M,S^2_M,S_M^3)$ 
 to indicate the spinor angular-momentum operator on the space $\E_M^{S -\text{ext}}$.
 Similarly, for the antimode $\ov M$, one has a space $\E_{\ov M}^{S -\text{ext}}$
 and an operator $\bS_{\ov M}$. 
 A key point is that $\E_{\ov M}^{S -\text{ext}}$ is the complex conjugate of $\E_{M}^{S -\text{ext}}$.

 Below, we show that Eq.(\ref{sk=-sk'}) is valid in this generic case.
 Consider an arbitrary element of the group $SU(2)$,
 which is written as ${\cal R}_M$ when acting on the spinor space $\E_M^{S -\text{ext}}$,
 while, is written as $\R_{\ov M}$ for $\E_{\ov M}^{S -\text{ext}}$.
 As is well known, $\R_M$ is characterized by three real parameters $\theta_k$
 and is written as
\begin{gather}\label{R-ang}
 {\cal R}_M = e^{-i\theta_k S_M^k}.
\end{gather}
 For the operator $\R_{\ov M}$, on one hand, it should be given by the complex conjugate of $\R_M$,
 i.e., $\R_{\ov M} = e^{ i\theta_k {\ov S}^k_M}$;
 on the other hand, it should be characterized by the same parameters $\theta_k$
 with respect to the operator $\bS_{\ov M}$,
 that is,  written as $\R_{\ov M}= e^{-i\theta_k S_{\ov M}^k}$.
 Then, one sees that the angular-momentum operators of these two modes should satisfy
 an equation like Eq.(\ref{sk=-sk'}), i.e.,
\begin{gather}\label{sk-sk'}
 \bS_{\ov M}= - \ov \bS_{M}.
\end{gather}

\subsection{Fermionic modes}\label{sect-BM-layer-modes}

 In this section, we discuss states of multilayer fermionic modes.
 Layers of these modes are either $b$-layer or $\ov b$-layer.

 We use $f$ to indicate a fermionic mode, whose first layer lies in a space $\WW$.
 Under a given momentum $\bp$ and a given label $\alpha$,  we write the top-layer's spinor as $|u^\alpha(\bp)\ra$.
 Then, the state of the mode, denoted by $|f^{\alpha}_{ \bp  }\ra$, is written as
\begin{gather}
 |f^{\alpha}_{ \bp  }\ra  =  |\bp\ra |\cs_f^{\alpha }(\bp)\ra, \label{|f-bp-al>}
\end{gather}
 where $|\cs_f^{\alpha }(\bp)\ra$ indicates its spinor part, 
\begin{gather}\label{Sf}
 |\cs_{f}^{\alpha }(\bp)\ra 
 = \frac{1}{\sqrt{n_f}} \left( \begin{array}{c}  e^{i\vartheta_1^f} |u^\alpha(\bp)\ra
 \\ e^{i\vartheta_2^f} |\LL^f_2\ra 
 \\ \vdots    \\ e^{i\vartheta_{n_f}^f} |\LL^f_{n_f}\ra \end{array} \right)
\end{gather}
 with $n_f$ the number of layers of $f$.
 According to BA2, layer spinors of other layers with $\WW$ 
 should be written as $|u^\alpha(\bp)\ra$, too, while,  those with $\ov \WW$ as $|\ov v^\alpha(\bp)\ra$.
 Thus, we have 
\begin{align}\label{LLf}
 & |\LL^f_m \ra = |u^\alpha(\bp) \ra \ \text{or} \ |\ov v^\alpha(\bp)\ra \quad 
 (|\LL^f_1 \ra = |u^\alpha(\bp) \ra). 
\end{align}

 The corresponding state of the antimode $\ov f$ is written as 
 $|\ov f^{\alpha}_{ \bp  }\ra  =  |\bp\ra |\cs_{\ov f}^{\alpha }(\bp)\ra  \label{|ovf-bp-al>}$.
 For $|\cs_{\ov f}^{\alpha }(\bp)\ra$ and $|\cs_{f}^{\alpha }(\bp)\ra$ to have the same helicity,
 one needs to set the first layer spinor of $|\cs_{\ov f}^{\alpha }(\bp)\ra$ as $|\ov v^\alpha(\bp)\ra$.
 Thus,  we write
\begin{gather} \label{Sovf}
|\cs_{\ov f}^{\alpha }(\bp)\ra 
 = \frac{1}{\sqrt{n_f}} \left( \begin{array}{c}  e^{i\vartheta_1^{\ov f} } |\ov v^\alpha(\bp)\ra
 \\ e^{i\vartheta_2^{\ov f}} |\LL^{\ov f}_2\ra 
 \\ \vdots  \\ e^{i\vartheta_{n_f}^{\ov f}} |\LL^{\ov f}_{n_f}\ra  \end{array} \right),
\end{gather}
 where $ |\LL^{\ov f}_m \ra = |\ov v^\alpha(\bp)\ra  \ \text{or} \ |u^\alpha(\bp) \ra$.

 Now, we discuss ACC-bras of the above states.
 Note that the previously given definitions of antimode-counterpart state
 and of ACC operation are valid for an arbitrary mode $M$. 
 When taking the antimode-counterpart state of $|\cs_f^{\alpha }(\bp)\ra$,
 the layer phases should be changed to those of its antimode $\ov f$.
 This gives that
\begin{gather}\label{Sf-anti}
 |\cs_{f}^{\alpha }(\bp)\ra_{\rm ac}
 = \frac{1}{\sqrt{n_f}} \left( \begin{array}{c}  e^{i\vartheta_1^{\ov f}} |\LL^f_1\ra_{\rm ac}
 \\ \vdots    \\ e^{i\vartheta_{n}^{\ov f}} |\LL^f_{n_f}\ra_{\rm ac} \end{array} \right).
\end{gather}
 Then,  one gets that
\begin{gather}\label{}
 \la \wh{\cs}^{\alpha }_f(\bp)| = \frac{1}{\sqrt{n_f}}
 \left(e^{-i\vartheta_1^{\ov f}} \la \wh\LL^f_1| , e^{-i\vartheta_2^{\ov f}} \la \wh\LL^f_2| , \cdots   \right).
\end{gather}
 For these states to possess the same scalar products as those in Eq.(\ref{IP-uv}), 
 the two modes of $f$ and $\ov f$ must share the same layer phases, i.e., 
\begin{align}\label{theta-f-ovf}
 \vartheta_m^{f}= \vartheta_m^{\ov f} \qquad \forall m. 
\end{align}
 Then,  one gets the following scalar product, 
 \footnote{In the case of a two-layer mode of $\varrho=+$, with one $b$-layer and one $\ov b$-layer, 
 it is straightforward to check that this product is effectively 
 the same as the ordinary inner product for Dirac spinors.}
\begin{gather}\label{<S|S>-BMlayer}
   \la \wh \cs_f^\beta(\bp) |\cs_f^\alpha(\bp) \ra = \varrho(\alpha) \delta^{\beta \alpha }.
\end{gather}
 It is not difficult to see that the above results are also valid for the antimode $\ov f$.

 To summarize, for an arbitrary fermionic mode $F$, one has
\begin{gather}\label{SP-pb-b}
 \la \wh{F}^{\beta}(\bq)|F^{\alpha}(\bp)\ra
 = \varrho(\alpha) |\bp| \delta^3(\bp-\bq)\delta^{\alpha \beta}.
\end{gather}
 This ACC-scalar product is an inner product for $\varrho=+$,
 while, it is a negative inner product for $\varrho=-$ (Appendix \ref{app-inner-product}).

\subsection{Bosonic modes}\label{sect-bosonic-modes}

 In this section, we discuss spinor states of multilayer bosonic modes,
 the layers of which lie in $\VV$, or $\SP$, or $ \ov\SP$.

 In the construction of a bosonic mode, further attention should be paid to the space $\VV$.
 In fact, although $\VV$ is its own complex conjugate,
 the helicity states $|\varepsilon^\lambda(\bk) \ra $ of $\lambda =1,2$ 
 are not equal to their complex conjugates $|\ov\varepsilon^\lambda(\bk) \ra$ [Eq.(\ref{ov-varepsilon})].
 We assume that these two cases should be distinguished,
 more exactly, there may exist two types of four-component vector layers within one bosonic mode, 
 which we indicate as \emph{$z$-layer} and \emph{$\ov z$-layer}, respectively.

 We use $|\varepsilon^\lambda(\bk) \ra $ to indicate the spinor state of a $z$-layer in a bosonic mode. 
 According to BA2, the layer spinor of an $\ov z$-layer in this mode
 should be described by the vector $|\varepsilon^\lambda(\bk) \ra $, too. 
 To make a distinction, we use $|\varpi^\lambda(\bk)\ra$ to indicate the spinor state of the $\ov z$-layer, with
\begin{align}\label{varpi}
 |\varpi^\lambda(\bk)\ra \equiv |\varepsilon^\lambda(\bk)\ra.
\end{align}
 One should note that $z$-layer and $\ov z$-layer have a relative difference only
 and one needs to distinguish between them only when they appear within a same bosonic mode. 
 For example, there is no essential difference between the two configurations of 
 $\left( \begin{array}{c} |\varepsilon^\lambda(\bk) \ra \\ |\varepsilon^\lambda(\bk) \ra \end{array} \right)$
 and $\left( \begin{array}{c} |\varpi^\lambda(\bk) \ra \\ |\varpi^\lambda(\bk) \ra \end{array} \right)$.

 Thus, the spinor state of a bosonic mode $|B\ra$, whose first, second, and third layers are 
 $z$-layer, $\ov z$-layer, and $s$-layer, respectively, is written as
\begin{gather}\label{cs-B}
 |\cs_{B}^{\lambda }(\bk)\ra = \frac{1}{\sqrt{n_V}}
 \left( \begin{array}{c} e^{i\vartheta_1^B}|\varepsilon^{\lambda}(\bk)\ra 
 \\ e^{i\vartheta_2^B} |\varpi^{\lambda}(\bk)\ra
 \\  |s\ra \\ \vdots   \end{array} \right),
\end{gather}
 where $n_V$ indicates the number of the $z(\ov z)$-layers. 
 The corresponding ACC-bra is written as
\begin{align} \label{wh-S-Bk}
 \la \wh{\cs_{B}^{\lambda}}(\bk)|  = \frac{1}{\sqrt{n_V}} \big( e^{-i\vartheta_1^{\ov B}} \la \wh{ \varepsilon^{\lambda}}(\bk)| ,  
    e^{-i\vartheta_2^{\ov B}} \la \wh{ \varpi^{\lambda}}|,    \la \wh s|, ... \big). 
\end{align}
 Similarly to Eq.(\ref{theta-f-ovf}), one has $\vartheta_m^{B}= \vartheta_m^{\ov B}$ for all $m$.
 Then, making use of Eqs.(\ref{<wh-vep|vep>}) and (\ref{<s|s>=0}), one finds that
\begin{gather}\label{<S|S>-VSlayer}
 \la \wh{\cs_{B}^{\lambda }}(\bk)|\cs_{B}^{\lambda' }(\bk) \ra =\wh g^{\lambda \lambda'}.
\end{gather}

\subsection{Identity operators}\label{sect-identity}

 In this section, we discuss identity operators that act on state spaces of one and two modes. 
 We extend usage of the index $\alpha$ to a generic mode $M$, defined by
\begin{align}\label{}
 \alpha := \left\{            \begin{array}{ll}
              (r,\varrho) & \hbox{for a fermionic mode,} \\
              \lambda & \hbox{for  a bosonic mode.}
            \end{array}          \right.
\end{align}
 States of  $M$ are then written as
\begin{gather}\label{state-M}
 |M^\alpha_{\bp }\ra = |\bp\ra |\cs_M^{\alpha}(\bp)\ra.
\end{gather}
 From Eqs.(\ref{IP-pb}), (\ref{<S|S>-BMlayer}), and (\ref{<S|S>-VSlayer}), one gets that
\begin{align}\label{Mstate-SP}
 \la \wh M^\alpha_{\bp }|M^\beta_{\bq }\ra = |\bp| \delta^3(\bp-\bq) \Upsilon^{\alpha \beta},
\end{align}
 where $\Upsilon^{\alpha\beta}$ indicates a symbol defined by
\begin{align}\label{Upsilon-ab}
 \Upsilon^{\alpha \beta} :=
 \left\{   \begin{array}{ll}
     \varrho(\alpha) \delta^{\alpha \beta} &  \hbox{for fermionic modes,} \\
     \wh g^{\alpha \beta} &  \hbox{for bosonic modes.}
   \end{array} \right.
\end{align} 
 It is easy to verify that $ \label{} \Upsilon^{\alpha \beta} = \Upsilon^{\beta \alpha }$ and
\begin{align} \label{Up-delta}
  \Upsilon^{\alpha \beta} = \Upsilon(\alpha, \alpha) \delta^{\alpha \beta},
 \ \  \left( \Upsilon(\alpha, \alpha) \right)^2 =1, 
\end{align}
 where $\Upsilon(\alpha, \alpha)$ indicates $\Upsilon^{\alpha \alpha}$ without taking summation. 
 Furthermore, introducing a symbol $\Upsilon_{\alpha\beta}$, 
 which by definition has the same matrix as $\Upsilon^{\alpha\beta}$,
 one may raise and lower the index $\alpha$. 
 For example, 
\begin{align}\label{alpha-raise}
 |\psi_\alpha\ra = \Upsilon_{ \alpha \beta} |\psi^\beta\ra,
 \quad |\psi^\alpha\ra = \Upsilon^{\alpha \beta} |\psi_\beta\ra.
\end{align}
 Note that, to get the relation between a label $\alpha$ in the lower position (e.g., in $|\psi_\alpha\ra$)
 and a helicity label $r$, one first needs to relate it to a label $\beta$ in the upper position,
 then, makes use of Eq.(\ref{uv-alpha}). 
 To avoid potential confusion, when the notation of $\alpha$ is used, 
 we never change position of the label $r$ involved.

 It is easy to check that
\begin{equation}\label{F-ab}
  {X_{\ldots}^{\ \ \ \alpha}\ (Y)^{  \cdots }}_{ \alpha } =  {X_{\ldots \alpha}\ (Y)^{\cdots \alpha}},
\end{equation}
 i.e., exchange of positions of repeated $\alpha$ brings no change.
 Moreover, using Eq.(\ref{Up-delta}), one finds that 
\begin{align}\label{Upsilon-delta}
 \Upsilon_{\alpha}^{\ \ \beta} = \Upsilon^{\alpha' \beta } \Upsilon_{\alpha' \alpha} = \delta_\alpha^\beta,
 \quad \Upsilon^{\alpha}_{\ \  \beta} =  \delta^\alpha_\beta. 
\end{align}
 and this implies that
\begin{align}\label{Mstate-IP}
 \la \wh M_{\bp \alpha }|M^\beta_{\bq }\ra = |\bp| \delta^3(\bp-\bq) \delta_\alpha^\beta.
\end{align}

 Making use of properties discussed above, 
 it is straightforward to verify that the identity operator that acts on $\E_M$, denoted by  $I_M$, is written as
\begin{align} \label{IM}
 I_{M}   =  \int d\ww p  |M^\alpha_{\bp }\ra \la \wh{M}_{\bp \alpha}|.
\end{align}
 For the space $\E_M  \otimes \E_{M'}$ of two different modes $M$ and $M'$, 
 the identity operator is written as
\begin{align} \label{IMM'}
 I_{MM'}   =  \int d\ww p d\ww p' |M^\alpha_{\bp } {M'}^{\alpha'}_{\bp' }\ra 
 \la {\wh{M'}}_{\bp' \alpha' } \wh{M}_{\bp \alpha}|.
\end{align}

\section{H-operators}\label{sect-$H$-operator}

 In this section, we discuss $H$-operators of fundamental processes,
 which map the state spaces of incoming modes to those of outgoing modes in  straightforward ways. 
 In Sec.\ref{sect-H-generic}, we give generic discussions. 
 Expressions of $H$-operators of free process and of FIP are derived in Secs.\ref{sect-free-process} and \ref{sect-FIP}, respectively. 
 In Sec.\ref{sect-concise-H-FIP}, we write the $H$-operator of FIP in a concise form. 
 $H$-operators of VF and of BPIPs are discussed in Sec.\ref{sect-H-BPIP}.
 Finally, field expressions of the $H$-operators of free processes and of $H_{\rm int}$ are derived in Sec.\ref{sect-Hint-field}.

 \subsection{Generic form of H-operators}\label{sect-H-generic}

 We use $X$ to indicate the set of modes that enters into a fundamental process
 and use $Y$ for the outgoing set.
 States of modes in the two sets are written as $|X^\eta_\bP \ra = |\bP\ra |\cs_{X}^{ \eta}(\bP) \ra$
 and $|Y^\xi_\bQ \ra = |\bQ\ra |\cs_{Y}^{ \xi}(\bQ) \ra$, 
 lying in spaces denoted by $\E_X$ and $\E_Y$, respectively, 
 where $\bP = \bp$ for one mode and $\bP = (\bp,\bp')$ for two modes, similar for $\bQ$, 
 and $\eta$ and $\xi$ represent the corresponding spinor indices.
 Exact meanings of  $X$ and $Y$ are as follows.
\begin{itemize}
  \item Free process: $X=Y=M$.
  \item VF: $X=|0\ra $ and $Y = (f,\ov f)$ for emergence of a VF pair,
  where $|0\ra $ indicates the vacuum state, and  the reverse is for vanishing of a VF pair. 
  \item FIP: $X=f \ov f'$ and $Y=B$ for an FIP-$1$, and the reverse is for an FIP-$2$. 
\end{itemize}
 Thus, more exactly, we write $H_{\rm FP}$ as $H^{X\to Y}_{\rm FP}$, which maps $\E_X$ to $\E_Y$.

 Below in this section, we discuss $H^{X\to Y}_{\rm free }$ and $H^{X\to Y}_{\rm FIP}$. 
 To carry out the maps they describe, these two $H$-operators should be written in a form like
\begin{align}\label{|Y><X|}
 |Y^\xi_\bQ \ra \cdots \la \wh{X}^\eta_\bP |,
\end{align}
 where the bra is written as an ACC-bra, because we need it to annihilate a state vector in the space $\E_X$.  
 The simplest operator that includes the ACC-bra in (\ref{|Y><X|}) is the identity operator that acts on the space $\E_X$,  
 which we denote by $I_X$.
 According to Eqs.(\ref{IM})-(\ref{IMM'}), it is written as
\begin{align}\label{IX}
 I_X  
 = \int d\ww \bP |\bP\ra |\cs_{X}^{ \eta}(\bP) \ra  \la\wh\cs_{X \eta}(\bP)| \la \bP|.
\end{align}
 We use $P_Y$ to indicate the operator that includes the ket $|Y^\xi_\bQ \ra$ in (\ref{|Y><X|}) and write
\begin{align}\label{H-gfp-origin}
 H^{X\to Y}_{\rm FP} = P_Y \C^{X\to Y} I_X,
\end{align}
 where $\C^{X\to Y}$ is an operator that accomplishes the main mapping function.

 For $Y$ as a single mode $M$, it is natural to take $P_Y$ as the identity operator in Eq.(\ref{IM}),
 i.e., $P_Y = I_Y$. 
 While, for $Y$ as two fermionic modes, although the identity operator in Eq.(\ref{IMM'}) is a simple choice, 
 one may not say that it is the only natural choice,
 as to be discussed later at the end of Sec.\ref{sect-FIP}.

 It is natural to assume that $\C^{X\to Y}$ is the product of a momentum part
 denoted by $\C^{X\to Y}_{\rm mom}$ and a spinor part denoted by $\C^{X\to Y}_{\rm sp}$, i.e., 
\begin{align}\label{C-XY}
 \C^{X\to Y} =\C^{X\to Y}_{\rm mom} \C^{X\to Y}_{\rm sp}.
\end{align}
 Then, for $Y$ as a single mode, we get that
\begin{align}\label{H-operator}
 H^{X\to Y}_{\rm FP} = \int d\ww \bP d\ww \bQ |{Y}^{\xi}_{\bQ} \ra \ C^{X\to Y}_{\bP \bQ}  C^{X\to Y}_{ \eta \xi}
 \la {{\wh X}}_{\bP}^{ \eta} |,
\end{align}
 where 
\begin{subequations}\label{C-ms}
\begin{align}\label{}
 & C^{X\to Y}_{\bP \bQ} = \la \bQ | \C^{X\to Y}_{\rm mom} |\bP\ra,
 \\ & C^{X\to Y}_{ \eta \xi} = \la \wh{\cs}_{ Y \xi}(\bQ) | \C^{X\to Y}_{\rm sp} |\cs_{X \eta}(\bP) \ra . \label{C-XY-etaxi}
\end{align}
\end{subequations}
 When computing the momentum and spinor parts of the $H$-operator amplitude in Eq.(\ref{C-ms}), 
 regarding commutability of mode states [cf.~Eq.(\ref{commut-FB})], 
 the most natural assumption is that spinor states satisfy Eq.(\ref{SAB-ov-commu-two}) like that in the 
 case of unit mode, while, momentum states are commutable. 
 
\subsection{H-operator for free process}\label{sect-free-process}

 For brevity, we use $H^0_M$ to indicate $H^{M\to M}_{\rm free}$.
 Under momentum conservation, we write $\C^{X\to Y}_{\rm mom}$ as
\begin{align}\label{C-momentum-for-H0-1}
 \C^{M\to M}_{\rm mom} = \int d\ww p d\ww q |\bq\ra  \lambda_C |\bp|
 \delta^3 \left( \bp - \bq \right)  \la \bp|.
\end{align}
 For $\C^{M\to M}_{\rm mom}$ to have the dimension of momentum,
 the simplest choice of $\lambda_C$ is that $\lambda_C=|\bp|$.
 Then, we get that
\begin{align}\label{C-momentum-for-H0}
 \C^{M\to M}_{\rm mom} = \int d\ww p |\bp\ra   |\bp|  \la \bp|.
\end{align}
 Meanwhile,  the simplest choice for the spinor part is that $\C^{M\to M}_{\rm sp}=1$,
 which gives that $C^{M\to M}_{\alpha \beta} = \Upsilon_{\beta \alpha}$.
 As a result, by direct derivation (Appendix \ref{app-derive-eqs-free-process}),  from Eq.(\ref{H-gfp-origin}) one gets that
\begin{gather}\label{H0M}
 H_M^0  = \int d\ww p \   |M^\alpha_{\bp }\ra  |\bp| \la  \wh M_{\bp \alpha}|.
\end{gather}

 Let us compute the value of $p^0$ for a mode $M$ in a state $|M^\alpha_\bp\ra$, as defined in Eq.(\ref{p0-def}).
 Due to the $\delta$-function in the normalization condition of $|M^\alpha_\bp\ra$ in Eq.(\ref{Mstate-SP}),
 $p^0$ should be written as
\begin{align}\label{p0-M}
 p^0 = \int d\ww q \la \wh M^\beta_\bq|H^0_M|M^\alpha_\bp\ra \quad \text{with $\beta = \alpha$}.
\end{align}
 Making use of Eq.(\ref{Mstate-SP}), direct computation gives that  (Appendix \ref{app-derive-eqs-free-process})
\begin{align}\label{p0-M-1}
 p^0 = \Upsilon(\alpha, \alpha)  |\bp|.
\end{align}
 For a fermionic mode $F$ with $\Upsilon(\alpha, \alpha)$ given in Eq.(\ref{Upsilon-ab}), 
 one finds that $p^0 = \varrho(\alpha) |\bp|$ and, hence, 
\begin{align}\label{}
 \text{sign of $p^0$} = \varrho \quad \text{for fermionic modes}.
\end{align}

 \subsection{H-operators for  FIPs}\label{sect-FIP}

 In this section, we discuss $H$-operators of  FIPs. 
 To satisfy momentum conservation,  the operator $\C^{X\to Y}_{\rm mom}$ of an FIP should contain the term of
 $|\bP_\Sigma| \delta^3(\bP_\Sigma - \bQ_\Sigma)$, where $\bP_\Sigma = \bp$
 for one mode and  $\bP_\Sigma =\bp+\bp'$ for two modes, and similar for $\bQ_\Sigma$.
 With such a term, $\C^{X\to Y}_{\rm mom}$ automatically gets the dimension of momentum.
 Hence, its simplest form is written as
\begin{align}\label{C-momentum}
 \C^{X\to Y}_{\rm mom} = \int d\ww\bP d\ww\bQ |\bQ\ra  |\bP_\Sigma|
 \delta^3 \left( \bP_\Sigma - \bQ_\Sigma \right)  \la \bP|.
\end{align}

 Let us first discuss FIP-$1$. 
 Consider an FIP-$1$ of $f\ov f' \to B$, 
 with both $f$ and $f'$ possessing $n_f$ layers and $B$ possessing $n_B$  layers. 
 For this process, $Y$ is a single mode $B$ and, hence, $P_Y = I_B$.
 Then, making use of Eqs.(\ref{H-operator}), (\ref{C-XY-etaxi}), and (\ref{C-momentum}), 
 from Eq.(\ref{H-gfp-origin}) one gets that
\begin{align}
 H_{\rm FIP}^{f \ov f'\to B}  =  \int d\Omega \
 |B^\lambda_\bk\ra \delta^3_P C^{f \ov f'\to B}_{\alpha \alpha' \lambda }
 \la \wh{\ov f'}_{\bp'}^{ \alpha'}| \la   \wh{f}_{\bp}^{ \alpha}|, \label{H-FIP1}\
\end{align}
  where  $  d\Omega = |\bk| d\ww p d\ww p' d\ww k$,  $\delta^3_P = \delta^3(\bp + \bp' - \bk)$, and
\begin{gather}
 C^{f \ov f'\to B}_{\alpha \alpha' \lambda }  = \la \wh{\cs}_{B\lambda}(\bk)| \C^{f \ov f'\to B}_{\rm sp}
 |\cs_{f \alpha}(\bp) \cs_{\ov f' \alpha'}(\bp') \ra.   \label{Cs}
\end{gather}

 The function of  the operator $\C^{f \ov f'\to B}_{\rm sp}$ is to project a spinor
 $|\cs_{f \alpha}(\bp) \cs_{\ov f' \alpha'}(\bp') \ra$ to $|{\cs}_{B\lambda}(\bk)\ra $.
 To carry out this function, the simplest form is written as
\begin{align}\label{GG-ff'B}
 \C^{f \ov f'\to B}_{\rm sp} =  {T}^{f \ov f'\to B} \overrightarrow D.
\end{align}
 Here, $\overrightarrow D$ indicates an operation that changes $|\cs_{f \alpha}(\bp) \cs_{\ov f' \alpha'}(\bp') \ra$ 
 to an $n_f$-row and one-column matrix; more exact, 
 with Eqs.(\ref{Sf}) and (\ref{Sovf}), $\overrightarrow D$ acts as
\begin{align}\label{F-right}
 \overrightarrow D  |\cs_{f \alpha}(\bp) \cs_{\ov f' \alpha'}(\bp') \ra
 = \frac{1}{n_f}
 \left( \begin{array}{c}  e^{i(\vartheta^f_1+\vartheta^{f'}_1)} |\LL^{f}_1\ra |\LL^{\ov f'}_1\ra
 \\ e^{i(\vartheta^f_2+\vartheta^{f'}_2)} |\LL^{f}_2\ra |\LL^{\ov f'}_2\ra
 \\ \vdots  \\ e^{i(\vartheta^f_{n_f}+\vartheta^{f'}_{n_f})} |\LL^{f}_{n_f}\ra |\LL^{\ov f'}_{n_f}\ra
  \end{array} \right).
\end{align}
 The symbol ${T}^{f \ov f'\to B}$  projects the result of $\overrightarrow D$  to spinor states of $B$. 
 The simplest form of ${T}^{f \ov f'\to B}$ is written as 
\begin{align}\label{T-ff'B}
 {T}^{f \ov f'\to B} = {P}^{f \ov f'\to B} T_k.
\end{align}
 Here, $T_k$ changes an $n_f$-row and one-column matrix
 to  an $n_B$-row and one-column matrix, written as an $n_B \times n_f$ matrix,
\begin{align}\label{T-ff'B-k}
 {T}_k = \left( \begin{array}{ccc} T_{11} & \cdots & T_{1 n_f} \\
 \vdots & & \\ T_{n_B 1} & \cdots & T_{n_B n_f}  \end{array} \right).
\end{align}
 For different classes  of FIPs,
 the exact meaning of which will be discussed later in Sec.\ref{sect-physical-model} in a specific model of modes,
 $T_k$ may take different forms indicated with $k=1,2,\ldots$.

 On the rhs of Eq.(\ref{T-ff'B}), 
 the function of the symbol ${P}^{f \ov f'\to B}$ is to change direct products of Weyl spinors,
 which lie in rows of the result of $T_k \overrightarrow D$, to
 bosonic spinor states that match the corresponding layers of $B$. 
 To express this symbol explicitly, we introduce an index $\Theta$ for spinors, defined as 
\begin{align}\label{}
 \Theta := \left\{
            \begin{array}{ll}
              1, & \hbox{for spinors in $\VV$ or $\WW \otimes \ov\WW$;} 
          \\    2, & \hbox{for spinors in $\SP$ or $ \WW \otimes \WW$;}
           \\   3, & \hbox{for spinors in $\ov\SP$ or $\ov\WW \otimes \ov\WW$.}
            \end{array}
          \right.
\end{align}
 To get the above-mentioned matching, 
 ${P}^{f \ov f'\to B}$ should be proportional to $ \delta_{\Theta^B_l \Theta^{TD}_l}$,
 where  $\Theta^B_l$ indicates the $\Theta$ value of the $l$-th layer of the mode $B$
 and $\Theta^{TD}_l$ for the $l$-th row of the result of $T_k \overrightarrow D$.
 Furthermore,  we note that $s(\ov s)$-layers of $B$ should give no contribution to the physical amplitude
 due to Eq.(\ref{<s|s>=0}).
 As a result, for brevity one may set 
\begin{align}\label{Tnm=0-fors}
 T_{n m} =0, \ \text{if the $n$th layer of $B$ is an $s(\ov s)$-layer}.
\end{align}
 Then, ${P}^{f \ov f'\to B}$ includes contributions only from $z(\ov z)$-layers of $B$.
 The only method of linking spinors in $ \WW \otimes \ov\WW$ to those in $\VV$
 is through the operator $\sigma$ in Eq.(\ref{sigma}).
 Thus, with the contribution from $s(\ov s)$-layers already excluded by Eq.(\ref{Tnm=0-fors}), 
 ${P}^{f \ov f'\to B}$ is written as
\begin{align}\label{PffB}
 {P}^{f \ov f'\to B} = \sigma \prod_{l=1}^{n_B} \delta_{\Theta^B_l \Theta^{TD}_l}.
\end{align}
 Clearly, a nontrivial FIP should satisfy 
\begin{align}\label{d-Theta=1}
 \prod_{l=1}^{n_B} \delta_{\Theta^B_l \Theta^{TD}_l} =1.
\end{align}

 Next, we discuss FIP-$2$. 
 In the FIP-$2$ of $B\to f \ov f'$, $Y$ contains two fermionic modes ($f$ and $\ov f'$).
 If one requires that the bra that matches $|Y^\xi_\bQ \ra$ in (\ref{|Y><X|})
 should be an ACC-bra, the simplest choice of $P_Y$ is the identity operator $I_Y$. 
 However, presently it is unclear whether this is the only natural choice.
 The main reason is that, under the momentum-spin-separated form of the operator $ \C^{X\to Y}$ in Eq.(\ref{C-XY}),
 the operator $\sigma^T$ in its spinor part, which is to be used
 in a way similar to $\sigma$ in Eq.(\ref{PffB}), is a purely mathematical operator
 and does not unnecessarily require an ACC-bra for the spinor part of $P_Y$.

 We observe that the above-discussed uncertainty in $P_Y$ may be circumvented, 
 if one adopts a well known rule in quantum mechanics, which states that the time evolution operator should be Hermitian. 
 In fact, this rule directly determines $H_{\rm FIP}^{B\to f \ov f'}$ from $H_{\rm FIP}^{f \ov f' \to B}$ by the relation that
\begin{align}\label{H-FIP12-Hermitain}
 H_{\rm FIP}^{B\to f \ov f'} = \left( H_{\rm FIP}^{f \ov f' \to B} \right)^+,
\end{align}
 where the superscript ``$+$'' indicates adjoint operator defined with respect to the ACC-scalar product. 
 (See Appendix \ref{app-adjoint-operator} for detailed discussions on adjoint operators.)

\subsection{A concise expression of FIP amplitude}\label{sect-concise-H-FIP}

 In this section, we derive a concise expression for the FIP amplitude
 $ C^{f \ov f'\to B}_{\alpha \alpha' \lambda }$.
 Substituting the fermionic state in Eq.(\ref{F-right}),  bosonic state as in Eq.(\ref{cs-B}), 
 and the operator $\C^{f \ov f'\to B}_{\rm sp}$ in Eq.(\ref{GG-ff'B}) into Eq.(\ref{Cs}), 
 and making use of Eqs.(\ref{whvep=ovvep}), (\ref{varpi}), (\ref{T-ff'B-k}), and (\ref{PffB}), 
 one gets that
\begin{align}\notag
 C^{f \ov f'\to B}_{\alpha \alpha' \lambda } = &  \frac{1}{n_f\sqrt{n_V}} 
 \left( \prod_{l=1}^{n_B} \delta_{\Theta^B_l \Theta^{TD}_l} \right)
 \sum_{n=1}^{n_B}  \sum_{m=1}^{n_f}  \la\la \ov{\varepsilon}_{\lambda}(\bk)|  \sigma T_{nm} 
 \\ & \times
 \exp \Big[ i(\vartheta^f_m+\vartheta^{f'}_m - \vartheta_n^B) \Big] |\LL^{f}_m\ra |\LL^{\ov f'}_m\ra. \label{Cfovf'B-expre-1}
\end{align}

 On the rhs of Eq.(\ref{Cfovf'B-expre-1}), due to Eq.(\ref{Tnm=0-fors}), the label $n$ effectively runs only over
 $z(\ov z)$-layers of $B$. 
 Hence, under the nonvanishing condition of FIP in Eq.(\ref{d-Theta=1}), 
 only two types of the product $|\LL^{f}_m\ra |\ov\LL^{f'}_m\ra$ contribute, 
 i.e., $|u^f_\alpha(\bp)\ra |\ov v^{\ov f'}_{{\alpha'}}(\bp')\ra$ and 
 $|\ov v^f_{\alpha}(\bp)\ra |u^{\ov f'}_{\alpha'}(\bp')\ra$.
 We use $\M_z$ and $\M_{\ov z}$ to denote the sets of the layer number $m$
 of $f$ and $\ov f'$, corresponding to these two cases, respectively, i.e., 
\begin{subequations}\label{Mz-Movz}
\begin{align}\label{Mz}
 & \M_z := \{ m: |\LL^{f}_m\ra |\ov\LL^{f'}_m\ra =|u_\alpha(\bp)\ra |\ov v_{{\alpha'}}(\bp')\ra \},
 \\ & \M_{\ov z} := \{ m: |\LL^{f}_m\ra |\ov\LL^{f'}_m\ra =|\ov v_{\alpha}(\bp)\ra |u_{\alpha'}(\bp')\ra \}. \label{Movz}
\end{align}
\end{subequations}
 Then, subject to the condition (\ref{d-Theta=1}),  $C^{f \ov f'\to B}_{\alpha \alpha' \lambda }$ is written as
\begin{align}\notag
 C^{f \ov f'\to B}_{\alpha \alpha' \lambda } = \frac{1}{n_f\sqrt{n_V}}  \la\la \ov{\varepsilon}_{\lambda}(\bk)|
 & \sigma  \Big( c |u^{f}_\alpha(\bp)\ra |\ov v^{\ov f'}_{{\alpha'}}(\bp')\ra
 \\ & + d  |\ov v^{f}_{\alpha}(\bp)\ra |u^{\ov f'}_{\alpha'}(\bp')\ra \Big), \label{Cfovf'B-expre-cd}
\end{align}
 where $c$ and $d$ are $c$-numbers determined as follows, 
\begin{subequations}\label{cd}
\begin{align}\label{c}
 c =  \sum_{n=1}^{n_B} \sum_{m\in \M_z} T_{nm} e^{i(\vartheta^f_m+\vartheta^{f'}_m  - \vartheta_n^B)},
 \\ d = \sum_{n=1}^{n_B}  \sum_{m\in \M_{\ov z}} T_{nm} e^{i(\vartheta^f_m+\vartheta^{f'}_m  - \vartheta_n^B)}.
\end{align}
\end{subequations}

 Substituting the operator $\sigma$ in Eq.(\ref{sigma}) into Eq.(\ref{Cfovf'B-expre-cd}), 
 after some derivation (Appendix \ref{app-FIP-amplitudes}), 
 we get the following concise expression of the FIP-$1$ amplitude under Eq.(\ref{d-Theta=1}), i.e., 
\begin{align}\label{Cfovf'B-Dirac}
 C^{f \ov f'\to B}_{\alpha \alpha' \lambda }
 = {\ov\varepsilon}_{\lambda \mu}(\bk) V_{\alpha'}^{ \dag}(\bp')  \Gamma^{f \ov f'\to B} \gamma^\mu U_{\alpha }(\bp),
\end{align}
 where $\ov{\varepsilon}_{\lambda \mu}(\bk) =\la\la \ov{\varepsilon}_{\lambda}(\bk)|T_\mu\ra
 = {\varepsilon}^*_{\lambda \mu}(\bk)$ and $\gamma^\mu$ is given in Eq.(\ref{gamma-mu}).
 On the rhs of Eq.(\ref{Cfovf'B-Dirac}), 
 $U^{\alpha }(\bp), V^{\alpha }(\bp)$, and $\Gamma^{f \ov f'\to B}$ are matrices defined below, 
\begin{align}\label{UV-def}
 &  U^{\alpha }(\bp) := \left( \begin{array}{c}   u^{\alpha  A}(\bp) \\  \ov v_{B'}^{\alpha}(\bp)   \end{array} \right), \ 
  V^{\alpha }(\bp) := \left( \begin{array}{c}   v^{\alpha  B}(\bp) \\  - \ov u_{A'}^{\alpha}(\bp)  \end{array} \right),
 \\ \label{Gamma1}
 & \Gamma^{f \ov f'\to B}  := \frac{1}{n_f \sqrt {2n_V}} \Gamma_{cd},
\end{align}
 where $\Gamma_{cd}$ indicates a $2\times 2$ matrix to be used often in later discussions, defined by
\begin{align}
 & \Gamma_{cd} := \left( \begin{array}{cc} 0 & c \\ d  & 0 \end{array} \right). \label{Gamma-cd}
\end{align}


\subsection{H-operators for VF and BPIPs}\label{sect-H-BPIP}

 In this section, we discuss $H$-operators for VF and use them to 
 construct $H$-operators for BPIPs.

 By definition, in a VF pair, if one fermionic mode lies in a state $|f^\alpha_{\bp}\ra$ with $\alpha=(r,\varrho)$,
 then, the other mode should lie in a state $|\ov f^{\ov \alpha}_{\ov\bp}\ra$
 with $\ov\alpha = ( r,-\varrho)$ and $\ov \bp =-\bp$ [cf.~Eq.(\ref{ov-def-nonspinor})].
 Hence, the $H$-operator $H_{\rm VF}$ for the emergence of a VF pair 
 basically consists of $|f^\alpha_{\bp}\ra |\ov f^{\ov \alpha}_{\ov\bp}\ra$,
 meanwhile, that for the vanishing of a VF pair consists of
 $\la \wh{\ov f}^{\ov \alpha}_{\ov\bp}| \la \wh{f}^\alpha_{\bp}|$.
 As discussed previously, 
 in the construction of a BPIP from an FIP that contains a negative-$p^0$ fermionic state,
 the function of VF is to make such a state appear as an inside state.
 This implies that the exact form of  $H_{\rm VF}$ should be expressed with respect to $H_{\rm FIP}$.

 Let us first discuss BPIPs that are constructed from an FIP-$1$ of $f \ov f' \to B$. 
 There are four such BPIPs, corresponding to the four $(\varrho,\varrho')$ pairs, i.e., 
 $(\varrho,\varrho') = (+,+), (+,-), (-,+)$, and $(-,-)$.
 We use $H^{f \ov f' \to B}_{{\rm BPIP},i}$ of $i=1,2,3$, and $4$ to indicate $H$-operators of these four BPIPs,
 respectively.
 Clearly, for $i=1$, the FIP is already a BPIP.

 Within each of the rest three BPIPs, 
 the FIP contains at least one negative-$p^0$ fermionic mode.
 For such a fermionic mode not to appear as an incoming/outgoing mode, 
 it should belong to some VF pair.
 Mathematically, this point is described by that the two negative-$p^0$ states form a scalar product. 
 We use  $H_{\rm VF}^{ |0\ra \to f\ov f } \left( H_{\rm FIP}^{f\ov f' \to B} \right)$
 and $ H_{\rm VF}^{ f\ov f \to |0\ra } \left( H_{\rm FIP}^{B \to f\ov f' } \right)$ to indicate 
 such combinations of VFs and FIPs.
 Explicitly, they are written as
\begin{subequations}\label{VK}
\begin{align}\label{VK1}
 H_{\rm VF}^{ |0\ra \to f\ov f } \left( H_{\rm FIP}^{f\ov f' \to B} \right)
  & := \int d\ww q |{\ov f}_{\ov \bq \ov \beta} \ra \left( H_{\rm FIP}^{f\ov f' \to B} \right)^{\curvearrowleft} 
  |{f}_{\bq \beta} \ra,
 \\ H_{\rm VF}^{ f\ov f \to |0\ra } \left( H_{\rm FIP}^{B \to f\ov f' } \right)
 &  := \int d\ww q  \la \wh{f}_{\bq  \beta}|^{\curvearrowright}
  \left( H_{\rm FIP}^{B \to f\ov f' } \right) \la \wh{\ov f}_{\ov \bq  \ov \beta}|, \label{VK2}
\end{align}
\end{subequations}
 where $\beta =(s,-)$
 and the two symbols of ``$^{\curvearrowleft}$'' and ``$^{\curvearrowright}$'' are defined as
\begin{subequations}\label{CL-R}
\begin{align}\label{CL}
 \left(  \cdots \la \wh \psi_{f}| \cdots  \right)^{\curvearrowleft} |\phi_f \ra
 : = \cdots \la \wh\psi_f| \phi_f\ra \cdots,
 \\ \la \wh\phi_f |^{\curvearrowright}  \left(  \cdots | \psi_{f}\ra \cdots  \right)
 : = \cdots \la \wh\phi_f | \psi_{f}\ra \cdots. \label{CR}
\end{align}
\end{subequations}
 In Eq.(\ref{VK1}), the positive-$p^0$ state in the VF pair is put on the left-hand side
 of $H_{\rm FIP}$, because it is for an outgoing mode; while, it is put on the rhs of $H_{\rm FIP}$ in Eq.(\ref{VK2})
 in the case that it is for an incoming mode.

 For $i=2$, the negative-$p^0$ mode in the contained FIP is  $\ov f'$ and hence the VF pair consists of a pair of
 $(f', \ov f')$. 
 For $i=3$, the VF pair is $(f,\ov f)$ with a negative-$p^0$ mode $f$. 
 For $i=4$, both fermionic modes in the FIP have negative $p^0$ and, hence, two VF pairs are involved. 
 Making use of $H_{\rm FIP}^{f \ov f'\to B}$ and Eq.(\ref{VK1}), 
 the four BPIP $H$-operators are written as follows, 
\begin{subequations}\label{H-type1-BPIP}
\begin{align}
 H^{f \ov f' \to B}_{{\rm BPIP},1}  & =  H_{\rm FIP}^{f \ov f'\to B},
 \\ \label{H-type1-BPIP2}
 H^{f \ov f' \to B}_{{\rm BPIP},2} & = H_{\rm VF}^{|0\ra \to f'\ov f' } \left( H_{\rm FIP}^{f \ov f'\to B} \right),
 \\ \label{H-type1-BPIP3}
 H^{f \ov f' \to B}_{{\rm BPIP},3} & = H_{\rm VF}^{|0\ra \to f\ov f } \left( H_{\rm FIP}^{f \ov f'\to B} \right),
 \\  \label{H-type1-BPIP4}
 H^{f \ov f' \to B}_{{\rm BPIP},4} & = H_{\rm VF}^{|0\ra \to f\ov f }
 \left( H_{\rm VF}^{|0\ra \to f'\ov f' }\big( H_{\rm FIP}^{f \ov f'\to B} \big) \right).
\end{align}
\end{subequations}

 Now, we discuss physical meanings of the above-derived $H$-operators $H^{f \ov f' \to B}_{{\rm BPIP},i}$ of $i=2,3,4$.
 In the case of $i=2$,  a mode $f$ combines with the negative-$p^0$ mode of $\ov f'$ in a VF pair $(f', \ov f')$ and
 forms to a mode $B$, with the positive-$p^0$ mode $f'$ in the pair leaving as an outgoing mode. 
 The net effect is that a mode $f$ emits a mode $B$ and changes to a mode $f'$.
 Similarly, the net effect of $H^{f \ov f' \to B}_{{\rm BPIP},3}$ is that a mode $\ov f'$ emits a mode $B$
 and changes to a mode $\ov f$.
 In a process of $H^{f \ov f' \to B}_{{\rm BPIP},4}$,  
 two VF pairs appear from the vacuum, which contains two negative-$p^0$ modes of $(f,\ov f')$,
 and these two modes form a bosonic mode $B$;
 the net effect is that three modes of $(\ov f, f', B)$ appear from the vacuum. 

 Next, we discuss BPIPs that are constructed from an FIP-$2$ of $B \to f \ov f' $. 
 Following a procedure similar to that given above, four corresponding BPIPs are constructed
 with $H$-operators indicated by $H^{B \to f \ov f'}_{{\rm BPIP},i}$.
 One finds that 
\begin{subequations}\label{}
\begin{align}
 H^{B \to f \ov f'}_{{\rm BPIP},1} & =  H_{\rm FIP}^{B \to f \ov f'},
 \\ \label{H-type2-BPIP2}
 H^{B\to f \ov f' }_{{\rm BPIP},2} & = H_{\rm VF}^{f'\ov f' \to |0\ra} \left( H_{\rm FIP}^{B\to f \ov f'} \right),
 \\ \label{H-type2-BPIP3}
 H^{B\to f \ov f' }_{{\rm BPIP},3} & = H_{\rm VF}^{f\ov f \to |0\ra} \left( H_{\rm FIP}^{B\to f \ov f'} \right),
 \\  \label{H-type2-BPIP4}
 H^{B\to f \ov f' }_{{\rm BPIP},4} & = 
 H_{\rm VF}^{f\ov f \to |0\ra}\left( H_{\rm VF}^{f'\ov f' \to |0\ra} \big( H_{\rm FIP}^{B\to f \ov f'} \big) \right).
\end{align}
\end{subequations}
 The net effect of $H^{B\to f \ov f' }_{{\rm BPIP},2}$ is that a mode $f'$ absorbs a mode $B$ and changes to a mode $f$,
 that of $H^{B\to f \ov f' }_{{\rm BPIP},3}$ is that a mode $\ov f$ absorbs a mode $B$ and changes to a mode $\ov f'$,
 and that of $H^{B\to f \ov f' }_{{\rm BPIP},4}$ is the vanishing of three modes of  $(\ov f, f', B)$ into the vacuum. 
 It is straightforward to check that the Hermitian relationship still holds for BPIPs, i.e., 
\begin{align}\label{H-BPIP-Hermitain}
 H^{B \to f \ov f'}_{{\rm BPIP},i} = \left ( H^{f \ov f' \to B}_{{\rm BPIP},i} \right)^+.
\end{align}

\subsection{Field expression of interaction H-operator}\label{sect-Hint-field}

 In this section, we derive a concise expression 
 for the time evolution operator in Eq.(\ref{H0-Hint}),
 written as a function of quantum fields given by creation and annihilation operators of free modes.

 Creation operators for a mode $M$, denoted by $b_M^{\alpha \dag}(\bp)$, are defined by the following relation,
\begin{gather}\label{b-dag-def}
 b_M^{\alpha \dag}(\bp)|\Psi \ra \equiv |M^\alpha_\bp\ra |\Psi\ra.
\end{gather}
 where $|\Psi\ra$ indicates an arbitrary state vector.
 Annihilation operators are defined as the adjoint operators of creation operators with respect to the ACC-scalar product,
 that is, 
\begin{gather}\label{b-M^alpha}
 b_M^{\alpha} (\bp)  := \left[ b_M^{\alpha \dag}(\bp) \right]^+.
\end{gather}
 After some derivation (Appendix \ref{app-adjoint-operator}), one finds that
\begin{align}\label{b-express}
 & \la \wh\Phi| b_M^{\alpha} (\bp)|\Psi\ra =  \la \wh\Phi|  \la \wh M^{\alpha}_\bp| |\Psi\ra,
 \\ & b_M^{\alpha \dag} (\bp) = \left[b_M^{\alpha } (\bp) \right]^+  , \label{b+-bdag}
\end{align}
 where $|\Phi\ra$ indicates another arbitrary state vector.
 The above relations, together with Eq.(\ref{A++=A}), implies that 
 \emph{the two superscripts of $\dag$ and $+$ are equivalent with respect to ACC-scalar products}. 
 Moreover, Eq.(\ref{b-express}) implies that $b_M^{\alpha} (\bp)$ is effectively equivalent to $ \la \wh M^{\alpha}_\bp|$.
 In terms of creation and annihilation operators, the free $H$-operator $H_M^0$ in Eq.(\ref{H0M}) is written as
\begin{align}
 H^0_M  =  \int d\ww p \ |\bp| \ b_M^{\alpha \dag}(\bp) b_{M\alpha} (\bp). \label{H0M-bbdag}
\end{align}
 Making use of Eqs.(\ref{A+-|><|}) and (\ref{F-ab}), one checks that $H_M^0$ is Hermitian,
 i.e.,  $H_M^0 = (H_M^0)^+$.

 Substituting the FIP $H$-operator in Eq.(\ref{H-FIP1}) into Eq.(\ref{H-type1-BPIP}),
 and writing the results with creation and annihilation operators, 
 one gets that (Appendix \ref{app-derive-eqs-BPIP})
\begin{subequations}\label{H-BPIP-1-4}
\begin{align}
 H^{f f' \to B}_{{\rm BPIP},1} & = \int d\Omega b^{\lambda \dag}_{B}(\bk)  \delta^3_P 
 C^{f \ov f'\to B}_{\alpha \alpha' \lambda }
 b^{ \alpha'}_{\ov f'}(\bp')  b^{ \alpha}_{f}(\bp), \label{H1}
 \\  H^{f f' \to B}_{{\rm BPIP},2}  & = \int d\Omega b^{\lambda \dag}_{B}(\bk) \delta^3_P  
 C^{f \ov f'\to B}_{\alpha \alpha' \lambda } 
  b^{\ov \alpha' \dag }_{f'}(\ov \bp') b^{ \alpha}_{f}(\bp), \label{H2}
 \\ H^{f f' \to B}_{{\rm BPIP},3} & =  \int d\Omega b^{\lambda \dag}_{B}(\bk)  \delta^3_P 
 C^{f \ov f'\to B}_{\alpha \alpha' \lambda } 
   b^{\ov \alpha \dag }_{\ov f}(\ov \bp) b^{ \alpha'}_{\ov f'}(\bp'), \label{H3}
 \\ H^{f f' \to B}_{{\rm BPIP},4} & =  \int d\Omega b^{\lambda \dag}_{B}(\bk) \delta^3_P 
 C^{f \ov f'\to B}_{\alpha \alpha' \lambda }
  b^{\ov \alpha \dag }_{\ov f}(\ov \bp) b^{\ov \alpha' \dag }_{ f'}(\ov \bp').   \label{H4}
\end{align}
\end{subequations}
 From these expressions, the previously-discussed net effects of BPIPs  are seen clearly;
 for example, $H^{f f' \to B}_{{\rm BPIP},2}$ is for a net process of $f\to f'  B$.

 Examing the rhs of Eqs.(\ref{H2})-(\ref{H4}), one sees that
 not all the mode-species labels and state labels 
 in the amplitudes exactly match those in the creation and annihilation operators.
 In fact, the mismatching in mode-species labels is not serious, because 
 in the explicit expression of amplitude in Eq.(\ref{Cfovf'B-Dirac}) 
 the Dirac spinors are in fact independent of mode species
 and the mode-species dependence lies only in the matrix $\Gamma^{f \ov f' \to B}$.
 Meanwhile, the mismatching in state labels may be moved out
 by making use of certain helicity-state relations for opposite momenta,
 particularly those given in Eq.(\ref{uv-al-p=ovq}) of Appendix \ref{app-spinor-ov-p}.
 After some derivations and with some adjustments of the spinor labels, 
 one gets that  (Appendix \ref{app-derive-eqs-BPIP})
 \begin{widetext}
\begin{subequations}\label{H-BPIP-1-4-UV}
\begin{align}
 H^{f f' \to B}_{{\rm BPIP},1} = &  \int d\ww p d\ww q d\ww k \delta^3(\bk-\bq-\bp) \
 {\ov\varepsilon}_{\lambda \mu}(\bk)  \ 
 V_{\beta}^{  \dag}(\bq)  \ \Gamma^{f \ov f'\to B} \gamma^\mu \ 
 U_{\alpha }(\bp) b^{\lambda \dag}_{B}(\bk) b^{ \beta}_{\ov f'}(\bq) b^{ \alpha}_{f}(\bp), \label{H1-UV}
 \\ H^{f f' \to B}_{{\rm BPIP},2} = &  \int d\ww p d\ww q d\ww k
 \delta^3(\bk+\bq-\bp)  \ov{\varepsilon}_{\lambda \mu}(\bk) \
 U_{\beta }^{ \dag }(\bq) \ \Gamma^{f \ov f'\to B} \gamma^\mu 
 \ U_{\alpha }(\bp)  b^{\lambda \dag}_{B}(\bk)  b^{\beta \dag }_{f'}(\bq)  b^{ \alpha}_{f}(\bp), \label{H2-UV}
 \\  H^{f f' \to B}_{{\rm BPIP},3} = & -\int d\ww p d\ww q d\ww k
 \delta^3(\bk+\bp-\bq)  \ov{\varepsilon}_{\lambda \mu}(\bk) \
 V_{\beta}^{ \dag }(\bq) \ \Gamma^{f \ov f'\to B} \gamma^\mu \ 
 V_{\alpha }(\bp)  b^{\lambda \dag}_{B}(\bk)  b^{\alpha \dag }_{\ov f}(\bp) b^{ \beta}_{\ov f'}(\bq),  \label{H3-UV}
 \\  H^{f f' \to B}_{{\rm BPIP},4} = & - \int d\ww p d\ww q d\ww k
 \delta^3(\bk+\bp + \bq)  \ov{\varepsilon}_{\lambda \mu}(\bk)  \
 U_{\beta }^{ \dag }(\bq) \ \Gamma^{f \ov f'\to B} \gamma^\mu \ V_{\alpha }(\bp) 
  b^{\lambda \dag}_{B}(\bk) b^{\alpha \dag }_{\ov f}(\bp) b^{\beta  \dag }_{ f'}(\bq).  \label{H4-UV}
\end{align}
\end{subequations}
\end{widetext}

 To write $H_{\rm int}$ in a concise form, we use quantum fields as defined below, 
\begin{subequations}\label{psif-phiB-expan}
\begin{align}\label{psi-f-explicit}
 \psi_{f}(\bx) = & \int d\ww p e^{ i \bp \cdot \bx }   U_{\alpha }(\bp) b^{ \alpha}_{f}(\bp) 
 + e^{ -i \bp \cdot \bx } V_{\alpha }(\bp) b^{\alpha \dag }_{\ov f}(\bp),
 \\ \psi_{f}^\dag(\bx) = & \int d\ww p e^{ -i \bp \cdot \bx }   U_{\alpha}^{ \dag}(\bp) b^{ \alpha \dag}_{f}(\bp) 
 + e^{ i \bp \cdot \bx } V_{\alpha}^{ \dag}(\bp) b^{\alpha }_{\ov f}(\bp),
 \\ \phi_{B \mu}(\bx) & =  \int d\ww k    e^{i \bk \cdot \bx } {\varepsilon}_{\lambda \mu}(\bk) b^{\lambda }_{B}(\bk),
 \\ \phi_{B \mu}^\dag(\bx) & =  \int d\ww k  e^{-i \bk \cdot \bx } {\ov\varepsilon}_{\lambda \mu}(\bk) 
 b^{\lambda \dag}_{B}(\bk).
\end{align}
\end{subequations}
 It is straightforward to check that  (Appendix \ref{app-derive-eqs-BPIP})
\begin{align}\label{HffB-BPIP-fields-1}
 \sum_{i=1}^4 H^{f \ov f' \to B}_{{\rm BPIP},i} 
 =  \frac1{(2\pi)^{3}}  \int d^3\bx  :\psi^\dag_{f'} \Gamma^{f \ov f'\to B} \gamma^\mu \psi_f \phi_{B\mu}^\dag :,
\end{align}
 where ``$: \ldots :$'' indicates a norm product.
 Now, $H_{\rm int}$ in Eq.(\ref{H-int}) is written as
\begin{align}\label{}
 H_{\rm int} = \sum_{\text{$(f,\ov f',B)$}} H^{f\ov f'B},
\end{align}
 where the summation is over all possible combinations of $(f,\ov f',B)$ and 
 $H^{f\ov f'B}$ is an interaction $H$-operator defined by
\begin{align}\label{Hfovf'B-def}
 H^{f\ov f'B} := \sum_{i=1}^4 H^{f \ov f' \to B}_{{\rm BPIP},i}  + H^{B \to f \ov f'}_{{\rm BPIP},i}.
\end{align}
 Making use of Eqs.(\ref{HffB-BPIP-fields-1}) and (\ref{H-BPIP-Hermitain}), one writes that
\begin{align}\label{H-ff'B}  
 & H^{f\ov f'B}  = \frac1{(2\pi)^{3}}  \int d^3\bx  : \HH^{f\ov f'B}(\bx) :,
\end{align}
 where $\HH^{f\ov f'B}(\bx) $ is the $H$-operator density,
\begin{align} \label{HH-1}
  \HH^{f\ov f'B}(\bx) &  = \psi^\dag_{f'}(\bx)  \Gamma^{f \ov f'\to B} \gamma^\mu \phi_{B\mu}^\dag(\bx)  \psi_f(\bx)
  + \text{h.c.}.
\end{align}

\section{A model of modes}\label{sect-physical-model}

 The five BAs discussed above are too general to fix a specific model of modes. 
 In this section, by imposing several further assumptions, we construct and study a specific model of modes.
 We give assumptions for the model in Sec.\ref{sect-assump-model}.
 Then, we discuss two-layer modes in Sec.\ref{sect-two-layer-modes}
 and discuss six-layer fermionic modes in Secs.\ref{sect-six-layer-modes-I} and \ref{sect-six-layer-modes-II}. 
 Finally, we derive interaction $H$-operators  in Sec.\ref{sect-inter-H-model}.

\subsection{Model assumptions}\label{sect-assump-model}

 In order to get a specific model, three things need to be fixed: 
 (i) layer configurations of those modes that are taken as physical modes,
 \footnote{Here, ``physical modes'' refer to modes that have the potential of being associated
 with independent physical entities.}
 (ii) relative phases between neighboring layers, and (iii) exact expressions of the matrices $T_k$. 
 In this section, we introduce three assumptions correspondingly, 
 which we denote by ${\rm MA}_{\cdots}^{\cdots}$
 with ``MA'' standing for ``model assumption'' and with super/subscripts indicating their features.

 Firstly, we discuss physical modes. 
 The layer configuration of a physical mode should be stable in some sense.
 Specifically, we assume that it should be \emph{closed} and be relatively short. 
 Here, closedness means that the layer sequence of a mode $M$,
 which is written as $1\to 2 \to \cdots \to n_M$ from top to bottom in the matrix form, 
 in fact forms a loop as
\begin{align}\label{intrinsic-direction}
  1\to 2 \to \cdots \to n_M \to 1.
\end{align}
 Thus, the shortest physical modes have two layers.

 More exactly, we introduce the following assumption for physical modes,
 denoted by ${\rm MA}_{\rm mode}^{\rm species}$.
\begin{itemize}
  \item ${\rm MA}_{\rm mode}^{\rm species}$.  The model includes the following modes
  as physical modes.
  \\ (i) All two-layer closed fermionic modes.
  \\ (ii) Those six-layer fermionic modes, which are constructed from modes in ``(i)''
  by replacing their layers by three-layer closed fermionic modes, with the interlayer  relationship kept. 
  \\ (iii) Bosonic modes, whose layers directly correspond to the layers of 
  $\overrightarrow D  |\cs_{f } \cs_{\ov f' }\ra$
 for two-layer or three-layer closed fermionic modes of $f$ and $\ov f'$.
\end{itemize}
 Here, ``interlayer relationship'' refers to whether the spinor spaces of two layers are of a same type
 or are the complex conjugate of each other. 
 We call three-layer closed fermionic modes, which lie inside a six-layer mode, 
 \emph{sections} of the six-layer mode.

 Secondly, we discuss layer phase.
 As is well known, corresponding to a $2\pi$ rotation in the real space, 
 a Weyl spinor gets a phase $\pi$, while, a vector spinor gets a phase $2\pi$. 
 In analog to this, 
 we assume that a phase $\pi$ may be accumulated when circling the layer loop of a fermionic mode,
 starting from the top layer;
 and a phase $2\pi$ be accumulated for a bosonic mode.
 This assumption implies that the top layer should play a special role for a fermionic mode, 
 while, unnecessarily so for a bosonic mode.
 More exactly, we introduce the following assumption, denoted by ${\rm MA}^{\rm layer}_{\rm phase}$. 
\begin{itemize}
 \item ${\rm MA}^{\rm layer}_{\rm phase}$. 
 Going from the top to the bottom layer along the layer loop in an $n$-layer mode $M$,
 the layer phase increases by $\delta_\theta^M$ at each step,
 where $\delta_\theta^M = 0$ if all the layers are of a same type, otherwise,
 $\delta_\theta^M = \pi/n$ for a fermionic mode  and $\delta_\theta^M = 2\pi/n$ for a bosonic mode.
\end{itemize}
 We recall that there are six types of layers, i.e., $b$, $\ov b$, $z$, $\ov z$, $s$, and $\ov s$.

 From ${\rm MA}^{\rm layer}_{\rm phase}$ and Eq.(\ref{theta-f-ovf}),
 one sees that $\delta_\theta^M= \delta_\theta^{\ov M}$.
 One may take the top layer's phase $\vartheta^M_1$ as a global phase.
 Then, e.g., one writes the spinor state in Eq.(\ref{Sf}) as
\begin{gather}\label{Sf-relative-phase}
 |\cs_{f}^{\alpha }(\bp)\ra 
 = \frac{e^{i\vartheta_1^f}}{\sqrt{n_f}} \left( \begin{array}{c}   |u^\alpha(\bp)\ra
 \\ e^{i\delta_\theta^f} |\LL^f_2\ra \\ e^{i2\delta_\theta^f} |\LL^f_3\ra 
 \\ \vdots    \\ e^{i(n_f-1)\delta_\theta^f} |\LL^f_{n_f}\ra \end{array} \right).
\end{gather}
 We have a freedom in setting global phases of modes.
 We set 
\begin{align}\label{}
 \vartheta^{B}_1 =0 \quad \text{for all bosonic modes $B$},
\end{align}
 while,  global phases of fermionic modes are mode-dependent and to be set in the next section.

 Thirdly, we discuss the matrices $T_k$ in ${T}^{f \ov f'\to B}$ in Eq.(\ref{T-ff'B}), which are crucial
 for explicit expressions of interaction $H$-operators.
 In view of the method of constructing bosonic modes, which is given in ${\rm MA}_{\rm mode}^{\rm species}$-(iii), 
 a preliminary form for $T_k$-matrices may be expected as a unit matrix.
 For two reasons, modifications are needed. 
 One reason is Eq.(\ref{Tnm=0-fors}),
 which was introduced for the purpose of writing ${P}^{f \ov f'\to B}$ in the simple form in Eq.(\ref{PffB}),
 with $s(\ov s)$-layer-related elements of $T_k$ set zero. 
 The other reason is that $n_f$ is unnecessarily equal to $n_B$.

 We use $T_1$ to indicate the matrix for the case of $n_f =n_B=2$ or $3$.
 It is a square matrix with a dimension $n_f$. 
 Under Eq.(\ref{Tnm=0-fors}), its elements are given by
\begin{align}\label{T1}
 & ({T}_{1})_{nm}  = \left\{                       \begin{array}{ll}
                         \delta_{nm}, & \hbox{if $m \in \M_z$ or $ \M_{\ov z}$;} \\
                         0, & \hbox{otherwise.}
                       \end{array}                      \right.
\end{align}
 In the case of $n_f \ne n_B$, further modifications are needed due to specific layer configurations involved.
 We are to give exact forms of the related $T_k$ matrices in later sections
 and, here, merely mention that there are three such $T_k$.
 Thus, we introduce the third assumption, denoted by ${\rm MA}^{T}_{\rm matrix}$.
\begin{itemize}
  \item ${\rm MA}^{T}_{\rm matrix}$. There are four matrices of $T_k$,
 where $T_1$ is applicable to FIPs that contain only two-layer modes or only three-layer modes,
 and $T_{2,3,4}$ are for FIPs that involve six-layer modes. 
\end{itemize}

\subsection{Two-layer modes and their FIPs}\label{sect-two-layer-modes}

 In this section, we discuss two-layer physical modes and FIPs that they participate in. 
 States of these modes are determined by the two
 assumptions of ${\rm MA}_{\rm mode}^{\rm species}$ and ${\rm MA}^{\rm layer}_{\rm phase}$.

 Firstly, we discuss two-layer  fermionic modes. 
 There are four two-layer configurations that may be constructed from $b/\ov b$-layers.
 We call them \emph{$e$-mode, $\ov e$-mode, $\nu$-mode}, and \emph{$\ov\nu$-mode}.
 Spinor states of $\nu$ and $\ov \nu$-modes are written as
 \begin{subequations}\label{S-novn}
\begin{align} \label{S-nu} |\cs_{\nu}^{\alpha }(\bp)\ra = \frac{\exp( i\vartheta^\nu_1)}{\sqrt 2}
 \left( \begin{array}{c} |u^\alpha(\bp)\ra \\ |u^\alpha(\bp)\ra  \end{array} \right),
 \\ |\cs_{\ov \nu}^{\alpha }(\bp)\ra = \frac{\exp( i\vartheta^\nu_1)}{\sqrt 2}
 \left( \begin{array}{c} |\ov v^\alpha(\bp)\ra \\ |\ov v^\alpha(\bp)\ra \end{array} \right). \label{S-ovnu}
\end{align}
\end{subequations}
 where $\delta^\nu_\theta =0$ has been used, as predicted by ${\rm MA}^{\rm layer}_{\rm phase}$.
 And, spinor states of $e$ and $\ov e$-modes are written as
 \begin{subequations}\label{S-eove}
\begin{align}\label{S-e}
 |\cs_{e}^{\alpha }(\bp)\ra = \frac{\exp( i\vartheta^e_1)}{\sqrt 2}
 \left( \begin{array}{c} |u^\alpha(\bp)\ra \\ i |\ov v^\alpha(\bp)\ra  \end{array} \right),
 \\ |\cs_{\ov e}^{\alpha }(\bp)\ra = \frac{\exp( i\vartheta^e_1)}{\sqrt 2}
 \left( \begin{array}{c} |\ov v^\alpha(\bp)\ra \\ i |u^\alpha(\bp)\ra \end{array} \right), \label{S-ove}
\end{align}
\end{subequations}
 where  the factors ``$i$'' in the second layers come from $e^{i\delta^e_\theta}$ with $\delta^e_\theta =\pi/2$.
 We set zero the global phases of these modes, i.e., 
\begin{align}\label{theta-enu=0}
 \vartheta^e_1=\vartheta^\nu_1=0.
\end{align}

 Next, we discuss two-layer bosonic modes.
 According to ${\rm MA}_{\rm mode}^{\rm species}$-(iii), 
 only four bosonic modes may be constructed, corresponding to
 the four pairs of $(\nu, \ov \nu), (e, \ov e), (\nu, \ov e)$, and $(e, \ov \nu)$. 
 We call them \emph{$Z$-mode}, \emph{$A$-mode}, \emph{$W$-mode},  and \emph{$\ov W$-mode}, respectively.
 Since the two layers of a $\nu$-mode have identical states and the same for an $\ov \nu$-mode, 
 the two layers of a $Z$-mode are identical, too, both being a $z$-layer.
 While, the two layers of an $e$-mode are different and, hence, an $A$-mode has a $z$-layer and an $\ov z$-layer. 
 Specifically, spinor states of these modes are written as follows,
\begin{subequations}\label{S-ZAW}
\begin{align} \label{S-Z}
      &    |\cs_{Z}^\lambda(\bk) \ra = \frac{1}{\sqrt 2} \left( \begin{array}{c} |\varepsilon^\lambda(\bk) \ra  \\
 |\varepsilon^\lambda(\bk) \ra \end{array} \right),
 \\ \label{S-A}
        &  |\cs_A^\lambda(\bk) \ra = \frac{1}{\sqrt 2} \left( \begin{array}{c} |\varepsilon^\lambda(\bk) \ra \\
 -|\varpi^\lambda(\bk) \ra \end{array} \right),
 \\ &  |\cs_{W}^\lambda(\bk) \ra = \left( \begin{array}{c} |\varepsilon^\lambda(\bk) \ra  \\
 -  |s \ra \end{array} \right),
  \\ & |\cs_{\ov W}^\lambda(\bk) \ra =  \left( \begin{array}{c} |\varepsilon^\lambda(\bk) \ra  \\
 - |\ov s \ra \end{array} \right),
\end{align}
\end{subequations}
 where $\delta_\theta^{Z} = 0$ and $\delta_\theta^{A} =\delta_\theta^W = \pi$ have been used.  
 It is seen that both the $Z$-mode and the $A$-mode are their own antimodes,
 while,  the $W$-mode and $\ov W$-mode are the antimode of each other.

 Finally, we compute the $\Gamma^{f \ov f'\to B}$ matrices in Eq.(\ref{Gamma1}), as predicted by $T_1$
 for two-layer modes with $n_f= n_B=2$.
 Under  the condition of Eq.(\ref{d-Theta=1}),
 only six combinations of the above-discussed modes are allowed for forming FIP, 
 i.e.,  $e \ov e \to A$, $e \ov e \to Z$, $\nu \ov \nu \to Z$, $\nu \ov \nu \to A$, $\nu \ov e \to W$, 
 and $e\ov \nu  \to \ov W$.
 For these FIPs, the quantities $c$ and $d$ in Eq.(\ref{cd}) can be computed by 
 making use of Eqs.(\ref{T1})-(\ref{S-eove}) and (\ref{S-ZAW}) (Appendix \ref{app-derive-G-enu}),
 giving the following matrices $\Gamma^{f \ov f' \to B}$,
\begin{subequations}\label{Gamma-enu}
\begin{align}
  \Gamma^{\nu \ov \nu \to Z} & =  \frac{1}{2}\Gamma_{10},  \label{Gamma-nnZ}
 \\ \Gamma^{\nu \ov \nu \to A} & =  0,  \label{Gamma-nnA}
 \\ \Gamma^{e \ov e \to Z}  &  = \frac{1}{4}  \Gamma_{1-1}, \label{Gamma-eeZ}
 \\  \Gamma^{e \ov e \to A}  & =  \frac{1}{4}  \Gamma_{11},  \label{Gamma-eeA}
 \\ \Gamma^{\nu \ov e \to W} & = \Gamma^{e \ov \nu \to \ov W}  = 
  \frac{1}{2\sqrt 2} \Gamma_{10}.  \label{Gamma-neW}
\end{align}
\end{subequations}
 According to the definition of $\Gamma_{cd}$ in Eq.(\ref{Gamma-cd}), one has
 $ \Gamma_{10} = \left( \begin{array}{cc} 0 & 1 \\ 0  & 0 \end{array} \right)$,
 $\Gamma_{11} = \left( \begin{array}{cc} 0 & 1 \\ 1  & 0 \end{array} \right)$,
 and  $\Gamma_{1-1} = \left( \begin{array}{cc} 0 & 1 \\ -1  & 0 \end{array} \right)$.
 Note that, due to the vanishing $\Gamma$-matrix in Eq.(\ref{Gamma-nnA}), 
 in fact, there is no FIP of $\nu \ov \nu \to A$.

\subsection{Six-layer fermionic modes and their interactions with two-layer bosonic modes}\label{sect-six-layer-modes-I}

 In this section, we discuss six-layer fermionic modes.
 Based on two structural features of their spinor states,
 we propose explicit forms of two matrices of $T_2$ and $T_3$, 
 which describe FIPs that include two-layer bosonic modes.

 Firstly, according to ${\rm MA}_{\rm mode}^{\rm species}$, 
 six-layer modes are built of three-layer fermionic modes as  blocks. 
 Clearly, three-layer modes with $b$/$\ov b$-layers have totally eight configurations.
 Here, we consider six of them, each including both $b$-layer and $\ov b$-layer,
 and shall discuss the rest two in Sec.\ref{sect-dark-matter}.
 We call these six modes \emph{$t$-modes} with ``$t$'' standing for triple,
 more exactly, \emph{$t_i$-modes} and $\ov t_i$-modes with $i=1,2,3$.
 Spinor states of $t_i$-modes are written as
\begin{subequations}\label{Sqi}
\begin{gather}\label{Sq1}
 |\cs_{t_1}^{\alpha }(\bp)\ra = \frac{\exp( i\vartheta^{t_1}_1)}{\sqrt 3}
 \left( \begin{array}{c} |u^\alpha(\bp)\ra \\ e^{i\delta_\theta^t} |\ov v^\alpha(\bp)\ra
 \\ e^{i2\delta_\theta^t} |\ov v^\alpha(\bp)\ra \end{array} \right),
 \\ \label{Sq2} |\cs_{t_2}^{\alpha }(\bp)\ra = \frac{\exp( i\vartheta^{t_2}_1)}{\sqrt 3}
 \left( \begin{array}{c}    |u^\alpha(\bp)\ra \\ e^{i\delta_\theta^t}  |u^\alpha(\bp)\ra
 \\ e^{i2\delta_\theta^t} |\ov v^\alpha(\bp)\ra  \end{array} \right),
 \\ \label{Sq3} |\cs_{t_3}^{\alpha }(\bp)\ra =\frac{\exp( i\vartheta^{t_3}_1)}{\sqrt 3}
 \left( \begin{array}{c}  |u^\alpha(\bp)\ra  \\ e^{i\delta_\theta^t}  |\ov v^\alpha(\bp)\ra
 \\ e^{i2\delta_\theta^t} |u^\alpha(\bp)\ra  \end{array} \right);
\end{gather}
\end{subequations}
 and those of $\ov t_i$-modes are
\begin{subequations}\label{Sovqi}
\begin{gather}\label{Sovq1}
 |\cs_{\ov t_1}^{\alpha }(\bp)\ra = \frac{\exp( i\vartheta^{t_1}_1)}{\sqrt 3}
 \left( \begin{array}{c} |\ov v^\alpha(\bp)\ra \\ e^{i\delta_\theta^t} |u^\alpha(\bp)\ra
 \\ e^{i2\delta_\theta^t} |u^\alpha(\bp)\ra \end{array} \right),
 \\ \label{Sovq2} |\cs_{\ov t_2}^{\alpha }(\bp)\ra = \frac{\exp( i\vartheta^{t_2}_1)}{\sqrt 3}
 \left( \begin{array}{c}    |\ov v^\alpha(\bp)\ra \\ e^{i\delta_\theta^t}  |\ov v^\alpha(\bp)\ra
 \\ e^{i2\delta_\theta^t} | u^\alpha(\bp)\ra  \end{array} \right),
 \\ \label{Sovq3} |\cs_{\ov t_3}^{\alpha }(\bp)\ra =\frac{\exp( i\vartheta^{t_3}_1)}{\sqrt 3}
 \left( \begin{array}{c}  |\ov v^\alpha(\bp)\ra  \\ e^{i\delta_\theta^t}  |u^\alpha(\bp)\ra
 \\ e^{i2\delta_\theta^t} |\ov v^\alpha(\bp)\ra  \end{array} \right).
\end{gather}
\end{subequations}
 According to ${\rm MA}^{\rm layer}_{\rm phase}$, one has $\delta_\theta^t = \pi/3$.

 Now, we are ready to discuss six-layer fermionic modes,
 which according to ${\rm MA}_{\rm mode}^{\rm species}$-(ii) are constructed from the four modes of
 $e, \ov e, \nu$, and $ \ov \nu$, by changing the $b/\ov b$-layers in them to $t_i/\ov t_i$-modes.
 We use the name of \emph{$u$-mode} to refer to those that are constructed from the $\nu$-mode, 
 and use \emph{$d$-mode} for those from the $e$-mode.
 More exactly, we say \emph{$u_i$-modes} and \emph{$d_i$-modes}, which have the following spinor states, 
\begin{subequations}\label{S-ud}
\begin{align}\label{}
\label{Sui}
 & |\cs_{{u}_j}^{\alpha }(\bp)\ra = \frac{1}{\sqrt 2}
 \left( \begin{array}{c}   |\cs_{t_j}^{\alpha }(\bp)\ra  \\   |\cs_{t_j}^{\alpha }
 (\bp)\ra \end{array} \right),
 \\ \label{Sdi} & |\cs_{{d}_j}^{\alpha }(\bp)\ra = \frac{1}{\sqrt 2}
 \left( \begin{array}{c}   |\cs_{t_j}^{\alpha }(\bp)\ra  \\ i
 |\cs_{\ov t_j}^{\alpha }(\bp)\ra \end{array} \right).
\end{align}
\end{subequations}
 Spinor states of their antimodes, i.e., of $\ov u_i$-modes and $\ov d_i$-modes, are written as
\begin{subequations}\label{S-ovud}
\begin{align}\label{Sovui}
 & |\cs_{{\ov u}_j}^{\alpha }(\bp)\ra = \frac{1}{\sqrt 2}
 \left( \begin{array}{c}   |\cs_{\ov t_j}^{\alpha }(\bp)\ra  \\
  |\cs_{\ov t_j}^{\alpha }(\bp)\ra \end{array} \right),
 \\ & \label{Sovdi} |\cs_{{\ov d}_j}^{\alpha }(\bp)\ra =\frac{1}{\sqrt 2}
 \left( \begin{array}{c}  |\cs_{\ov t_j}^{\alpha }(\bp)\ra  \\ i
 |\cs_{t_j}^{\alpha }(\bp)\ra \end{array} \right).
\end{align}
\end{subequations}

 Next, we discuss interactions of six-layer fermionic modes and two-layer bosonic modes.
 The main task is to determine $T_k$-matrices for such FIPs,
 which should have the dimension of $2 \times 6$, converting the six-layer matrix of 
 $\overrightarrow D  |\cs_{f \alpha}(\bp) \cs_{\ov f' \alpha'}(\bp') \ra$ to a two-layer configuration.
 We do this, based on two structural features of the six-layer modes.

 The first feature is seen in the case of $f=f'$, with $f$ and $\ov f'$ as antimodes of each other. 
 It is that $\overrightarrow D  |\cs_{f \alpha}(\bp) \cs_{\ov f' \alpha'}(\bp') \ra$
 contains only two types of layers:  $|u^f_\alpha(\bp)\ra |\ov v^{\ov f'}_{{\alpha'}}(\bp')\ra$
 and $|\ov v^f_{\alpha}(\bp)\ra |u^{\ov f'}_{\alpha'}(\bp')\ra$.
 Obviously, putting layers of each type together, one may change the six-layer configuration to a two-layer configuration.
 We assume that FIPs may happen, corresponding to this feature.
 Such an FIP is described by $T_2$, whose matrix elements are written as $(T_2)_{1m} = \delta_m$ and 
 $(T_2)_{2m} = 1- \delta_m$, where $\delta_m =1$ for $m \in \M_z$ and $\delta_m =0$ for $m \in \M_z$,
 with $\M_{z}$ and $\M_{\ov z}$ defined in Eq.(\ref{Mz-Movz}).
 Written explicitly, 
\begin{align}\label{T-zm}
 & {T}_{2}   = \left( \begin{array}{llllll} \delta_1 & \delta_2 & \delta_3 & \delta_4 & \delta_5 & \delta_6
    \\ 1-  \delta_1 & 1- \delta_2 &  1- \delta_3 & 1-\delta_4 & 1- \delta_5 &  1- \delta_6
 \end{array} \right).
\end{align}

 Under the condition in Eq.(\ref{d-Theta=1}), an FIP-$1$ with $T_2$ may generate a $Z$-mode, or an $A$-mode.
 Direct derivation shows that the matrix $T_2$ predicts the following matrices $\Gamma_{cd}$ for 
 FIPs of $u_i \ov u_i \to A,Z$ and $d_i \ov d_i \to A,Z$ (Appendix \ref{proof-G-cd-ud-AZ}),
\begin{subequations}\label{Gamma-cd-ud-AZ}
\begin{align} \notag
  \Gamma_{cd} = 2\exp(  i2\vartheta^{t_1}_1 & )  \Gamma_{11} / \Gamma_{1-1} 
 \\ & \text{for} \ u_1  \ov u_1 \ \& \ d_1  \ov d_1 \to A/Z,  \label{Gamma-cd-ud-AZ-1}
 \\ \notag \Gamma_{cd} = 2\exp( i2\vartheta^{t_2}_1 & + i\delta_\theta^t)   \Gamma_{11} / \Gamma_{1-1} 
  \\ & \text{for $u_2  \ov u_2  \ \& \ d_2 \ov d_2 \to A/Z$}, 
 \\ \notag \Gamma_{cd} = 2\exp( i2\vartheta^{t_3}_1 & -i\delta_\theta^t)  \Gamma_{11} / \Gamma_{1-1} 
  \\ & \text{for $u_3  \ov u_3  \ \& \ d_3 \ov d_3 \to A/Z$},
\end{align}
\end{subequations}
 where the slashes mean that $\Gamma_{11}$ are for the $A$-mode and $\Gamma_{1-1}$ for the $Z$-mode. 
 The above matrices have the simplest expressions under the following  choices of the 
 global phases $\vartheta^{t_i}_1$, i.e.,
\begin{align}\label{theta-t123}
 \vartheta^{t_1}_1 =0 , \quad \vartheta^{t_2}_1 = - \delta_\theta^t/2,
 \quad \vartheta^{t_3}_1 =  \delta_\theta^t/2.
\end{align}
 Then, substituting the above $\Gamma_{cd}$ into Eq.(\ref{Gamma1})
 and noting that $n_f=6$, $n_V=2$, one gets that 
\begin{subequations}\label{Gamma-ud-AZ-T2}
\begin{align}   
 & \Gamma^{u_i \ov u_i \to Z}_{(T_2)} = \Gamma^{d_i \ov d_i \to Z}_{(T_2)} =   \frac{1}{ 6} \Gamma_{1-1},
  \label{Gamma-uiiZ}
 \\ \label{Gamma-uiiA}
 & \Gamma^{u_i \ov u_i \to A}_{(T_2)} = \Gamma^{d_i \ov d_i \to A}_{(T_2)} =   \frac{1}{ 6} \Gamma_{11}.
\end{align}
\end{subequations}

 The second feature is related to the section structure of the six-layer modes.
 In fact, reverting the constructing procedure of the six-layer modes,
 which is stated in ${\rm MA}_{\rm mode}^{\rm species}$-(ii),
 one would get two-layer fermionic modes from the six-layer modes,
 with the conversions of  $u_i \to \nu$ and $d_i \to e$. 
 We assume that each of the resulting two-layer fermionic modes may get a new phase of 
 $\theta_{\rm con} = \pi/2$.
 After the above-discussed change, an FIP like that discussed in Sec.\ref{sect-two-layer-modes} may follow.
 Thus, an FIP may happen, in which two six-layer fermionic modes change to a two-layer bosonic mode.

 We use $T_3$ to indicate the $T_k$-matrix for the above-discussed FIP. 
 From the explicit expressions of spinor states of the $u$/$d$-modes given previously, 
 it is easy to see that one gets the conversions of $u_i \to \nu$ and $d_i \to e$, 
 if replacing all their sections by the first layers of the corresponding  $t$-modes, 
 apart from the global phases of the $t$-modes. 
 This enables one to write the following explicit expression of $T_3$,
\begin{align}\label{T-shrink}
 & {T}_{3}  =3 \exp( 2i\theta_{\rm con} - 2i\vartheta^{t}_1 ) T_1 \times \left( \begin{array}{ll} 1,0,0,0,0,0
    \\  0,0,0,1,0,0
 \end{array} \right),
\end{align}
 where $T_1$ is a $2 \times 2$ matrix, 
 the term $\exp( -2i\vartheta^{t}_1 )$ is used to eliminate the global phases of old sections, 
 and the number $3$ is used for compensating that each section has been changed to one layer. 
 It is easy to check that $T_3$ predicts the same $\Gamma$-matrices as those given by $T_1$ 
 in Eq.(\ref{Gamma-enu}), apart from a minus sign,  i.e., 
\begin{subequations}\label{Gamma-ud-shrink}
\begin{align}\label{}
  & \Gamma^{u_j \ov u_j \to Z}_{(T_3)} =  -\frac{1}{2}\Gamma_{10}, 
 \\ & \Gamma^{u_j \ov u_j \to A}_{(T_3)} =  0,  \label{Gamma-uuA-T3}
 \\ & \Gamma^{d_j \ov d_j \to Z}_{(T_3)} =   - \frac 14  \Gamma_{1-1}, 
 \\ & \Gamma^{d_j \ov d_j \to A}_{(T_3)} =   - \frac 14  \Gamma_{11},  \label{Gamma-ddA-T3}
 \\ & \Gamma^{u_j \ov d_j \to W}  = \Gamma^{d_j \ov u_j \to \ov W}  = 
 -\frac{1}{2\sqrt 2}  \Gamma_{10}.  \label{Gamma-udW}
\end{align}
\end{subequations}

 Finally, putting contributions from $T_2$ and $T_3$ given in Eqs.(\ref{Gamma-ud-AZ-T2}) and
 (\ref{Gamma-ud-shrink}) together, we get the following
 $\Gamma$-matrices for FIPs that contain $u$-$d$-modes and $A$-$Z$-modes, 
\begin{subequations}\label{}
\begin{align}\label{Gamma-uuZ}
  & \Gamma^{u_j \ov u_j \to Z} = \frac{1}{ 6} \Gamma_{1-1} -\frac{1}{2}\Gamma_{10},
 \\ & \Gamma^{u_j \ov u_j \to A} = \frac{1}{ 6} \Gamma_{11}, \label{Gamma-uuA}
 \\ & \Gamma^{d_j \ov d_j \to Z} =  \frac{1}{ 6} \Gamma_{1-1} - \frac 14  \Gamma_{1-1} = - \frac{1}{12}  \Gamma_{1-1},
 \label{Gamma-ddZ}
 \\ & \Gamma^{d_j \ov d_j \to A} = \frac{1}{ 6} \Gamma_{11}  - \frac 14  \Gamma_{11}= - \frac{1}{12}  \Gamma_{11}.
 \label{Gamma-ddA}
\end{align}
\end{subequations}

\subsection{Interactions of six-layer fermionic modes with three-layer bosonic modes}\label{sect-six-layer-modes-II}

 In this section, we discuss  FIPs in which the six-layer fermionic modes interact
 with three-layer bosonic modes.

 We first discuss three-layer bosonic modes predicted by ${\rm MA}_{\rm mode}^{\rm species}$-(iii). 
 From $t_i$-modes of $i=1,2,3$ and $\ov t_j$-modes  of $j=1,2,3$, nine bosonic modes may be constructed, 
 which we call \emph{$g$-modes}, more exactly, \emph{$g_{i \ov j}$-modes}.
 Spinor states of $g_{i \ov j}$-modes, denoted by $|\cs_{g_{i\ov j}}^\lambda(\bk) \ra$,
 can be directly obtained from the spinor states of $t$-modes in Eqs.(\ref{Sqi})-(\ref{Sovqi}), i.e., 
\begin{subequations}\label{S-giovi}
\begin{align}\label{}
  & |\cs_{g_{1\ov 1}}^\lambda(\bk) \ra = \frac{1}{\sqrt 3}
  \left( \begin{array}{c} |\varepsilon^\lambda(\bk) \ra \\  e^{i\delta_\theta^{g}}|\varpi^\lambda(\bk) \ra
 \\  e^{i2\delta_\theta^{g}}|\varpi^\lambda(\bk) \ra \end{array} \right), \
 \\ & |\cs_{g_{2\ov 2}}^\lambda(\bk) \ra = \frac{1}{\sqrt 3}
  \left( \begin{array}{c} |\varepsilon^\lambda(\bk) \ra \\  e^{i\delta_\theta^{g}} |\varepsilon^\lambda(\bk) \ra
 \\ e^{i2\delta_\theta^{g}} |\varpi^\lambda(\bk) \ra \end{array} \right),
 \\ & |\cs_{g_{3\ov 3}}^\lambda(\bk) \ra = \frac{1}{\sqrt 3}
  \left( \begin{array}{c} |\varepsilon^\lambda(\bk) \ra  \\ e^{i\delta_\theta^{g}} |\varpi^\lambda(\bk) \ra  
  \\ e^{i2\delta_\theta^{g}} |\varepsilon^\lambda(\bk) \ra \end{array} \right), \label{S-gii}
\end{align}
\end{subequations}
 and 
\begin{subequations}\label{S-g3ovi}
\begin{align}\label{}
  &  |\cs_{g_{1\ov 2}}^\lambda(\bk) \ra = \frac{1}{\sqrt 2}
  \left( \begin{array}{c} |\varepsilon^\lambda(\bk) \ra \\ e^{i\delta_\theta^{g}} |\ov s\ra
 \\ e^{i2\delta_\theta^{g}} |\varpi^\lambda(\bk) \ra \end{array} \right),
 \\ & |\cs_{g_{2\ov 1}}^\lambda(\bk) \ra = \frac{1}{\sqrt 2}
  \left( \begin{array}{c} |\varepsilon^\lambda(\bk) \ra \\  e^{i\delta_\theta^{g}} |s\ra
 \\ e^{i2\delta_\theta^{g}} |\varpi^\lambda(\bk) \ra \end{array} \right),
 \\  &  |\cs_{g_{1\ov 3}}^\lambda(\bk) \ra = \frac{1}{\sqrt 2}
  \left( \begin{array}{c} |\varepsilon^\lambda(\bk) \ra
 \\ e^{i\delta_\theta^{g}} |\varpi^\lambda(\bk) \ra  \\ e^{i2\delta_\theta^{g}} |\ov s\ra \end{array} \right),
 \\ & |\cs_{g_{3\ov 1}}^\lambda(\bk) \ra = \frac{1}{\sqrt 2}
  \left( \begin{array}{c} |\varepsilon^\lambda(\bk) \ra
 \\ e^{i\delta_\theta^{g}} |\varpi^\lambda(\bk) \ra \\ e^{i2\delta_\theta^{g}} |s\ra \end{array} \right),
 \\ &  |\cs_{g_{2\ov 3}}^\lambda(\bk) \ra = 
  \left( \begin{array}{c} |\varepsilon^\lambda(\bk) \ra
 \\ e^{i\delta_\theta^{g}} |s\ra  \\ e^{i2\delta_\theta^{g}} |\ov s\ra \end{array} \right),
  \\ & |\cs_{g_{3\ov 2}}^\lambda(\bk) \ra =
  \left( \begin{array}{c} |\varepsilon^\lambda(\bk) \ra \\  e^{i\delta_\theta^{g}} |\ov s\ra
 \\  e^{i2\delta_\theta^{g}} |s\ra \end{array} \right), \label{S-gij}
\end{align}
\end{subequations}
 where $\delta_\theta^{g} = 2\delta_\theta^{t} = 2\pi/3$. 
 Due to the closed feature of the modes and the fact that $3\delta_\theta^{g}=2\pi$, 
 the two states of  $|\cs_{g_{2\ov 2}}^\lambda(\bk) \ra$
 and $|\cs_{g_{3\ov 3}}^\lambda(\bk) \ra$ in Eq.(\ref{S-giovi}) differ only in their global phases. 
 This implies that the $g_{2\ov 2}$-mode should be equivalent to the $g_{3\ov 3}$-mode 
 and, as a result, there are totally eight independent $g$-modes.

 Direct derivation shows that $T_1$ predicts the following matrices of
 $\Gamma^{t_i \ov t_j \to g_{i\ov j}}$ (Appendix \ref{proof-G-qijg-all}),
\begin{subequations}\label{Gamma-qijg-all}
\begin{align}
  \Gamma^{t_1 \ov t_1 \to g_{1\ov 1}}  & =  \frac{1}{3\sqrt 6}\Gamma_{12},
  \\  \Gamma^{t_2 \ov t_2 \to g_{2\ov 2}} &  = \frac{\exp( -i\delta_\theta^t)}{3\sqrt 6}\Gamma_{21},
  \\ \Gamma^{t_3 \ov t_3 \to g_{3\ov 3}} & = \frac{\exp(  i \delta_\theta^t)}{3\sqrt 6}\Gamma_{21},  
  \label{Gamma-qiig}
 \\ \label{Gamma-t12g}
 \Gamma^{t_1 \ov t_2 \to g_{1\ov 2}}  & = \Gamma^{t_2 \ov t_1 \to g_{2\ov 1}}  
  = \frac{\exp( -i\delta_\theta^t/2)}{6} \Gamma_{11}, 
 \\ \Gamma^{t_1 \ov t_3 \to g_{1\ov 3}}  & = \Gamma^{t_3 \ov t_1 \to g_{3\ov 1}} 
   = \frac{\exp( i\delta_\theta^t/2)}{6} \Gamma_{11}, 
 \\ \label{Gamma-t23g}
 \Gamma^{t_2 \ov t_3 \to g_{2\ov 3}}  & = \Gamma^{t_3 \ov t_2 \to g_{3\ov 2}} =  \frac{1}{3\sqrt 2} \Gamma_{10}.
\end{align}
\end{subequations}
 Here, the following facts have been used, i.e., $n_f=3$, $n_V=3$ for $g_{i\ov i}$-modes, $n_V=2$ for $g_{i\ov j}$-modes
 with either $i=1$ or $j=1$, and $n_V=1$ for $g_{2\ov 3}$-mode and $g_{3\ov 2}$-mode.

 Based on the section feature of six-layer modes,
 we assume another type of FIP, in which the interaction may break connections between sections, 
 with the three-layer fermionic modes retained. 
 However, according to the assumption ${\rm MA}_{\rm mode}^{\rm species}$, 
 three-layer fermionic modes are not physical modes that may exist independently.
 Hence, the generated three-layer fermionic modes should immediately form $g$-modes (described by $T_1$), otherwise, 
 no interaction may take place. 
 We use $T_4$ to indicate the $T_k$ matrix for such an FIP.
 Further assuming that upper sections only combine with upper sections and the same for lower sections,
 one gets the following expression of $T_4$,
\begin{align}\label{T-section}
 T_4 =  T_1 \times \left(1, 1  \right)_{\rm section},
\end{align}
 where $T_1$ is a $3 \times 3$ matrix and the subscript ``section'' means that the matrix $(1,1)$ applies to sections.

 Below, we derive $\Gamma$-matrices predicted by $T_4$. 
 Firstly, a $u$-mode, whose upper and lower sections are identical [Eq.(\ref{Sui})], 
 may interact with an $\ov u$-mode. 
 In such an interaction, the action of $T_4 \overrightarrow D$
 on the two modes gives the same result as $T_1  \overrightarrow D$ for the corresponding $t$-type modes, i.e., 
 $T_2  \overrightarrow D  |\cs_{u_i \alpha}(\bp) \cs_{\ov u_j \alpha'}(\bp') \ra
 = T_1  \overrightarrow D  |\cs_{t_i \alpha}(\bp) \cs_{\ov t_j \alpha'}(\bp') \ra$. 
 This implies that
\begin{align}\label{G-uug}
 \Gamma^{u_k \ov u_j \to g_{k\ov j}} =  \Gamma^{t_k \ov t_j \to g_{k\ov j}}. 
\end{align}

 Secondly, for the interaction between a $d_i$-mode and an $\ov d_j$-mode,  one gets that
\begin{align}\label{}\notag
  & T_4  \overrightarrow D  |\cs_{d_i \alpha}(\bp)  \cs_{\ov d_j \alpha'}(\bp')  \ra
 \\ & = \frac 12 T_1  \overrightarrow D  |\cs_{t_i \alpha}(\bp) \cs_{\ov t_j \alpha'}(\bp') \ra
  - \frac 12 T_1  \overrightarrow D  | \cs_{\ov t_i \alpha}(\bp) \cs_{t_j \alpha'}(\bp') \ra \notag
 \\ & = \frac 12 T_1  \overrightarrow D  |\cs_{t_i \alpha}(\bp) \cs_{\ov t_j \alpha'}(\bp') \ra
   + \frac 12 T_1  \overrightarrow D  | \cs_{t_j \alpha'}(\bp')  \cs_{\ov t_i \alpha}(\bp) \ra.
\end{align}
 Then, one finds that
\begin{subequations}\label{G-ddg}
\begin{align} \label{G-ddg1}  
 \Gamma^{d_i \ov d_i \to g_{i\ov i}} & =  \Gamma^{t_i \ov t_i \to g_{i\ov i}},
 \\  \Gamma^{d_i \ov d_j \to g_{i\ov j}} &   = \frac 12 \Gamma^{t_i \ov t_j \to g_{i\ov j}} \quad (i\ne j),
 \\   \Gamma^{d_i \ov d_j \to g_{j\ov i}} &  = \frac 12 \Gamma^{t_j \ov t_i \to g_{j\ov i}} \quad (i\ne j).
  \label{G-ddg2}
\end{align}
\end{subequations}
 Note that $\Gamma^{t_i \ov t_j \to g_{i\ov j}} 
 = \Gamma^{t_j \ov t_i \to g_{j\ov i}} $ according to (\ref{Gamf=Gam-ovf}).

 Finally, no FIP with $T_4$ may happen between a $u$-type mode and a $d$-type mode. 
 In fact, it is straightforward to check that, for such a pair of six-layer fermionic modes, 
 either their upper sections or their lower sections may not form a $g$-mode. 
 For example, for a $u_i$-mode and an $\ov d_j$-mode,
 their lower sections are a $t_i$-mode and a $t_j$-mode, respectively, from which no $g$-mode may be formed. 

 \subsection{Interaction H-operators}\label{sect-inter-H-model}

 In this section, we derive field-expressions of $\HH^{f\ov f'B}$,
 which are predicted by the FIPs of the model discussed in previous sections. 
 In this derivation, the following relation is to be used,
\begin{gather}\label{Gam-gam-exc-1}
  \left(\Gamma^{f \ov f'\to B } \gamma^{\mu} \right)^\dag =  \left( \Gamma^{f \ov f'\to B }\right)^* \gamma^\mu,
\end{gather}
 the proof of which is given in Appendix \ref{app-derive-Gam-dag}.

 To write the interaction $H$-operators in concise forms, 
 it proves convenient to introduce fields as discussed below. 
 Firstly, like the total fermionic field $\psi_f$ in Eq.(\ref{psi-f-explicit}), we define a total field for a bosonic mode $B$,
\begin{align}\label{psi-Bmu}
  \psi_{B\mu} & := \phi^\dag_{\ov B \mu}(\bx) + \phi_{B\mu}(\bx).
\end{align}
 Secondly, we use a tilde to indicate a fermionic field, which is given by the complex conjugate of a 
 given fermionic field multiplied by the matrix $\Gamma_{11}$, i.e., 
 \footnote{In the chiral representation of the $\gamma^\mu$-matrices,  which we in fact use in this paper, 
 $\gamma^0$ has the same $2\times 2$ form as $\Gamma_{11}$, i.e., $\left( \begin{array}{cc} 0 & 1 \\ 1 & 0\end{array}\right)$. 
 Hence, $\ww\psi$ used here has a form similar to $\ov \psi=\psi^\dag \gamma^0$ in the ordinary notation of QFT.
 }
\begin{align}\label{}
  \label{ww-psi} \ww\psi_{f} & := \psi^\dag_{f} \Gamma_{11}.
\end{align}
 Thirdly, similar to left-handed and right-handed fields used in the SM, we define
\begin{align}
 \psi^L_f & := \left( \begin{array}{cc} 1 & 0 \\ 0 & 0\end{array}\right) \psi_f,
 \quad \psi^R_f := \left( \begin{array}{cc} 0 & 0 \\ 0 & 1 \end{array}\right) \psi_f.
\end{align}
 Combining the above two notations, we write
\begin{subequations}\label{ww-psi-LR}
\begin{align}\label{}
 \ww\psi^{L}_f & = \psi^{L \dag}_f \Gamma_{11} 
 = \psi^{\dag}_f \left( \begin{array}{cc} 1 & 0 \\ 0 & 0\end{array}\right) \Gamma_{11}, 
 \\ \ww\psi^{R}_f & = \psi^{R \dag}_f \Gamma_{11}
  = \psi^\dag_f \left( \begin{array}{cc} 0 & 0 \\ 0 & 1 \end{array}\right) \Gamma_{11}.
\end{align}
\end{subequations}
 Clearly, 
\begin{align}\label{}
 & \psi_f = \psi^L_f + \psi^R_f, \qquad \ww \psi_f = \ww \psi^L_f + \ww \psi^R_f.
\end{align}
 From Eq.(\ref{ww-psi-LR}), it is direct to see that
\begin{align}\label{}
 \ww\psi^{L }_f &  = \psi^\dag_f  \Gamma_{10}, \quad
    \ww\psi^{R }_f  = \psi^\dag_f  \Gamma_{01}.
\end{align}

 Now, we are ready to discuss interaction $H$-operators. 
 We first discuss $\HH^{f\ov f'B}$ of those BPIPs that involve $A$-mode.
 Fermionic modes, which are involved in these BPIPs, are $f=f'=e,u$, and $d$.
 Their $\Gamma$-matrices, given in Eqs.(\ref{Gamma-eeA}), (\ref{Gamma-uuA}), and (\ref{Gamma-ddA}),
 are real and proportional to $\Gamma_{11}$.
 Due to the relation in Eq.(\ref{Gam-gam-exc-1}), these $\Gamma$-matrices satisfy 
 $(\Gamma^{f \ov f'\to B } \gamma^{\mu} )^\dag = \Gamma^{f \ov f'\to B } \gamma^\mu$.
 Substituting these $\Gamma$-matrices into Eq.(\ref{HH-1}), one gets the following $H$-operator densities,
\begin{subequations}\label{H-A}
\begin{align}\label{}
  \HH^{e\ov eA}  & = \frac{1}{4}  \ww\psi_{e} \gamma^\mu  \psi_e \psi_{A\mu}, \label{H-eeA}
 \\  \HH^{u_i\ov u_iA} &  = \frac{1}{4} \ww\psi_{u_i} \gamma^\mu  ( \frac 23 )  \psi_{u_i} \psi_{A\mu},
  \\   \HH^{d_i\ov d_iA}  & =  \frac{1}{4}  \ww\psi_{d_i} \gamma^\mu ( -\frac 13 ) \psi_{d_i} \psi_{A\mu},
\end{align}
\end{subequations} 
 where $\psi_{A\mu}(\bx)$ is defined in Eq.(\ref{psi-Bmu}).

 Next, we discuss BPIPs that involve $Z$-mode.
 Fermionic modes in them are $f=f'=\nu , e, u$, and $d$.
 Their $\Gamma$-matrices, given in Eqs.(\ref{Gamma-nnZ}), (\ref{Gamma-eeZ}), (\ref{Gamma-uuZ}), and (\ref{Gamma-ddZ}),
 are also real and, as a result, also satisfy
 $(\Gamma^{f \ov f'\to B } \gamma^{\mu} )^\dag = \Gamma^{f \ov f'\to B } \gamma^\mu$.
 Moreover, one checks that
\begin{subequations}\label{}
\begin{align}\label{}
 \Gamma_{10} \gamma^\mu = \Gamma_{10} \gamma^\mu  \left( \begin{array}{cc} 1 & 0 \\ 0 & 0\end{array}\right),
 \\ \Gamma_{01} \gamma^\mu = \Gamma_{10} \gamma^\mu  \left( \begin{array}{cc} 0 & 0 \\ 0 & 1\end{array}\right).
\end{align}
\end{subequations}
 With these relations, substituting the above $\Gamma$-matrices into Eq.(\ref{HH-1}), one gets that
\begin{subequations}\label{H-Z}
\begin{align}
  & \HH^{\nu \ov \nu Z}   =  \  \frac 12 \ww\psi^{L}_{\nu} \gamma^\mu \psi_\nu^L \psi_{Z\mu}, \label{H-Z-nu}
 \\ & \HH^{e\ov e Z}  =  \frac{1}{4} \left( \ww\psi^{L}_{e} \gamma^\mu  \psi^L_e 
 -  \ww\psi^{R}_{e} \gamma^\mu  \psi^R_e  \right) \psi_{Z\mu},  \label{H-Z-e}
 \\ \notag & \HH^{u_i\ov u_i Z} 
=  \Big( -\frac 12 \ww\psi^{L}_{u_i} \gamma^\mu \psi_{u_i}^L     
 \\  & \hspace{1.5cm} 
 +\frac{1}{6}   \ww\psi^{L}_{u_i} \gamma^\mu \psi^L_{u_i} 
 - \frac 16 \ww\psi^{R}_{u_i} \gamma^\mu \psi^R_{u_i}  \Big) \psi_{Z\mu}, \label{H-Z-u}
 \\ &  \HH^{d_i\ov d_i Z}  = - \frac{1}{12} \left( \ww\psi^{L}_{d_i} \gamma^\mu  \psi^L_{d_i} 
 - \ww\psi^{R}_{d_i} \gamma^\mu  \psi^R_{d_i}\right) \psi_{Z\mu}. \label{H-Z-d}
\end{align}
\end{subequations}
 Similarly,  for BPIPs that involve $W$-type modes, 
 from the $\Gamma$-matrices given in Eqs.(\ref{Gamma-neW}) and (\ref{Gamma-udW}), one gets that
\begin{subequations}\label{H-W}
\begin{align}
 & \HH^{\nu \ov e W}  = \frac{1}{2\sqrt 2}  \ww\psi^{L}_{e} \gamma^\mu \psi_\nu^L \psi_{W\mu},
 \\ & \HH^{e \ov \nu W}  = \frac{1}{2\sqrt 2}  \ww\psi^{L}_{\nu} \gamma^\mu \psi_e^L \psi_{\ov W\mu},
 \\ & \HH^{u_i\ov d_i W}   
  = -\frac{1}{2\sqrt 2}   \ww\psi^{L}_{d_i} \gamma^\mu  \psi_{u_i}^L \psi_{W\mu},
  \label{HH-ud-explicit}
 \\ & \HH^{d_i\ov u_i W}   
  = -\frac{1}{2\sqrt 2}   \ww\psi^{L}_{u_i} \gamma^\mu  \psi_{d_i}^L \psi_{\ov W\mu}.
  \label{HH-du-explicit}
\end{align}
\end{subequations}

 Finally, following a procedure similar to that given above, 
 one may derive expressions for the $H$-operator densities of those BPIPs that involve $u/d$-type modes and $g$-modes.
 This can be done in a straightforward way, 
 by making use of the $\Gamma$-matrices given in Eqs.(\ref{G-uug}), (\ref{G-ddg}), and (\ref{Gamma-qijg-all}),
 but, we are not to write them explicitly.
 
\section{Comparison with the SM}\label{sect-compare-SM}

 In this section, we give a comparison between the interaction $H$-operators of the above-discussed model 
 and interaction  Hamiltonians given in the SM. 
 We mainly discuss electroweak interactions and
 give a brief discussion for strong interactions at the end of this section.

 We use $\HH_{\rm gws}^{\rm int}$ to denote the part of the Hamiltonian density given in the SM,
 which is for interactions between the first-generation fermions (electron, electron neutrino, up quark, 
 down quark, and their anti particles) and vector bosons.
 In the ordinary notation (e.g., in Ref.\cite{Peskin}), $\HH_{\rm gws}^{\rm int}$ is written as
\begin{align}\label{}
 \HH_{\rm gws}^{\rm int} = -e A_\mu J^{\mu}_{EM}
 - g(W^+_\mu J^{\mu +}_W + W^-_\mu J^{\mu -}_W + Z^0_\mu J^{\mu}_Z), 
\end{align}
 where the currents are given by
\begin{subequations}\label{H-GWS}
\begin{align}\label{GWS-EM}
  & J^{\mu}_{EM}=\ov e \gamma^\mu (-1)e + \ov u \gamma^\mu (+\frac 23)u + \ov d \gamma^\mu (-\frac 13)d,
 \\ & J^{\mu }_Z  = \frac{1}{\cos \theta_{W}} \Big[ 
 J^{\mu }_{Z-\nu} + J^{\mu }_{Z-e} + J^{\mu }_{Z-u} + J^{\mu }_{Z-d} \Big], \label{GWS-Z}
 \\ \label{GWS-W+}
  & J^{\mu +}_W = \frac{1}{\sqrt 2} (\ov \nu_L \gamma^\mu e_L + \ov u_L \gamma^\mu d_L ),
 \\  & J^{\mu -}_W = \frac{1}{\sqrt 2} (\ov e_L \gamma^\mu \nu_L + \ov d_L \gamma^\mu u_L ). \label{GWS-W-}
\end{align}
\end{subequations}
 Current terms on the rhs of Eq.(\ref{GWS-Z}) are given by
\begin{subequations}\label{JZ-all}
\begin{align}\label{JZ-nu}
 & J^{\mu }_{Z-\nu} =  \ov \nu_L \gamma^\mu (\frac 12 ) \nu_L,
 \\ & J^{\mu }_{Z-e} =  \ov e_L \gamma^\mu (-\frac 12 + \sin^2 \theta_{W}) e_L 
 + \ov e_R \gamma^\mu (\sin^2 \theta_{W}) e_R, \label{JZ-e}
 \\  &  J^{\mu }_{Z-u} =   \ov u_L \gamma^\mu (\frac 12 -\frac 23 \sin^2 \theta_{W}) u_L 
 + \ov u_R \gamma^\mu (-\frac 23 \sin^2 \theta_{W}) u_R, \label{JZ-u}
 \\  &  J^{\mu }_{Z-d} =  \ov d_L \gamma^\mu (-\frac 12+\frac 13 \sin^2 \theta_{W}) d_L 
 + \ov d_R \gamma^\mu (\frac 13 \sin^2 \theta_{W}) d_R, \label{JZ-d}
\end{align}
\end{subequations}
 where $\theta_{W}$ indicates the Weinberg angle. 
 In Eqs.(\ref{H-GWS})-(\ref{JZ-all}), the overlines have the ordinary meaning;
 that is, e.g., $\ov \psi = \psi^\dag \gamma^0$.

 Below, we give a comparison between $\HH_{\rm gws}^{\rm int}$
 and the $H$-operator densities given in Eqs.(\ref{H-A})-(\ref{H-W}).

 Firstly, we discuss the electromagnetism interaction.
 There is no essential difference between the form of
 the term ``$-eA_\mu J^{\mu}_{EM}$'' with $J^{\mu}_{EM}$ given in Eq.(\ref{GWS-EM})
 and that of the densities given in Eq.(\ref{H-A}) for the $A$-mode. 
 In fact, there are only two differences in their forms:
 \\ (a) In the place where $\HH_{\rm gws}^{\rm int}$ has a coupling constant $e$, 
 the $H$-operator densities have a number factor ($1/4$).
 This a feature common to all terms and we are not to mention it below. 
 \\ (b) The up-quark and down-quark terms have negative and positive signs, respectively;
 meanwhile, the $u$-mode and $d$-mode terms have reverse signs, i.e., positive and negative signs, respectively. 
 Clearly, this difference in sign has no physical significance.

 Secondly, let us discuss interactions with $Z^0$-boson. 
 \\ (i) The current $J^{\mu }_{Z-\nu}$ in Eq.(\ref{JZ-nu}) for the neutrino field, 
 if added by the $Z^0$-boson field, has the same form as the $H$-operator density
 $\HH^{\nu \ov \nu Z}$ in Eq.(\ref{H-Z-nu}) for the $\nu$-mode field.
 \\ (ii) The current $J^{\mu }_{Z-e}$ in Eq.(\ref{JZ-e}) and the $H$-operator density $\HH^{e\ov e Z}$
 in Eq.(\ref{H-Z-e}) also have a similar formal structure.
 Their main difference lies in coefficients contained;
 in particular, the former contains the Weinberg angle, while, the latter does not. 
 Interestingly, they would have the same coefficients, if one replaces the term $\sin^2 \theta_{W}$ in the former
 by the number of $\frac 14$. 
 We observe that experimental results, interpreted within the framework of the SM, give a close value, i.e., 
 $\sin^2 \theta_{W} \simeq 0.23$ \cite{RMP13-Sirlin}.
  \\ (iii) The currents $J^{\mu }_{Z-u}$ and $J^{\mu }_{Z-d}$ in Eq.(\ref{JZ-all}) also have a structure formally
 similar to that of the $H$-operators $\HH^{u_i\ov u_i Z}$ and $\HH^{d_i\ov d_i Z}$ in Eq.(\ref{H-Z}).
 But, there exist some further differences in the coupling coefficients, 
 which are seen clearly in Table \ref{table-coeff}.
 Among the differences, the most significant one is between 
 the left-handed part of  $-J^{\mu }_{Z-d}$ and the corresponding part of $\HH^{d_i\ov d_i Z}$.

\begin{table}[!]
\begin{tabular}{|c|c|c|c | c|}
  \hline
   & $-J^{\mu }_{Z-u}$ & $\HH^{u_i\ov u_i Z}$  & $-J^{\mu }_{Z-d}$ & $\HH^{d_i\ov d_i Z}$ \\ 
 \hline  left-handed & $-\frac 12 + \frac{0.92}{6}$ & $-\frac 12 + \frac{1}{6}$ & $\frac 12 - \frac{0.23}{3}$ & $-\frac{1}{12}$ \\
  \hline right-handed & $\frac{0.92}{6}$  & $-\frac 16$ & $ - \frac{0.92}{12}$ & $\frac 1{12}$ \\
  \hline
\end{tabular}
  \caption{Coupling coefficients of the left-handed and right-handed parts
  of $-J^{\mu }_{Z-u}$ and $-J^{\mu }_{Z-d}$ in Eq.(\ref{JZ-all}), with $\sin^2 \theta_{W}$ taken as $ 0.23$, 
  in comparison with the corresponding ones of 
  the $H$-operator densities $\HH^{u_i\ov u_i Z}$ and $\HH^{d_i\ov d_i Z}$ in Eq.(\ref{H-Z}).}
  \label{table-coeff}
\end{table}

 Thirdly, let us discuss the currents $J^{\mu +}_W$ and $J^{\mu -}_W$  in Eqs.(\ref{GWS-W+})-(\ref{GWS-W-}),
 which are for interactions mediated by $W^\pm$-bosons given in the GWS theory.
 One sees that there is no essential difference between the form of this part of $\HH_{\rm gws}^{\rm int}$
 and that of the $H$-operator densities for the $W$-type modes given in Eq.(\ref{H-W}).

 In addition, there is a difference related to interactions among bosons. 
 In fact, $W^\pm$-bosons and $Z^0$-bosons may interact directly in the SM, while, 
 bosonic modes do not interact directly in a theory of modes. 
 This difference does not necessarily imply a real confliction, 
 because it is possible for bosonic particles to interact effectively in a mode model, 
 if the particles are interpreted as eigenstates of certain (sums of) $H$-operators.

 Finally, we give a brief discussion for strong interactions.
 In the SM, the first generation has two quarks --- the up and down quarks, 
 each possessing a color degree of freedom with three colors,
 and there are eight gluons that mediate interactions of quarks. 
 The studied model of modes has corresponding properties. 
 Specifically, it contains two types of six-layer fermionic modes
 --- $u$-type mode and $d$-type mode, each labelled by an index $i$ that takes three values, 
 and there are eight independent $g$-modes with which these two types of fermionic modes interact. 
 The index $i$ gives an interpretation to the color degree of freedom.

 Differences also exist between the two models.
 For example, interactions among $u$-modes behave differently from those among $d$-modes 
 [cf.~Eq.(\ref{G-uug}) and Eq.(\ref{G-ddg})], while, the $SU(3)$ group does not predict any difference
 between interactions among up quarks and those among down quarks. 
 However, detailed comparison of exact expressions for the strong interaction predicted by the two models
 is a complicated topic, which depends on the representations employed for particle/mode states.

\section{Summary and discussions}\label{sect-conclusion}

 In this section, we summarize results obtained in this paper and give some further discussions.
 We discuss the framework of mode theory in Sec.\ref{sect-summary-framework}
 and discuss the studied model of modes in Sec.\ref{sect-summary-model}.
 In Sec.\ref{sect-dark-matter}, we show that the studied model contains a candidate for dark matter. 
 
\subsection{About the framework of mode theory}\label{sect-summary-framework}

 In this paper, a framework has been proposed for developing theories 
 directly on a quantum ground for elementary physical entities, which we call modes. 
 State spaces of modes are built 
 by making use of representation spaces of the Lorentz group, more exactly, of the $SL(2,C)$ group.

 The framework is mainly based on five BAs (basic assumptions), which include the following key points.
 \\ (i) Modes differ  in their spinor-state spaces.
 \\ (ii) Spinor states of modes have a layer structure.
 \\ (iii) There are  three types of {fundamental processes}: free process, VF (vacuum fluctuation), 
 and FIP (fundamental interaction process, 
   as the change of two fermionic modes into one bosonic mode, or the reverse).
 \\ (iv) Fermionic modes with negative $p^0$ may exist instantly only.
 \\ (v) Finite time evolution of states obeys a Schr\"{o}dinger-type equation
 governed by a time evolution operator.

 In addition to the five BAs,  three important assumptions have been used in the proposed framework, as listed below:
 \\ (a)  Eq.(\ref{SAB-ov-commu-two}) for fermionicness of Weyl spinors within unit modes,
 as well as for computation of $H$-operator amplitudes;
 \\ (b) assumptions stated around Eq.(\ref{<s|s>=0}) about peculiar properties of $s(\ov s)$-modes;
 \footnote{ The peculiar properties of $s(\ov s)$-mode may be regarded as coming from Eq.(\ref{SAB-ov-commu-two}).
 }
 \\ (c) hermiticity of the time evolution operator.
 \\ Besides,  relatively ``minor'' assumptions have been used, which look natural from the viewpoint of the theory of spinors.

 Associated with each fundamental process, 
 an operator called $H$-operator is considered,
 which maps  the state space of the incoming mode(s) of the process to that of its outgoing mode(s).
 BPIPs (basic positive-$p^0$ interaction process),
 whose incoming and outgoing modes do not include any negative-$p^0$ fermionic mode,
 are constructed from FIPs and VF (if needed), with $H$-operators constructed accordingly. 
 The time evolution operator is given by the sum of $H$-operators of all 
 free modes ($p^0 > 0$ for fermionic modes) and of all BPIPs.
 It turns out that this operator is a function of 
 quantum fields, which are constructed from creation and annihilation operators for free-mode states. 
 Written with quantum fields, the interaction $H$-operator automatically gets a local feature.

 Compared with the framework supplied by QFT which is usually employed in the field of particle physics,
 the proposed framework has the following merits. 
 That is, (i) it is directly founded on a quantum ground and, hence, no procedure of quantizing
 classical field is needed. 
 (ii) Certain unification in the description of mode states is achieved at the fundamental level, 
 in the sense that  spinor-state spaces of all modes
 are built from the two smallest nontrivial representation spaces of the $SL(2,C)$ group.
 And, (iii) certain unification in the description of interactions is also achieved at the fundamental level,
 in the sense that interaction operators for fundamental processes have the common feature of
 mapping the state space of incoming mode(s) to that of outgoing mode(s).

\subsection{About the model of modes}\label{sect-summary-model}

 A model of modes has been constructed and studied within the above framework,
 with three model assumptions addressing three things:
 (a) exact layer structures of physical modes,  (b) layer phases,  and (c)  exact forms of $T_k$-matrices.
 These model assumptions are introduced, based on properties of spinor states of modes, 
 particularly on certain features of layer structures of modes.

 The studied model contains four types of fermionic modes --- $\nu$-mode, $e$-mode, $u$-mode, and $d$-mode,
 and four types of bosonic modes --- $A$-mode, $Z$-mode, $W$-mode, and $g$-mode (including their antimodes).
 These modes are in parallel with elementary particles in the SM,
 more exactly, with the four first-generation fermions of  electron neutrino, electron, up quark, and down quark, 
 and with the four bosons of photon, $Z^0$-boson, $W$-boson, and gluon, respectively.

 The time evolution operator of the model consists of three parts.
 The first part has a form similar to the electroweak interaction Hamiltonian (with fermions) in the SM, 
 with some differences in the coupling coefficients.
 The second part shows some global similarities to the interaction Hamiltonian given for strong interactions
 in the SM; for example, the number of bosonic modes that mediate the interactions is also eight.
 But, a full comparison of details of the second part with the strong-interaction part of the SM 
 is not available at the present stage for some technical reasons;
 in particular, (future) research work is needed
 in order to find representations of particle/mode states that are  appropriate for such a comparison. 
\footnote{In principle, by imposing sufficiently many model assumptions, 
 it should be possible to get a mode model whose time evolution operator has exactly the same form as 
 the corresponding part in the total Hamiltonian of the SM.
 But, such a possibility is not of particular interest, because  many assumptions would be needed 
 to achieve this goal. 
} 
 The third part was not discussed previously and is to be discussed in the next section.

 The studied model gives interesting predictions, which are not available in the SM. 
 Rephrased in the terminology of the SM, some major ones are listed below:
 (i) Up and down quarks should have $\frac 23$ and $\frac 13$ electronic charges, respectively, 
 in unit of the charge of electron. 
 (ii) Neutrino should have zero electronic charge.
 (iii) The Weinberg angle may correspond to $\sin^2 \theta_{W} = 0.25$. 
 \footnote{This prediction is not necessarily in confliction with the known result of $\sin^2 \theta_{W} \simeq 0.23$,
 because this value was computed within the theoretical framework of the SM. }
 And, (iv) the color degree of freedom of quarks may correspond to an index $i=1,2,3$,
 which labels certain layer configuration of mode states.

 At the present stage, it is too early to discuss direct comparison of predictions of the studied model 
 and experimental results of particles.
 In fact, for this purpose, an interpretation is needed to experimentally observed (free) particle states.
 In a natural interpretation, one may relate these states to eigenstates of certain (sums of) $H$-operators. 
 According to experiences obtained in the study of experimental data with the SM,  
 a long way may be expected ahead to get a real comparison for a model of modes.

 Anyway, it is possible for future study to show some need of modification to the mode model.
 In this case, one possible direction of modification is to find some other structural feature of spinor states involved and,
 based on it, introduce some new model assumption, e.g., for a new $T_k$ matrix. 
 Another direction is to make use of some symmetry consideration, perhaps, like gauge symmetries in the SM. 
 But, presently, the situation with the second direction is still obscure, 
 because gauge symmetries in the SM are first introduced for classical fields, 
 while, the proposed framework is already quantum mechanical in the beginning.

\subsection{A candidate for dark matter?} \label{sect-dark-matter}

 The studied model of modes has a part that have not been discussed. 
 This is related to the rest two three-layer fermionic modes mentioned in Sec.\ref{sect-six-layer-modes-I}
 (besides $t$-modes).
 In this section, we discuss six-layer modes that are constructed from these two three-layer modes,
 and show that they are candidates for dark matter.

 We call the two three-layer fermionic modes mentioned above
 \emph{$t_\nu$-mode} and \emph{$\ov t_\nu$-mode}.
 They have the following spinor states, 
\begin{subequations}\label{Snuq-ov}
\begin{gather}\label{Snuq}
 |\cs_{t_\nu}^{\alpha }(\bp)\ra = \frac{1}{\sqrt 3}
 \left( \begin{array}{c} |u^\alpha(\bp)\ra \\  |u^\alpha(\bp)\ra
 \\  |u^\alpha(\bp)\ra \end{array} \right),
 \\ \label{Sov-nuq}  |\cs_{\ov t_\nu}^{\alpha }(\bp)\ra = \frac{1}{\sqrt 3}
 \left( \begin{array}{c} |\ov v^\alpha(\bp)\ra \\ |\ov v^\alpha(\bp)\ra
 \\ |\ov v^\alpha(\bp)\ra \end{array} \right).
\end{gather}
\end{subequations}
 According to the assumption ${\rm MA}_{\rm mode}^{\rm species}$, the above two modes give rise to two
 six-layer fermionic modes, which we call $u_\nu$-mode and $d_\nu$-mode, respectively, with the following spinor states,
\begin{subequations}\label{S-ud-nu}
\begin{align}\label{}
\label{Su-nu}
 & |\cs_{{u}_\nu}^{\alpha }(\bp)\ra = \frac{1}{\sqrt 2}
 \left( \begin{array}{c}   |\cs_{t_\nu}^{\alpha }(\bp)\ra  \\   |\cs_{t_\nu}^{\alpha }
 (\bp)\ra \end{array} \right),
 \\ \label{Sd-nu} & |\cs_{{d}_\nu}^{\alpha }(\bp)\ra = \frac{1}{\sqrt 2}
 \left( \begin{array}{c}   |\cs_{t_\nu}^{\alpha }(\bp)\ra  \\ i
 |\cs_{\ov t_\nu}^{\alpha }(\bp)\ra \end{array} \right).
\end{align}
\end{subequations}
 Their antimodes are indicated as $\ov u_\nu$-mode and $\ov d_\nu$-mode, respectively, 
 with spinor states denoted by $|\cs_{{\ov u}_\nu}^{\alpha }(\bp)\ra$ and $|\cs_{{\ov d}_\nu}^{\alpha }(\bp)\ra$.

 Here, we are interested in interactions of the above six-layer modes with the $A$-mode only.
 It is the matrices $T_2$ and $T_3$ that may predict such interactions.
 Let us first discuss the $u_\nu$-mode. 
 The matrix $T_2$ effectively predicts a one-layer structure from  
 $ \overrightarrow D |\cs_{{u}_\nu}^{\alpha }(\bp)\ra|\cs_{{\ov u}_\nu}^{\alpha' }(\bp')\ra$,
 which does not correspond to any physical state.
 Meanwhile, the action of $T_3$ on
 $ \overrightarrow D |\cs_{{u}_\nu}^{\alpha }(\bp)\ra|\cs_{{\ov u}_\nu}^{\alpha' }(\bp')\ra$
 produces the same result as that for $ \overrightarrow D |\cs_{{u}_1}^{\alpha }(\bp)\ra|\cs_{{\ov u}_1}^{\alpha' }(\bp')\ra$
 and gives a vanishing $\Gamma$-matrix as in Eq.(\ref{Gamma-uuA-T3}). 
 Hence, one has
\begin{align}\label{}
 & \Gamma^{u_\nu \ov u_\nu \to A}_{(T_2)} = \Gamma^{u_\nu \ov u_\nu \to A}_{(T_3)} =0.
\end{align}

 Next, we discuss the $d_\nu$-mode. 
 Making use of the explicit expressions of spinor states given above,
 we note that the first three layers of 
 $\overrightarrow D |\cs_{{d}_\nu}^{\alpha }(\bp)\ra|\cs_{{\ov d}_\nu}^{\alpha' }(\bp')\ra$,
 which are obtained from the two upper sections, are identical; 
 and similar for the last three layers from lower sections.
 This implies that the action of $T_2$ on 
  $\overrightarrow D |\cs_{{d}_\nu}^{\alpha }(\bp)\ra|\cs_{{\ov d}_\nu}^{\alpha' }(\bp')\ra$
 should produce the same result as $T_1$ on 
 a pair of $e$-mode and $\ov e$-mode, namely on 
 $\overrightarrow D |\cs_{{e}}^{\alpha }(\bp)\ra|\cs_{{\ov e}}^{\alpha' }(\bp')\ra$.
 Hence, according to Eq.(\ref{Gamma-eeA}), one has 
 $\Gamma^{d_\nu \ov d_\nu \to A}_{(T_2)} =  \frac{1}{4}  \Gamma_{11}$.
 As for $T_3$, similar to the above-discussed case of $u_\nu$-mode, like Eq.(\ref{Gamma-ddA-T3}), one has 
 $\Gamma^{d_\nu \ov d_\nu \to A}_{(T_3)} =   - \frac 14  \Gamma_{11}$. 
 As a result,
\begin{align}\label{}
 \Gamma^{d_\nu \ov d_\nu \to A} = \Gamma^{d_\nu \ov d_\nu \to A}_{(T_2)}
 + \Gamma^{d_\nu \ov d_\nu \to A}_{(T_3)} =0. 
\end{align}
 Therefore, neither the $u_\nu$-mode nor the $d_\nu$-mode interacts with the $A$-mode.
 This is just a feature expected for dark matter.

 \acknowledgements

 The author is grateful to Yan Gu for valuable discussions and suggestions.
 This work was partially supported by the National Natural Science Foundation of China under Grant
 Nos.~11535011, 11775210, and 12175222.


\appendix

\section{Abstract notation for spinors}\label{app-abs-notation}

 In this appendix, we first recall basic properties of two-component Weyl spinors 
 \cite{Penrose-book,CM-book,Corson,Kim-group},
 written  in the abstract notation as discussed in Ref.\cite{pra16-commu}
 (Sec.\ref{sect-recall-Weyl-spinor}).
 Then, we recall basic properties of four-component vectors (Sec.\ref{sect-recall-vector}) 
 and write them also in the abstract notation (Sec.\ref{sect-vector-abstract}).
 
 Concerning bras to be used in this appendix, we discuss only spinor bras for scalar product,
 which are indicated as $\la\la \cdot |$ in the main text.
 But, for brevity, within this appendix, we use the ordinary notation of $\la \cdot |$ to indicate them. 
 
\subsection{Basic properties of two-component spinors}\label{sect-recall-Weyl-spinor}

 In the spinor theory, there are two smallest nontrivial representation spaces of the $SL(2,C)$ group,
 which are spanned by two types of two-component Weyl spinors, respectively,
 with the relationship of complex conjugation.
 The two spaces are denoted by $\WW$ and $\ov \WW$, 
 with overline indicating consequence of the operation of complex conjugation, as a convention in the theory of spinors.

 In terms of components, a Weyl spinor in $\WW$ is written as, say,
 $\kappa_A$ with an index $A=0,1$.
 In the abstract notation, a basis in the space $\WW$ is written as $|S^{A}\ra $
 and the above spinor is written as $|\kappa\ra$, with the expansion
\begin{equation}\label{|kappa>}
  |\kappa\ra = \kappa_A|S^{A}\ra,
\end{equation}
 with repeated index implying summation.
 \footnote{In Ref.\cite{pra16-commu}, Eq.(\ref{|kappa>}) is written as $|\kappa\ra = \kappa^A|S_{A}\ra
 =-\kappa_A|S^{A}\ra$,
 which would introduce a minus sign to the rhs of the two equalities in Eq.(\ref{kappa-A}).
  }
 One may introduce a space that is dual to $\WW$, composed of bras with a basis written as $\la S^A|$.
 In order to construct a product that is a scalar under $SL(2,C)$ transformations
 [see Eq.(\ref{ww-chi-kappa=chi-kappa})],
 the bra dual to the ket $|\kappa\ra$ should be written as
\begin{equation}\label{<kappa|}
  \la \kappa | =  \la S^{A}|  \kappa_A,
\end{equation}
 which has the same components as $|\kappa\ra$ in Eq.(\ref{|kappa>}),
 but not their complex conjugates.
 (See Appendix \ref{sect-SL2C-transf} for a brief discussion of basic properties of $SL(2,C)$ transformations.)

 By definition, one may set scalar products of the basis as
\begin{equation}\label{SA-SB}
  \la S^{A}|S^{B}\ra = \epsilon^{A B},
\end{equation}
 where $\epsilon^{AB}$ is a symbol defined by
\begin{equation}\label{epsilon}
 \epsilon^{AB} = \left( \begin{array}{cc} 0 & 1 \\ -1 & 0 \end{array} \right).
\end{equation}
 It proves convenient to introduce another matrix $\epsilon_{AB}$, which has the same
 elements as $\epsilon^{AB}$.
 These two matrices can be used to raise and lower indexes of components,  say,
\begin{equation}\label{A-raise}
  \kappa^A = \epsilon^{AB} \kappa_B, \quad \kappa_A = \kappa^B \epsilon_{BA},
\end{equation}
 as well as for the basis spinors, namely,
\begin{gather}\label{}
 |S^A\ra = \epsilon^{AB}|S_B\ra, \quad |S_A\ra = |S^B\ra \epsilon_{BA} .
\end{gather}
 It is not difficult to verify that (i) $\la S_{A}|S_{B}\ra = \epsilon_{A B}$; (ii)
\begin{equation}\label{f-AB}
  {f_{\ldots}^{\ \ \ A}\ (g)^{  \cdots }}_{ A } = - {f_{\ldots A}\ (g)^{\cdots A}},
\end{equation}
 i.e., a minus sign is introduced by the exchange of the positions of a pair of repeated two-component index;
 and (iii) the symbols $\epsilon_C^{\ \ A} =\epsilon^{BA} \epsilon_{BC}$ and
 $\epsilon^A_{\ \ C} =\epsilon^{AB} \epsilon_{BC}$ satisfy the relation:
\begin{gather}\label{eps-delta}
 \epsilon_C^{\ \ A} = -\epsilon^A_{\ \ C} = \delta^A_C,
\end{gather}
 where $\delta^A_B=1$ for $A=B$ and $\delta^A_B=0$ for $A \ne B$.
 ($\delta$-symbols for other types of indices are defined in the same way.)

 The scalar product of two generic spinors $|\chi\ra$ and $|\kappa\ra$, written as
 $\la\chi|\kappa\ra$, has the expression of
\begin{equation}\label{<chi|kappa>}
 \la\chi|\kappa\ra = \chi_A \kappa^A.
\end{equation}
 The anti-symmetry of $\epsilon^{AB}$ implies that
\begin{equation}\label{ck=-kc}
  \la \chi |\kappa\ra = -\la\kappa |\chi\ra
\end{equation}
 and, as a consequence, $\la \kappa |\kappa\ra =0$ for all $|\kappa\ra$.
 It is not difficult to verify the following properties:
 (a) The identity operator in the space $\WW$, denoted by  $I_{\WW}$, can be written as
\begin{eqnarray}\label{I-W}
 I_{\WW} = |S^{A}\ra \la S_{A}|,
\end{eqnarray}
 satisfying $I_{\WW}|\kappa\ra =|\kappa\ra $ for all $|\kappa\ra \in \WW$;
 and (b) the components of $|\kappa\ra$
 have the following expressions,
\begin{equation}\label{kappa-A}
  \kappa^A = \la S^{A}|\kappa\ra, \quad \kappa_A = \la S_{A}|\kappa\ra.
\end{equation}

 The operation of complex conjugation converts $\WW$ to $\ov\WW$,
 and changes spinors $|\kappa\ra \in \WW$ to spinors $|\ov\kappa\ra \in \ov\WW$.
 Corresponding to the basis $|S_{A}\ra \in \WW$, the space $\ov \WW$ has a basis
 denoted by $|\ov S_{A'}\ra$ with a primed index $A' = 0', 1'$.
 On the basis of $|\ov S_{A'}\ra$, $|\ov\kappa\ra$ is written as
\begin{equation}\label{ov-kappa}
  |\ov\kappa \ra = {\ov\kappa}_{A'}|\ov S^{A'}\ra,
\end{equation}
 where
\begin{equation}\label{}
  \ov\kappa^{A'} := (\kappa^A)^* .
\end{equation}
 Similar to the $\epsilon$ symbols discussed above,
 one introduces symbols $\epsilon^{A'B'}$ and $\epsilon_{A'B'}$,
 which have the same matrix elements as $\epsilon^{AB}$ and are used to raise and lower primed indices.
 The identity operator in the space $\ov\WW$, denoted by $I_{\ov\WW}$,
 has the following form 
\begin{align}\label{I-ovW}
 I_{\ov\WW} = |\ov S^{A'}\ra \la \ov S_{A'}|.
\end{align}
 When a spinor $\kappa^A$ is transformed by an $SL(2,C)$ matrix,
 the spinor $\ov\kappa^{A'}$ is transformed by the complex-conjugate matrix
 (see Appendix \ref{sect-SL2C-transf}).

 \subsection{Basic properties of four-component vectors}\label{sect-recall-vector}

 In this section, we recall basic properties of four-component vectors
 as special cases of spinors \cite{Penrose-book,Corson,CM-book}.
 We use the ordinary notation in this section and
 will discuss the abstract notation in the next section.

 A key point is a one-to-one mapping between a direct-product
 space $\WW \otimes\ov\WW$ and a four-dimensional space denote by $\VV$.
 The mapping is given by the so-called Enfeld-van der Waerden symbols, 
 in short \emph{EW-symbols}, denoted by $\sigma^{\mu AB'}$,
  with $\mu =0,1,2,3$, which are invariant under $SL(2,C)$ transformations.
 An often-used set of explicit expressions for these symbols is given by
 \footnote{Compared with Pauli matrices, the EW-symbols have a common prefactor $1/\sqrt 2$.
 This leads to the relations in Eq.(\ref{st-delta}), otherwise, the
 rhs of the two equalities in it should be multiplied by $2$.
 (This  prefactor $1/\sqrt 2$ was not used in Ref.\cite{pra16-commu}.)}
\begin{eqnarray}\notag
 & \sigma^{0AB'} = \frac{1}{\sqrt 2} \left(\begin{array}{cc} 1 & 0 \\ 0 & 1 \\ \end{array} \right),
 \sigma^{1AB'} = \frac{1}{\sqrt 2} \left(\begin{array}{cc} 0 & 1 \\ 1 & 0 \\ \end{array} \right), \ \ \ 
 \\ & \sigma^{2AB'} = \frac{1}{\sqrt 2} \left(\begin{array}{cc} 0 & -i \\ i & 0 \\ \end{array} \right),
 \sigma^{3AB'} = \frac{1}{\sqrt 2} \left(\begin{array}{cc} 1 & 0 \\ 0 & -1 \\ \end{array} \right).\ \
 \label{sigma^AB}
\end{eqnarray}
 Lowering the two-component indices like in Eq.(\ref{A-raise}), it is straightforward to verify that
\begin{eqnarray}\notag
 & \sigma^{0}_{AB'} = \frac{1}{\sqrt 2} \left(\begin{array}{cc} 1 & 0 \\ 0 & 1 \\ \end{array} \right),
 \sigma^{1}_{AB'} =  \frac{1}{\sqrt 2} \left(\begin{array}{cc} 0 & -1 \\ -1 & 0 \\ \end{array} \right),
 \\ & \sigma^{2}_{AB'} = \frac{1}{\sqrt 2} \left(\begin{array}{cc} 0 & -i \\ i & 0 \\ \end{array} \right),
 \sigma^{3}_{AB'} =  \frac{1}{\sqrt 2} \left(\begin{array}{cc} -1 & 0 \\ 0 & 1 \\ \end{array} \right). \label{sig-lower-AB}
\end{eqnarray}

 For example, a spinor $\phi_{AB'}$ in the space $\WW \otimes\ov\WW$ can be mapped to
 a vector $K^\mu$ in the space $\VV$ by the EW-symbols, 
\begin{equation}\label{map-WW-V}
  K^\mu = \sigma^{\mu AB'} \phi_{AB'}.
\end{equation}
 In the space $\VV$, of particular importance is a symbol $g^{\mu\nu}$, which is defined by
\begin{equation}\label{g-sig}
 g^{\mu\nu} := \sigma^{\mu AB'} \sigma^{\nu CD'} \epsilon_{AC} \epsilon_{B'D'}.
\end{equation}
 One may introduce a lower-indexed symbol $g_{\mu\nu}$, which by definition 
 has the same matrix elements as $g^{\mu\nu}$, namely, $[g^{\mu\nu}] = [g_{\mu\nu}]$.
 Then, like the $\epsilon$-symbols in the space $\WW$,
 these two symbols of $g$ can be used to raise and lower indexes, e.g.,
\begin{equation}\label{mu-raise}
  K_\mu = K^\nu g_{\nu \mu}, \quad K^\mu = g^{\mu\nu} K_\nu.
\end{equation}
 Making use of the anti-symmetry of the $\epsilon$ symbols,
 it is easy to verify that $g^{\mu\nu}$ is symmetric, i.e.,
\begin{equation}\label{g-sym}
 g^{\mu\nu} = g^{\nu\mu}.
\end{equation}
 Due to this symmetry, the upper-lower positions of a pair of repeated
 index  are exchangeable, namely
\begin{equation}\label{f-munu}
  {F_{\ldots}^{\ \ \ \mu}(f)^{ \cdots }}_{\mu } = {F_{\ldots \mu}(f)^{\cdots \mu}}.
\end{equation}

 The EW-symbols have the following properties,
\begin{equation}\label{st-delta}
 \sigma^{AB'}_\mu \sigma_{CD'}^\mu = \delta^{AB'}_{CD'}, \quad
 \sigma_{AB'}^\mu \sigma^{AB'}_\nu = \delta_\nu^\mu,
\end{equation}
 where 
 $ \delta^{AB'}_{CD'} := \delta^{A}_{C} \delta^{B'}_{D'}$.
 Using the relations in Eq.(\ref{st-delta}), it is not difficult to check that the map
 from $\WW\otimes\ov\WW$ to $\VV$ given in Eq.(\ref{map-WW-V}) is reversible.
 Making use of Eqs.(\ref{st-delta}) and (\ref{eps-delta}), 
 it is not difficult to verify the following relation,
\begin{gather}\label{ss-ee-2}
 \sigma_{\mu AB'} \sigma_{CD'}^\mu = \epsilon_{AC} \epsilon_{B'D'}.
\end{gather}
 Then, substituting the definition of $g^{\mu\nu}$ in Eq.(\ref{g-sig})
 into the product of $g^{\mu\nu} g_{\nu \lambda}$, after simple algebra, one gets
\begin{gather} \label{ggd}
 g^{\mu\nu} g_{\nu \lambda} = g^\mu_{\ \ \lambda} = g^{\ \ \mu}_{\lambda} = \delta^\mu_\lambda .
\end{gather}

 When a $SL(2,C)$ transformation is carried out on the space $\WW$,
 a related transformation is applied to the space $\VV$.
 Requiring invariance of the EW-symbols,
 transformations on the space $\VV$ can be fixed
 and turn out to constitute a (restricted) Lorentz group
 (see Appendix \ref{sect-SL2C-transf} for detailed discussions.).
 Therefore, the space $\VV$ is a four-component vector space.
 In fact, substituting the explicit expressions of the EW-symbols in Eq.(\ref{sigma^AB})
 into Eq.(\ref{g-sig}), one gets that
\begin{eqnarray} \label{gmunu}
 g^{\mu\nu} = \sigma^{\mu}_{A'B} \sigma^{\nu A'B}
 =\left(\begin{array}{cccc} 1 & 0 & 0 & 0 \\ 0 & -1 & 0 & 0 \\ 0&0 & -1 &0
 \\ 0 &0 &0 & -1 \end{array} \right),
\end{eqnarray}
 which is just the Minkovski's metric.

 As shown in Appendix \ref{sect-SL2C-transf}  [Eq.(\ref{<K|J>})], the following product, 
\begin{gather}\label{Kmu-Jmu}
  J_\nu K^\nu = J^\mu g_{\mu\nu} K^\nu,
\end{gather}
 is a scalar under Lorentz transformations.
 Physically, of more interest is a product, in which one of the two vectors takes a complex-conjugate form, say, 
\begin{gather}\label{Kmu-Jmu*}
 J^*_\nu K^\nu = J^{\mu *} g_{\mu\nu} K^\nu.
\end{gather}
 Similarly,  one finds that this product is also a scalar.

 \subsection{Abstract notation for four-component vectors}\label{sect-vector-abstract}

 In the abstract notation,
 a basis in the space $\VV$ is written as $|T^\mu\ra $ with $\mu =0,1,2,3$.
 The index of the basis can be lowered by $g_{\mu\nu}$,
 i.e., $|T_\mu\ra = g_{\mu\nu} |T^\nu\ra $, and similarly $ |T^{\mu}\ra = g^{\mu \nu}|T_\nu\ra $.
 A generic four-component vector $|K\ra$ in the space $\VV$ is expanded as
\begin{equation}\label{K-Kmu-T}
 |K\ra = K_\mu |T^\mu\ra =  K^\mu |T_\mu\ra .
\end{equation}
 In consistency with the expression of bra in Eq.(\ref{<kappa|}) for two-component spinors,
 the bra corresponding to $|K\ra$ is written as
\begin{gather}\label{<K|}
 \la K| = \la T_\mu| K^\mu .
\end{gather}
 To write the scalar product in Eq.(\ref{Kmu-Jmu}) in the form of $\la J|K\ra$, namely,
\begin{align}\label{}
 \la J|K\ra =  J_\nu K^\nu,
\end{align}
 the following requirement is needed, 
\begin{eqnarray}\label{Tmu-Tnu}
  \la T^\mu|T^\nu\ra = g^{\mu\nu},
\end{eqnarray}
 in a way similar to the case of two-component spinors [cf.~Eq.(\ref{SA-SB})].
 Due to the symmetry of $g^{\mu \nu}$ in Eq.(\ref{g-sym}), one sees that
\begin{align}\label{<J|K>=<K|J>}
 \la J|K\ra = \la K|J\ra.
\end{align}

 It is not difficult to verify the following properties.
 (i) Making use of Eq.(\ref{ggd}), one can verify that the identity operator
 in the space $\VV$ are written as
\begin{equation}\label{I-vector}
 I_{\VV} = |T_\mu\ra \la T^\mu | = |T^\mu\ra \la T_\mu|.
\end{equation}
 (ii) The components $K^\mu$ and $K_\mu$ have the following expressions,
\begin{equation}\label{K-mu}
  K^\mu = \la T^\mu |K\ra , \quad K_\mu = \la T_\mu |K\ra .
\end{equation}
 And, (iii) the symmetry of $g^{\mu\nu}$ implies that $\la T_\mu|T_\nu\ra = \la T_\nu|T_\mu\ra$,
 as a result,
\begin{eqnarray}\label{<K|J>=<J|K>}
  \la K|J\ra = \la J|K\ra
\end{eqnarray}
 for two arbitrary vectors $|K\ra$ and $|J\ra$ in the space $\VV$.

 Let us write basis spinors in the product space $\WW \otimes\ov\WW$ as
\begin{gather}\label{S-AB'}
 |S_{AB'}\ra := |S_A\ra | \ov S_{B'}\ra , \ \
 \la S_{B'A}| := \la \ov S_{B'}| \la S_A| .
\end{gather}
 For this product space, we assume 
 that Weyl spinors satisfy the anticommutability relations given in  Eq.(\ref{SAB-ov-commu-two}).
 It proves convenient to introduce an operator, which maps 
 the space $\WW \otimes \ov\WW$ to the space $\VV$ by the EM-symbols.
 It is just the operator $\sigma$ in Eq.(\ref{sigma}), i.e., 
\begin{equation}\notag \label{sigma-app}
 \sigma = \sigma^{\mu AB'} |T_\mu\ra \la S_{B'A}| \qquad \text{[Eq.(\ref{sigma}) of main text].}
\end{equation}
 For an arbitrary spinor $|\phi\ra = \phi_{AB'} |S^{AB'}\ra  \in \WW \otimes \ov\WW$,
 this operator $\sigma $ acts as
\begin{align}\label{sigma-phi}
 \sigma |\phi\ra =  \sigma^{\mu AB'} \phi_{AB'}   |T_\mu\ra. 
\end{align}
 Clearly,  for $|K\ra = \sigma |\phi\ra$, 
 this result is in agreement with Eq.(\ref{map-WW-V}). 
 Under the specific expressions of $\sigma^{\mu AB'}$ in Eq.(\ref{sigma^AB}),
 making use of Eq.(\ref{sigma-phi}) one finds that
\begin{subequations}\label{|T>-|SS'>}
\begin{align}\label{|T>-|SS'>1}
 |T^0\ra & = \frac{1}{\sqrt 2} \sigma ( |S^{11'}\ra + |S^{00'}\ra ),
 \\ |T^1\ra & = - \frac{1}{\sqrt 2} \sigma ( |S^{10'}\ra +|S^{01'}\ra ), \label{|T>-|SS'>2}
  \\  |T^2\ra & =   \frac{i}{\sqrt 2} \sigma(|S^{10'}\ra - |S^{01'}\ra),
 \\  |T^3\ra & = \frac{1}{\sqrt 2} \sigma (|S^{11'}\ra - |S^{00'}\ra).
\end{align}
\end{subequations}

 Since $\ov{|S_{AB'}\ra} =|S_{A'B}\ra= -|S_{BA'}\ra$, the operation of complex conjugation
 maps the space $\VV$ into itself.
 Hence, one may use the same index $\mu$ to index $\ov{|T_\mu\ra} $, the complex conjugates of $|T_\mu\ra $,
 and write it as $|\ov T_\mu\ra $.
 Making use of Eq.(\ref{sigma}), the complex conjugate of the operator $\sigma$, namely $\ov\sigma$,  is written as
\begin{equation}\label{ov-sigma}
 \ov\sigma = \ov\sigma^{\mu A'B} |\ov T_\mu\ra \la S_{BA'}|,
\end{equation}
 where $\ov\sigma^{\mu A'B} \equiv (\sigma^{\mu AB'})^*$.

 One sees that both $|T_\mu\ra$ and $|\ov T_\mu\ra$ give a basis for the space $\VV$.
 In fact, one has a freedom of imposing some relation between these two bases, 
 which would consequently implies certain relation between the two operators of $\sigma$ and $\ov \sigma$,
 and {\it vice versa}. 
 We found it convenient to set equivalence of $\sigma$ and $\ov \sigma$, i.e., 
\begin{equation}\label{ov-s=s}
 \ov \sigma = \sigma .
\end{equation}
 Moreover,  we note that the explicit expressions of the EM-symbols in Eq.(\ref{sigma^AB})
 imply hermiticity of the EW-symbols, i.e., 
\begin{equation}\label{sig-c}
 \ov\sigma^{\mu B'A}= \sigma^{\mu AB'}.
\end{equation}
 Then, we find that $\ov\sigma$ in Eq.(\ref{ov-sigma}) is written as
\begin{equation}\label{ov-sigma-2}
 \ov\sigma = \sigma^{\mu BA'} |\ov T_\mu\ra \la S_{BA'}|  = - \sigma^{\mu AB'} |\ov T_\mu\ra \la S_{B'A}|,
\end{equation}
 where Eq.(\ref{SA-ovSB-commu}) has been used in the derivation of the second equality.
 From Eqs.(\ref{sigma}), (\ref{ov-s=s}), and (\ref{ov-sigma-2}), one sees that
\begin{equation}\label{ovT=T}
 {|\ov T_\mu\ra }=-|T_\mu\ra \Longleftrightarrow \la \ov T_\mu|=- \la T_\mu| .
\end{equation}

 Let us write the scalar product in Eq.(\ref{Kmu-Jmu*}) in the abstract form.
 Making use of Eq.(\ref{ovT=T}), one finds that
\begin{subequations}\label{}
\begin{align}\label{|ovK>}
 |\ov K\ra = -K^{\mu *} |T_\mu\ra,
 \\ \la \ov J | = -\la T_\mu| J^{\mu *}.
\end{align}
\end{subequations}
 Then, one gets that
\begin{gather}\label{<Kmu-Jmu*>}
 \la \ov J |K\ra = - J^*_{\mu} K^\mu = -J^{\mu *} g_{\mu\nu} K^\nu.
\end{gather} 
 Clearly, the minus signs on the rhs of Eq.(\ref{<Kmu-Jmu*>}) comes from those in Eq.(\ref{ovT=T}).
 We recall that the same minus sign appears in the symbol $\wh g$  [see Eq.(\ref{wh-g-g})].
 In fact, using the hat symbol, one writes that 
\begin{gather}\label{<Kmu-Jmu*-ghat>}
 \la \ov J |K\ra = J^{\mu *} \wh g_{\mu\nu} K^\nu.
\end{gather} 
 Moreover, it is easy to verify that
\begin{gather}\label{JK-KJ*}
 \la\ov K|J\ra ^* = \la\ov J|K\ra.
\end{gather}

 It proves convenient to introduce an operation of \emph{transposition} for generic kets and bras,
 indicated by a superscript $T$, which is defined by
\begin{eqnarray}\label{transpose}
 |\phi \psi\ra^T := \la \psi \phi|, \ \la \psi \phi|^T := |\phi \psi\ra, \
 (|\phi\ra \la \psi|)^T := |\psi\ra \la \phi |. \hspace{0.9cm}
\end{eqnarray}
 For example, $|S_{AB'}\ra^T =\la S_{B'A}|$ and 
\begin{equation}\label{sigma-T}
 \sigma^{T} = \sigma^{\mu AB'}|S_{AB'}\ra \la T_\mu|.
\end{equation}
 Computing the product of $\sigma$ in Eq.(\ref{sigma}) and $\sigma^T$ in Eq.(\ref{sigma-T}),
 noting Eqs.(\ref{Tmu-Tnu}) and (\ref{st-delta}), it is easy to verify that
\begin{equation}\label{s-T-s}
  \sigma^T\sigma = I_{\WW} \otimes I_{\ov\WW}, \quad \sigma\sigma^T =I_{\VV},
\end{equation}
 where the identity operators are given in Eqs.(\ref{I-W}), (\ref{I-ovW}), and (\ref{I-vector}). 
 This implies that $\sigma^T$ is the reverse of $\sigma$, denote by $\sigma^{-1}$, 
\begin{equation}\label{sigma-reverse}
  \sigma^{-1} = \sigma^T.
\end{equation}
 We use $\sigma^\dag$ to denote the complex conjugate of the transposition of $\sigma$, 
\begin{align}\label{sigma-dag}
 \sigma^\dag := \ov{ \left(\sigma^T \right)}.
\end{align}
 Direction derivation shows that
\begin{align}\label{sdag=sT}
 \sigma^\dag = \sigma^T.
\end{align}

\subsection{Some detailed derivations}\label{app-derive-AppA}

 In this section, we give detailed derivations of Eqs.(\ref{sigma-phi}), (\ref{|T>-|SS'>}), (\ref{|ovK>}), (\ref{s-T-s})
 and (\ref{sdag=sT}).

\vspace{0.3cm}

\noindent Derivation of Eq.(\ref{sigma-phi}):
\begin{gather*}\notag
 \sigma |\phi\ra = \sigma^{\mu AB'} |T_\mu\ra \la S_{B'A}| \phi_{CD'} |S^{CD'}\ra
 \\ = \sigma^{\mu AB'} |T_\mu\ra \phi_{CD'} \delta^C_A \delta^{D'}_{B'}
 = \sigma^{\mu AB'} \phi_{AB'}   |T_\mu\ra. 
\end{gather*}

\noindent Derivation of Eq.(\ref{|T>-|SS'>}):
\\ Making use of the fact that $|S^{AB'}\ra = \phi_{CD'}|S^{CD'}\ra$, 
with $\phi_{CD'} = \delta^A_C \delta^{B'}_{D'}$, 
 and that $\sigma |\phi\ra = \sigma^{\mu AB'} \phi_{AB'}   |T_\mu\ra$, one finds that
\begin{gather*}
 \sigma  |S^{00'}\ra = \sigma^{\mu 00'} |T_\mu\ra = \frac{1}{\sqrt 2} (|T_0\ra + |T_3\ra)
 \\ =  \frac{1}{\sqrt 2} (|T^0\ra - |T^3\ra),
 \\ \sigma  |S^{01'}\ra = \sigma^{\mu 01'} |T_\mu\ra = \frac{1}{\sqrt 2} (|T_1\ra -i |T_2\ra)
 \\ =  \frac{1}{\sqrt 2} (-|T^1\ra +i |T^2\ra),
 \\ \sigma  |S^{10'}\ra = \sigma^{\mu 10'} |T_\mu\ra = \frac{1}{\sqrt 2} (|T_1\ra + i |T_2\ra)
 \\ =  \frac{1}{\sqrt 2} (-|T^1\ra  -i |T^2\ra),
 \\ \sigma  |S^{11'}\ra = \sigma^{\mu 11'} |T_\mu\ra = \frac{1}{\sqrt 2} (|T_0\ra - |T_3\ra)
 \\ =  \frac{1}{\sqrt 2} (|T^0\ra + |T^3\ra).
\end{gather*}

\noindent  Derivation of Eq.(\ref{|ovK>}):
\begin{gather*}
 \la \ov J |K\ra =  -\la T_\mu| J^{\mu *} K^{\nu } |T_\nu\ra
 =  - J^{\mu *} K^{\nu } g_{\mu\nu} = - J_{\nu}^{ *} K^{\nu }.
\end{gather*}

\noindent  Derivation of Eq.(\ref{s-T-s}):
\begin{gather*}
  \sigma^T\sigma = \sigma^{\mu AB'}|S_{AB'}\ra \la T_\mu|   \sigma^{\nu CD'} |T_\nu\ra \la S_{D'C}|
 \\ =  \sigma^{\mu AB'}  \sigma^{\nu CD'} g_{\mu\nu} |S_{AB'}\ra  \la S_{D'C}|
 \\ =  \sigma^{\mu}_{ AB'}  \sigma_{\mu}^{ CD'} |S^{AB'}\ra  \la S_{D'C}| = |S^{AB'}\ra  \la S_{B'A}|
\end{gather*}
\begin{gather*}
 \sigma  \sigma^T =   \sigma^{\nu CD'} |T_\nu\ra \la S_{D'C}| \sigma^{\mu AB'}|S_{AB'}\ra \la T_\mu| 
 \\ =  \sigma^{\nu CD'} \sigma^{\mu AB'} \epsilon_{CA} \epsilon_{D'B'} |T_\nu\ra  \la T_\mu| 
 \\ =  \sigma_{\nu}^{ CD'} \sigma^{\mu}_{ AB'}  |T^\nu\ra  \la T_\mu| = |T^\mu\ra  \la T_\mu| 
\end{gather*}

\noindent  Derivation of Eq.(\ref{sdag=sT}):
\begin{gather*}
 \sigma^\dag = \ov{ \left(\sigma^{\mu AB'}|S_{AB'}\ra \la T_\mu| \right)}
 = \ov\sigma^{\mu A'B}|S_{A'B}\ra \la \ov T_\mu|
 \\ = -\sigma^{\mu B A'}|S_{A'B}\ra \la T_\mu|  = \sigma^{\mu B A'}|S_{BA'}\ra \la T_\mu| = \sigma^T
\end{gather*}

\section{SL(2,C)  transformations}\label{sect-SL2C-transf}

 In this appendix, we recall basic properties of $SL(2,C)$ transformations,
 particularly, their relation to Lorentz transformations.
 We use tilde to indicate results of $SL(2,C)$ transformations.

 The group $SL(2,C)$ is composed of $2\times 2$ complex matrices with unit determinant
 \cite{Penrose-book,CM-book,Corson,pra16-commu,Kim-group}, written as
\begin{equation}\label{h-AB-app}
  h^{A}_{\ \ B} = \left( \begin{array}{cc} a & b \\ c & d \end{array} \right)
  \quad \text{with} \ ad-bc=1.
\end{equation}
 Under a transformation given by $h^{A}_{\ \ B}$,
 a two-component spinor $\kappa^A$ is transformed to
\begin{equation}\label{}
  \ww \kappa^A = h^{A}_{\ \ B} \kappa^B.
\end{equation}

 It is straightforward to verify that $\epsilon^{AB}$ is
 invariant under $SL(2,C)$ transformations, that is,
 $\ww\epsilon^{AB}= h^{A}_{\ \ C} h^{B}_{\ \ D} \epsilon^{CD}$
 has the same matrix form as $\epsilon^{AB}$ in Eq.(\ref{epsilon}).
 Direct derivation gives the following relations,
\begin{eqnarray} \label{h-property-1}
 & h^A_{\ \ B} h_{C}^{\ \ B} = h_{B}^{\ \ A} h^B_{\ \ C} = -\epsilon_C^{\ \ A}.
\\ & h_{AD} h^A_{\ \ C} = \epsilon_{DC}, \quad h^A_{\ \ B} h^{C B} = \epsilon^{AC}.
\label{h-property-2}
\end{eqnarray}
 It is not difficult to verify that the product $\chi_A \kappa^A$ is a scalar product,
 that is, 
\begin{align}\label{ww-chi-kappa=chi-kappa}
 \ww\chi_A \ww\kappa^A = \chi_A \kappa^A.
\end{align}
 Note that no complex conjugate is involved in the formation of this scalar product 
 and this is exactly the reason why the bra for Weyl spinor has the expansion given in Eq.(\ref{<kappa|}).

 When $\kappa^A$ is transformed by a matrix $h^{A}_{\ \ B}$,
 $\ov\kappa^{A'}$ is transformed by its complex-conjugate matrix, namely,
\begin{equation}\label{}
  \ww {\ov\kappa}^{A'} = \ov h^{A'}_{\ \ B'} \ov\kappa^{B'},
\end{equation}
 where
\begin{equation}\label{}
 \ov h^{A'}_{\ \ B'} := (h^{A}_{\ \ B})^* \quad \text{with} \ A=A', B= B'.
\end{equation}

 Now, we discuss relation between $SL(2,C)$ transformations and Lorentz transformations.
 Related to an $SL(2,C)$ transformation $h^{A}_{\ \ B}$, which is performed on a space $\WW$,
 we use $\Lambda^\mu_{\ \ \nu}$ to denote the corresponding transformation on an space $\VV$,
\begin{equation}\label{ww-K}
   \ \ww K^\mu = \Lambda^\mu_{\ \ \nu} K^\nu .
\end{equation}
 It proves convenient to require invariance of the EM-symbols under $SL(2,C)$ transformations,
 namely,
\begin{equation}\label{wwsig=sig}
 \ww \sigma^{\mu A'B} = \sigma^{\mu A'B},
\end{equation}
 where
\begin{gather}\label{ww-sig}
 \ww \sigma^{\mu A'B} = \Lambda^\mu_{\ \ \nu} \ov h^{A'}_{\ \ C'}
 h^{B}_{\ \ D} \sigma^{\nu C'D}.
\end{gather}
 This requirement imposes a restriction to $\Lambda^\mu_{\ \ \nu}$.
 In fact, substituting Eq.(\ref{ww-sig})
 into Eq.(\ref{wwsig=sig}) and rearranging positions of some indices, one gets that
\begin{gather}\label{sigma-Lam-int1}
 \sigma^{\mu}_{\ A'B} = \Lambda^\mu_{\ \ \nu} \ov h_{A' C'} h_{B D} \sigma^{\nu C'D}.
\end{gather}
 Multiplying both sides of Eq.(\ref{sigma-Lam-int1})
 by $ \ov h^{A'}_{\ \ E'}h^B_{\ \ F} \sigma_{\nu}^{E'F}$, its rhs gives
\begin{gather}\label{}\notag
 \Lambda^\mu_{\ \ \eta} \ov h_{A' E'} h_{B F} \sigma^{\eta E'F}
  \ov h^{A'}_{\ \ C'} h^B_{\ \ D} \sigma_{\nu}^{C'D}
  \\ = \Lambda^\mu_{\ \ \eta} \epsilon_{E'C'} \epsilon_{FD} \sigma^{\eta E'F} \sigma_{\nu}^{C'D}
  = \Lambda^\mu_{\ \ \nu},
\end{gather}
 where Eq.(\ref{h-property-2}) and Eq.(\ref{st-delta}) have been used.
 Then, one gets the following expression of $\Lambda^\mu_{\ \ \nu}$,
\begin{equation}\label{Lam-s-h}
  \Lambda^\mu_{\ \ \nu} = \sigma^{\mu}_{A'B} \ov h^{A'}_{\ \ C'} h^B_{\ \ D} \sigma_{\nu}^{C'D}.
\end{equation}

 Substituting Eq.(\ref{Lam-s-h}) into the product
 $\Lambda^\mu_{\ \ \eta} \Lambda^\nu_{\ \ \xi} g^{\eta\xi}$, one gets
\begin{gather*}
 \sigma^{\mu}_{A'B} \ov h^{A'}_{\ \ C'} h^B_{\ \ D} \sigma_{\eta}^{C'D}
 \sigma^{\nu}_{E'F} \ov h^{E'}_{\ \ G'} h^F_{\ \ H} \sigma_{\xi}^{G'H} g^{\eta\xi}.
\end{gather*}
 Using Eq.(\ref{ss-ee-2}), this gives
\begin{gather*}
 \sigma^{\mu}_{A'B} \ov h^{A'}_{\ \ C'} h^B_{\ \ D}
 \sigma^{\nu}_{E'F} \ov h^{E'C'} h^{FD}.
\end{gather*}
 Then, noting Eqs.(\ref{h-property-2}) and (\ref{g-sig}), one gets the first equality in
 the following two relations,
\begin{gather}\label{LLg=g}
 \Lambda^\mu_{\ \ \eta} \Lambda^\nu_{\ \ \xi} g^{\eta\xi} = g^{\mu\nu},
 \quad \Lambda^\mu_{\ \ \eta} \Lambda^\nu_{\ \ \xi} g_{\mu \nu} = g_{\eta \xi};
\end{gather}
 and, the second equality may be proved in a similar way. 
 Therefore, the transformations $\Lambda^\mu_{\ \ \nu}$ constitute the
 (restricted) Lorentz group
 and the space $\VV$ is composed of four-component vectors.

 The transformation $\Lambda$ and the matrix $g$ have the following properties.
 (i) The inverse transformation of $\Lambda^\mu_{\ \ \nu}$, denoted by $\Lambda^{-1}$
 has the simple expression of
\begin{equation}\label{Lambda-1}
 (\Lambda^{-1})^\nu_{\ \ \mu} = \Lambda_\mu^{\ \ \nu} \Longleftrightarrow
 (\Lambda^{-1})_{\nu \mu} = \Lambda_{\mu \nu}.
\end{equation}
 In fact, substituting Eq.(\ref{Lam-s-h}) into the product
 $\Lambda^\mu_{\ \ \nu} \Lambda_\lambda^{\ \ \nu}$ and making use of
 Eqs.(\ref{h-property-1}), (\ref{st-delta}), and (\ref{f-AB}), it is straightforward
 to verify Eq.(\ref{Lambda-1}).

 (ii) Equation (\ref{LLg=g}) implies that the matrix
 $g^{\mu\nu}$ is invariant under the transformation $\Lambda$, that is,
\begin{equation}\label{wwg=g}
  \ww g^{\mu\nu} = g^{\mu\nu}.
\end{equation}

 (iii) The product $K^\mu g_{\mu\nu} J^\nu = K_\mu J^\mu$ is a scalar under the
 transformation $\Lambda$, i.e.,
\begin{equation}\label{<K|J>}
 \ww K_\mu \ww J^\mu = K_\mu J^\mu,
\end{equation}
 which can be readily proved making use of Eq.(\ref{LLg=g}).

 (iv) Making use of Eq.(\ref{sig-c}), it is straightforward to show that
the transformation $\Lambda $ is real, namely,
\begin{equation}\label{real-Lam}
 \Lambda^\mu_{\ \ \nu} = (\Lambda^\mu_{\ \ \nu})^*.
\end{equation}
 Then, it is easy to check that $K^*_\mu J^\mu$ is also a scalar product.

\section{Helicity states of basic modes}\label{app-angular}

 In this appendix, we give detailed discussions on properties of helicity states of basic modes. 
 Specifically, in Sec.\ref{app-am-basic-mode} we derive  
 angular momentum operators, then, in Sec.\ref{app-helicity} discuss helicity states, 
 and, finally, in Sec.\ref{app-antimode-counterpart} discuss antimode-counterpart states and ACC-bras. 

\subsection{Angular momentum of basic modes}\label{app-am-basic-mode}

 In this section, we derive explicit expressions for the spinor angular-momentum operators
 of the $b$-mode and $\ov b$-mode, denoted by $\bs_b$ and $\bs_{\ov b}$, respectively,
 which act on the two spaces of $\WW$ and $\ov\WW$.
 This enables one to prove the relation in Eq.(\ref{sk=-sk'}).

 The operator $\bs_{b}$ is written as
 $\bs_b = (s_b^1, s_b^2, s_b^3) \equiv (s_b^{23},s_b^{31},s_b^{12})$, and similar for $\bs_{\ov b}$.
 In the ket-bra form, their components are written as
\begin{subequations}\label{sij-app}
\begin{align}\label{sij-app-1}
 s_b^{ij} & =  | S_A\ra (s_b^{ij})^{AB} \la\la S_B|,
 \\ s_{\ov b}^{ij} &  =  | S_{A'}\ra (s_{\ov b}^{ij})^{A'B'} \la\la S_{B'}|.
\end{align}
\end{subequations}
 To find expressions of $(s_b^{ij})^{AB}$ and $(s_{\ov b}^{ij})^{A'B'}$, 
 one may make use of knowledge about the angular-momentum operator for Dirac spinors, denoted by $\bf S$ with
\begin{equation}\label{J-vec-ten}
 {\bf S} =  (S^1,S^2,S^3) = (S^{23},S^{31},S^{12}).
\end{equation}
 The above-discussed three operators have the following relationship, 
\begin{gather}\label{S-ss'}
  S^{ij} = \left( \begin{array}{cc} s_b^{ij} & 0 \\ 0 & {s_{\ov b}}^{ij} \end{array}\right).
\end{gather}

 As is well known, in the chiral representation of $\gamma^\mu$-matrices, the component form of $\bf S$ satisfies
 $[S^{\mu\nu}] = \frac i4 [\gamma^\mu, \gamma^\nu]$.
 With the upper part of Dirac spinor associated with the space $\WW$
 and the lower part with $\ov \WW$, 
 the $\gamma^\mu$-matrices have the following expression \cite{Penrose-book,Kim-group,CM-book,Corson,pra16-commu}, 
\begin{gather} \label{gamma-mu}
 \gamma^\mu = \sqrt 2 \left( \begin{array}{cc} 0 & \sigma^{\mu AB'}
 \\ \ov\sigma^{\mu}_{A'B} & 0\end{array}\right).
\end{gather}
 Then, it is straightforward to find that the matrix $[S^{\mu\nu}]$ is written as
\begin{gather}\label{S-mu-nu}
  [S^{\mu\nu}] = \frac i2 \left( \begin{array}{cc} \sigma^{\mu\nu, A}_{\ \ \ \ \ B} & 0
  \\ 0 & -\ov\sigma^{\mu\nu, \ \ B'}_{\ \ \ A'}\end{array}\right),
\end{gather}
 where
 \begin{gather}\label{sig-mnAC}
 \sigma^{\mu\nu, A C} := \sigma^{\mu AB'}
 \ov\sigma^{\nu \  C }_{B'} - \sigma^{\nu AB'} \ov\sigma^{\mu \   C}_{B'}.
\end{gather}

 To translate the component expression in Eq.(\ref{S-mu-nu}) to an operator form
 and get expressions for the operators in Eq.(\ref{S-ss'}), let us consider
 an arbitrary Weyl spinor $|u\ra \in \WW$ and use $|\kappa\ra$ to indicate the action of $s_b^{ij}$ on it, i.e.,
\begin{align}\label{kap-s-u}
 |\kappa\ra = s_b^{ij}|u\ra.
\end{align}
 Substituting Eq.(\ref{sij-app-1}) into Eq.(\ref{kap-s-u}), then, 
 multiplying the result from left by $\la S^A|$ and making use of Eqs.(\ref{f-AB})-(\ref{eps-delta}) and (\ref{kappa-A}),
 one gets that
\begin{align} \notag
 \kappa^A & =\la\la S^A| \left( | S_C\ra (s_b^{ij})^{CB} \la S_B| \right)|u\ra  =\epsilon^A_{ \ C} (s_b^{ij})^{CB} u_B
 \\ & =- (s_b^{ij})^{AB} u_B =  (s_b^{ij})^{A}_{\ B} u^B.
 \label{kap-s-u-AB}
\end{align}
 Clearly, applying the abstract Dirac operator in Eq.(\ref{S-ss'}) to an abstract spinor,
 one should get a result equivalent to that obtained
 by applying the matrix form in Eq.(\ref{S-mu-nu}) to spinor components. 
 With Eq.(\ref{kap-s-u-AB}), this gives that
\begin{gather}\label{sb-ij-AB}
 (s_b^{ij})^{AB}  = \frac i2 \sigma^{ij, A B},
\end{gather}
 where positions of the index $B$ have been raised on both sides of the equation.

 The operator $s_{\ov b}^{ij}$ may be gotten by a similar method.
 In fact, similar to Eq.(\ref{kap-s-u-AB}), one gets that
 \begin{align} 
 \ov \kappa_{A'} & = (s_{\ov b}^{ij})_{A'  B'} u^{B'} = -  (s_{\ov b}^{ij})_{A'}^{\ \ B'} u_{B'},
 \label{ov-kap-s-u-AB}
\end{align}
 where positions of indices have been arranged for an easy comparison with Eq.(\ref{S-mu-nu}).
 Then, one sees that 
\begin{align}\label{sovb-ij-AB}
  (s_{\ov b}^{ij})^{A' B'} = \frac i2 \ov\sigma^{ij, A' B'}.
\end{align}
 From Eqs.(\ref{sb-ij-AB}) and (\ref{sovb-ij-AB}), one directly gets that
\begin{gather}\label{sk=-sk'-com}
 (s_{\ov b}^{ij})^{A' B'} = - \ov{({s}_b^{ij})^{A B}}.
\end{gather}
 This is just the component form of Eq.(\ref{sk=-sk'}).

\subsection{Helicity states}\label{app-helicity}

 In this section, we derive properties and expressions of helicity states of basic modes,
 particularly, those given in Eqs.(\ref {hr-b})-(\ref{ww-eps}).

 We recall that helicity states of a $b$-mode are written as $|w^r(\bp)\ra$, 
 satisfying $H^{\rm cy}_b |w^r(\bp)\ra = h^r_b|w^r(\bp)\ra$, with $H^{\rm cy}_b =\bs_b \cdot \bp/|\bp|$.
 To compute $|w^r(\bp)\ra$, let us consider the action of angular momentum on a generic spinor $|\chi\ra$. 
 Making use of Eqs.(\ref{sij-app-1}), (\ref{f-AB}), and (\ref{kappa-A}), one finds that
\begin{gather*}
 \\ s_b^k |\chi\ra = (s_b^k)_{AB}|S^A\ra \la\la S^B| \chi_C|S^C\ra
 \\ = (s_b^k)_{AB} \chi_C \epsilon^{BC} |S^A\ra = - (s_b^k)_{A}^{\ \ C} \chi_C  |S^A\ra,
\end{gather*}
 thus, 
\begin{gather}\label{sk-kappa}
 s_b^k |\chi\ra =- (s_b^k)_{A}^{\ \ B} \chi_B |S^A\ra.
\end{gather}
 Then, writing $|\kappa\ra = s_b^k|\chi\ra$ and expanding it as $|\kappa\ra = \kappa_A |S^A\ra$, one finds that
\begin{subequations}
\begin{gather}
   \kappa_A = - (s_b^k)_{A}^{\ \ B} \chi_B,
 \\ \kappa^A =(s_b^k)^{A}_{\ \ B} \chi^B. \label{sk-kappa-com}
\end{gather}
\end{subequations}

 We employ the explicit expressions of EW-symbols in Eqs.(\ref{sigma^AB})-(\ref{sig-lower-AB})
 to compute explicit expressions of angular-momentum operators.
 From Eq.(\ref{sb-ij-AB}), one finds the following well-known matrix forms of $s_b^{k}$, given by Pauli matrices, 
\begin{subequations}\label{sk-explicit}
\begin{gather}
 (s_b^{1})^{A}_{\ \ B} = \frac 12 \left(\begin{array}{cc} 0 & 1 \\ 1 & 0 \\ \end{array} \right),
 \\ (s_b^{2})^{A}_{\ \ B} = \frac 12 \left(\begin{array}{cc} 0 & -i \\ i & 0 \\ \end{array} \right),
 \\ (s_b^{3})^{A}_{\ \ B} = \frac 12 \left(\begin{array}{cc} 1 & 0 \\ 0 & -1 \\ \end{array} \right). \label{sk-explicit-3}
\end{gather}
\end{subequations}
 For example, for the third component, Eq.(\ref{sb-ij-AB}) gives that 
 $(s_b^{3})^{A}_{\ C}  = \frac i2 \sigma^{12, A}_{\ \ \ \ \  C}$ and, then, 
 one gets Eq.(\ref{sk-explicit-3}) by the following derivation for $\sigma^{12, A}_{\ \ \ \ \  C}$,
\begin{gather*}
 \sigma^{12, A}_{\ \ \ \ \  C} = \sigma^{1 AB'} \ov\sigma^{2 }_{B'C} - \sigma^{2 AB'} \ov\sigma^{1}_{B'C}
 \\ = \frac 12  \left(\begin{array}{cc} 0 & 1 \\ 1 & 0 \\ \end{array} \right)\left(\begin{array}{cc} 0 & i \\ -i & 0 \\ \end{array} \right)
 -\frac 12 \left(\begin{array}{cc} 0 & -i \\ i & 0 \\ \end{array} \right) \left(\begin{array}{cc} 0 & -1 \\ -1 & 0 \\ \end{array} \right)
 \\ = \frac 12  \left(\begin{array}{cc} -i & 0 \\ 0 & i \\ \end{array} \right)
 -\frac 12 \left(\begin{array}{cc} i & 0 \\ 0 & -i \\ \end{array} \right) = \left(\begin{array}{cc} -i & 0 \\ 0 & i \\ \end{array} \right).
\end{gather*}

 Let us consider a momentum $\bp_0 = (0,0,|\bp_0|)$ along the $z$-direction.
 Making use of Eqs.(\ref{sij-app-1}) and  (\ref{sk-explicit-3}), one finds that
\begin{align}\label{}
 H^{\rm cy}_b =  s_b^{3} = - | S_A\ra (s_b^{ij})^{A}_{\ \ B} \la\la S^B|
 = | S_1\ra \la\la S^1| - | S_0\ra \la\la S^0|.
\end{align}
 Using Eq.(\ref{eps-delta}), one then gets that $s_b^{3} |S_0\ra = |S_0\ra$ and $s_b^{3} |S_1\ra = -|S_1\ra$.
 Thus, under the well-known spin-angular-momentum expression in Eq.(\ref{sk-explicit}), 
 the helicity states of $|w^r(\bp_0)\ra$ of $b$-mode are in fact given by the basis spinors $|S_A\ra$.

 More exactly, one may set components of $|w^r(\bp_0)\ra$ on the basis of $|S_A\ra$ as
\begin{gather}\label{urASM-RP0}
 [ w^{rA}(\bp_0)] =  \left\{
   \begin{array}{ll}
     \left(\begin{array}{c} 1 \\ 0 \\ \end{array} \right) & \hbox{for $r=0$}, \\
     \left(\begin{array}{c} 0 \\ 1 \\ \end{array} \right) & \hbox{for $r=1$.}
   \end{array} \right.
\end{gather}
 Correspondingly, the helicities are written as
 \begin{gather}\label{hc-$b$-mode}
 h^r_b = \frac{(-1)^r}{2} \quad (r=0,1).
\end{gather}
 Then, noting that $|w^r(\bp_0)\ra = - w^{rA}(\bp_0) |S_A\ra$, one gets
\begin{align}\label{wr-SA}
  |w^r(\bp_0)\ra =  \left\{   \begin{array}{ll}
    -|S_0\ra  & \hbox{for $r=0$}, \\
     -|S_1\ra & \hbox{for $r=1$}.
   \end{array} \right.
\end{align}
 One directly gets Eq.(\ref{ww-eps}) from Eq.(\ref{wr-SA}).

 Further, the helicity states $|w^r(\bp)\ra$ can be obtained by a Lorentz transformation that brings 
 $\bp_0$ to $\bp$.
 Clearly, the helicity is equal to $\frac 12$ for the state $|w^0(\bp)\ra$, while, equal to $-\frac 12$ for $|w^1(\bp)\ra$.

 For an $\ov b$-mode, from Eq.(\ref{sk=-sk'}), one sees that its
 helicity states are given by the complex conjugates of $|w^r(\bp)\ra$, i.e., by $|\ov w^r(\bp)\ra$,
 with opposite eigenvalues.
 That is,
\begin{align}\label{}
 & H^{\rm cy}_{\ov b} |\ov w^r(\bp)\ra = h^r_{\ov b} |\ov w^r(\bp)\ra,
 \\ \label{hc-anti-$b$-mode-app}
 & h^r_{\ov b} =  \frac{ (-1)^{r+1}}{2}   = \left\{
          \begin{array}{ll}
            -\frac 12, & \hbox{for $|\ov w^0(\bp)\ra$,} \\
            \frac 12, & \hbox{for $|\ov w^1(\bp)\ra$.}
          \end{array}
        \right.
\end{align}

\subsection{Antimode-counterpart states and ACC-bras}\label{app-antimode-counterpart}

 In this section, we give detailed discussions on properties of antimode-counterpart states,  as well as
 on ACC-bras and ACC-scalar products for spinor states of basic modes.
 In particular, we derive Eqs.(\ref{bm-ac2}) and (\ref{hat-w-w-main})-(\ref{IP-ww-main}).

 Making use of Eq.(\ref{bm-ac}), which gives states $|{w^r_\varrho}(\bp)\ra_{\rm ac}$ of $\varrho=+$,
 one may compute $|{w_{r\varrho}}(\bp)\ra_{\rm ac}$ of $\varrho=+$.
 To this end, we note that the relation of $|w_{r\varrho}(\bp)\ra = |w^s_\varrho(\bp)\ra \epsilon_{sr} $
 implies that $|w_{r\varrho}(\bp)\ra_{\rm ac} = |w^s_\varrho(\bp)\ra_{\rm ac} \epsilon_{sr} $.
 Substituting Eq.(\ref{bm-ac+}) into the rhs of this result
 and making use of the relation of $|\ov w_{s\varrho} (\bp)\ra = |\ov w^t_\varrho (\bp)\ra \epsilon_{ts}$, 
 one gets that
\begin{align}\label{}
 |w_{r\varrho}(\bp)\ra_{\rm ac} = \sum_s |\ov w^t_\varrho (\bp)\ra \epsilon_{ts} \epsilon_{sr} .
\end{align}
 Further, we note that 
\begin{align}\label{epep-delta}
 \sum_s \epsilon_{ts} \epsilon_{sr} = -\delta_{tr}, 
\end{align}
 which can be easily checked by direct computation.
 Then, one gets that $|w_{r\varrho}(\bp)\ra_{\rm ac} = -|\ov w^r_\varrho (\bp)\ra$.
 Similarly, one computes $|w^r_{\varrho}(\bp)\ra_{\rm ac}$ of $\varrho=-$ and gets that 
\begin{align}\label{}\notag
 |w^r_{\varrho}(\bp)\ra_{\rm ac} = \epsilon^{rs}|w_{s\varrho}(\bp)\ra_{\rm ac} 
 =  \epsilon^{rs}|\ov w^s_{\varrho}(\bp)\ra
 =  \epsilon^{rs}\epsilon^{st}|\ov w_{t\varrho}(\bp)\ra.
\end{align}
 To summarize, one gets the following relations, 
\begin{subequations}\label{bm-ac2-app}
\begin{align}\label{}
 \label{bm-ac+2-app} & |w_{r\varrho}(\bp)\ra_{\rm ac} = -|\ov w^r_\varrho (\bp)\ra \qquad \text{$(\varrho =+)$, }
 \\ \label{bm-ac-2-app} & |w^r_{\varrho}(\bp)\ra_{\rm ac} = -|\ov w_{r\varrho} (\bp)\ra \qquad \text{$(\varrho =-)$. }
\end{align}
\end{subequations}
 which are just those in Eq.(\ref{bm-ac2}).

 Then, making use of Eqs.(\ref{bm-ac}) and (\ref{bm-ac2}), one finds that
\begin{subequations}\label{hat-w-w+}
\begin{gather}\label{hat-w-w+1}
 | {w^r_{\varrho}}(\bp)\ra_{\rm ACC} = |w_{r\varrho}(\bp)\ra \qquad (\varrho=+),
 \\  | { w_{r\varrho}}(\bp)\ra_{\rm ACC} = - | w^r_{\varrho}(\bp)\ra \qquad (\varrho=+),
\end{gather}
\end{subequations}
 and
\begin{subequations}\label{hat-w-w-}
\begin{gather}\label{hat-w-w-1}
 | {w^r_{\varrho}}(\bp)\ra_{\rm ACC} = -| w_{r\varrho}(\bp)\ra \qquad (\varrho=-),
 \\  | { w_{r\varrho}}(\bp)\ra_{\rm ACC} =   | w^r_{\varrho}(\bp)\ra \qquad (\varrho=-).
\end{gather}
\end{subequations}
 Clearly, Eq.(\ref{hat-w-w-main}) is just a concise form of the above equations.
 Further, making use of Eqs.(\ref{ww-eps}), (\ref{eps-delta}), and (\ref{hat-w-w+})-(\ref{hat-w-w-}),
 it is direct to compute the ACC-scalar products, getting that
\begin{subequations}\label{IP-ww+}
\begin{align}\label{IP-ww-+}
 & \la \wh {w^r_{\varrho}}(\bp)| {w^s_{\varrho}}(\bp)\ra = \delta^{rs} \qquad (\varrho=+),
 \\ &  \la \wh{w_{r\varrho}}(\bp)|{w_{s\varrho}}(\bp)\ra = \delta_{rs} \qquad (\varrho=+),
\end{align}
\end{subequations}
 and
\begin{subequations}\label{IP-ww-}
\begin{align}\label{IP-ww-+}
 & \la \wh {w^r_{\varrho}}(\bp)| {w^s_{\varrho}}(\bp)\ra = -\delta^{rs} \qquad (\varrho=-),
 \\ &  \la \wh{w_{r\varrho}}(\bp)|{w_{s\varrho}}(\bp)\ra = -\delta_{rs} \qquad (\varrho=-),
\end{align}
\end{subequations}
 as summarized in Eq.(\ref{IP-ww-main}).

 Finally, we discuss commutability of 
 taking antimode-counterpart state and taking complex conjugation  within an ACC operation.
 For example, for a state $| {w^r_\varrho}(\bp)\ra$ with $\varrho=+$, reversing the two suboperations, one first gets 
 $|\ov{w^r_\varrho}(\bp)\ra$, then, gets $|\ov{w^r_\varrho}(\bp)\ra_{\rm ac} = |w_{r\varrho} (\bp)\ra$
 according to Eq.(\ref{bm-ac+});
 this is just what is given to $ | {w^r_\varrho}(\bp)\ra_{\rm ACC}$ in Eq.(\ref{hat-w-w+1-main}). 
 Meanwhile, for a state $| {w^r_\varrho}(\bp)\ra$ with $\varrho=-$, reversing the two suboperations, one first gets 
 $|\ov{w^r_\varrho}(\bp)\ra$, then, gets $|\ov{w^r_\varrho}(\bp)\ra_{\rm ac} = -|w_{r\varrho} (\bp)\ra$,
 where Eq.(\ref{bm-ac-2}) has been used;
 this result is just what is given to $ | {w^r_\varrho}(\bp)\ra_{\rm ACC}$ in Eq.(\ref{hat-w-w-1-main}).

\section{ACC operation and ACC-scalar product of generic spinors}\label{app-derive-properties}

 In this appendix, we give detailed discussions on some properties of
 ACC-scalar products of generic spinors (Appendix \ref{app-inner-product}), and on
 some relations about ACC operations (Appendix \ref{app-double-ACC}).

\subsection{Properties of ACC-scalar products}\label{app-inner-product}

 Firstly, we show that, for two arbitrary single-mode states $|\psi\ra $ and $|\phi\ra$, one has
\begin{align}\label{<wh|>*}
 \ov{\la \wh\psi |\phi\ra} \equiv \left( \la \wh\psi |\phi\ra  \right)^* =  \la \wh\phi |\psi\ra.
\end{align}
 To this end, let us write
\begin{subequations}\label{psi-phi-Map}
\begin{align}\label{psi-Map}
 |\psi\ra = \int d\ww p C_\alpha (\bp) |M^\alpha_{\bp }\ra,
 \\ |\phi\ra = \int d\ww p D_\beta (\bp) |M^\beta_{\bp }\ra,
\end{align}
\end{subequations}
 where $C_\alpha (\bp)$ and $D_\alpha (\bp)$ are expanding coefficients.
 The ACC-bra of $|\phi\ra$ is written as
\begin{gather}\label{}
 \la \wh{\phi}| = \int d\ww p  D_\beta^*(\bp) \la \wh{M}^{\beta}_\bp|.
\end{gather}
 It is then easy to get that
\begin{align}\label{}\notag
 \la \wh\phi |\psi\ra = \int d\ww p d\ww q D^*_\beta (\bq) C_\alpha (\bp)  \la \wh M^\beta_{\bq } |M^\alpha_{\bp }\ra
 \\ = \int d\ww p d\ww q D^*_\beta (\bq) C_\alpha (\bp)  |\bp| \delta^3(\bp-\bq) \Upsilon^{ \beta \alpha}. 
 \label{<wh-phi|psi>}
\end{align}
 Similarly, one has
\begin{align}\label{}\notag
 \la \wh\psi |\phi\ra =  \int d\ww p d\ww q D_\beta (\bq) C^*_\alpha (\bp)  |\bp| \delta^3(\bp-\bq) \Upsilon^{ \alpha  \beta}.
\end{align}
 Then, one gets Eq.(\ref{<wh|>*}).

 Secondly, we show that the ACC-scalar product is an inner product for a generic fermionic mode $F$ with $\varrho=+$,
 while, it is a negative inner product for $F$ with $\varrho=-$.
 Indeed, for $|\psi\ra$ in Eq.(\ref{psi-Map}) with  $M=F$, 
 noting that $\Upsilon^{\alpha \beta} = \varrho(\alpha) \delta^{\alpha \beta}$, 
 from Eq.(\ref{<wh-phi|psi>}) with $|\psi\ra = |\phi\ra$ one gets that
\begin{align}
  \la \wh{\psi}|\psi\ra  & = \varrho(\alpha) \int d\ww p \sum_{r(\alpha)} |C_\alpha(\bp)|^2. \label{IP-psi-psi}
\end{align}

 Finally, we discuss the ACC-scalar product $\la \wh \Phi|\Psi\ra$ of two arbitrary vectors of $|\Psi\ra$ and $|\Phi\ra$. 
 If the product is a $c$-number, then, 
 it is straightforward to check that the relation in Eq.(\ref{<wh|>*}) still holds for $\la \wh \Phi|\Psi\ra$,
 that is, 
\begin{align}\label{<Ph|Ps>^*-c}
 \la \wh \Phi|\Psi\ra^* =\la \wh \Psi|\Phi\ra
 \quad \text{for $\la \wh \Phi|\Psi\ra$ as a $c$-number}.
\end{align}
 Otherwise, one may divide both $\la \wh\Phi|$ and $|\Psi\ra$ into two parts, 
 written as $\la \wh\Phi| = \la \wh\Phi_2| \la \wh\Phi_1|$ and $|\Psi\ra =|\Psi_1\ra |\Psi_2\ra $, 
 such that $\la \wh\Phi_1|$ forms an ACC-scalar product with $|\Psi_1\ra$, 
 meanwhile, $\la \wh\Phi_2|$ and $|\Psi_2\ra$ remain purely vectors, i.e., 
\begin{align}\label{<wh-Phi|Psi>-gen-1}
 \la \wh \Phi|\Psi\ra = \la \wh\Phi_2| \left( \la \wh\Phi_1|\Psi_1\ra \right) |\Psi_2\ra. 
\end{align}
 In this case, making use of Eq.(\ref{<Ph|Ps>^*-c}), one gets that
\begin{align}\label{}
 \la \wh \Phi|\Psi\ra^* = \ov{ \la \wh\Phi_2|} \left( \la \wh\Psi|\Phi_1\ra \right) \ov{|\Psi_2\ra}. 
\end{align}

\subsection{Some relations for ACC operations}\label{app-double-ACC}

 In this section, we derive some relations for ACC operations,
 particularly, those for double ACC operations. 
 For brevity, for ket, we also use a hat to indicate the result of an ACC operation, i.e., 
\begin{align}\label{}
 |\wh \Psi\ra \equiv |\Psi\ra_{\rm ACC}. 
\end{align}

 By definition, for a state $|\Psi\ra$, the ACC operation gives $|\wh\Psi\ra$. 
 Double ACC operations bring no change to momentum states [see Eq.(\ref{p-ACC})]. 
 It is natural to assume that $|\wh{\wh{s}}\ra  = |s\ra$. 
 These two properties, together with Eq.(\ref{whvep=ovvep}), imply that
\begin{gather}\label{hat-hat-B}
 |\wh{\wh{B^{\lambda }_\bk}}\ra =  |B^{\lambda }_\bk\ra,
\end{gather}
 i.e., double application of ACC operation does not change states of bosonic modes.

 The situation is a little more complex with spinor states of fermionic modes.
 For such a state, making use of Eq.(\ref{whu-v}), one gets that
\begin{gather}\label{hat-hat-f}
 |\wh{\wh {\cs_f^\alpha}}(\bp)\ra = - |{ {\cs}}_f^\alpha(\bp)\ra.
\end{gather}
 i.e., double application of ACC operation generates a minus sign.
 Therefore, for a generic vector $|\Psi\ra$, one has
\begin{align}\label{whwh-Psi}
 |\wh{\wh \Psi}\ra = (-1)^{N_f(\Psi)} |\Psi\ra,
\end{align}
 where $N_f$ is the number of fermionic modes contained in $|\Psi\ra$.

 Below, we discuss ACC operation on a term of the form of $\la \wh \Phi|\Psi\ra$.
 Let us first consider single modes.
 In this case, $\la \wh \Phi|\Psi\ra$ is a $c$-number and, by definition, the ACC operation 
 gives its complex conjugate. 
 For bosonic modes, one gets that
\begin{align}\label{ACC-<wh|>-B}
 \wh{\la \wh{B^\alpha_{\bp }}|B^\beta_{\bq}\ra} = \la \wh{B^\alpha_{\bp }}|B^\beta_{\bq}\ra^*
 = \la \wh{B^\beta_{\bq}}| {B^\alpha_{\bp }}\ra,
\end{align}
 where Eq.(\ref{<wh|>*}) has been used when deriving the last equality. 
 Consistently,  one may directly compute with Eq.(\ref{hat-hat-B}) and get 
\begin{align}\label{}
 \wh{\la \wh{B^\alpha_{\bp }}|B^\beta_{\bq}\ra} 
 = \la {B^\alpha_{\bp }}|\wh{B^\beta_{\bq}}\ra
 = \la \wh{B^\beta_{\bq}}| {B^\alpha_{\bp }}\ra,
\end{align}
 where Eq.(\ref{<J|K>=<K|J>}) has been used when deriving the last equality.

 For spinor states of fermionic modes, say, for $\la \wh{\cs_f^{\alpha}} | \cs_f^{\beta}\ra$, one compute that
\begin{align}\label{ACC-<wh|>-f}
 \wh{ \la \wh{\cs_f^{\alpha}} | \cs_f^{\beta}\ra} = \left( \la \wh{\cs_f^{\alpha}} | \cs_f^{\beta}\ra \right)^*
 = \la \wh{\cs_f^{\beta}}| {\cs_f^{\alpha}} \ra;
\end{align}
 and, consistently, with Eq.(\ref{hat-hat-f}),
\begin{align}\label{}
  \wh{ \la \wh{\cs_f^{\alpha}} | \cs_f^{\beta}\ra} =  \la \wh{\wh{\cs_f^{\alpha}}} | \wh{\cs_f^{\beta}}\ra
  = - \la \cs_f^{\alpha} | \wh{\cs_f^{\beta}}\ra = \la \wh{\cs_f^{\beta}}| {\cs_f^{\alpha}} \ra,
\end{align}
 where the anti-symmetry of scalar product of Weyl spinors in Eq.(\ref{ck=-kc})
 has been used in the derivation of the last equality.

 Next, it is straightforward to generalize the above arguments to 
 a generic $c$-number scalar product $\la \wh \Phi|\Psi\ra$.
 As a generalization of Eqs.(\ref{ACC-<wh|>-B}) and (\ref{ACC-<wh|>-f}), one gets that
\begin{align}\label{wh-<wh|>-c-num}
 \wh{\la \wh \Phi|\Psi\ra} = \la \wh \Psi|\Phi\ra
 \quad \text{for $\la \wh \Phi|\Psi\ra$ as a $c$-number}.
\end{align}

 Finally, we discuss $\la \wh \Phi| \Psi\ra$ of two arbitrary vectors, as written in Eq.(\ref{<wh-Phi|Psi>-gen-1}).
 By definition, one writes
\begin{align}\label{}
 \wh{\la \wh \Phi|\Psi\ra} = \la \wh{\wh{\Phi}}_2| \left( \la \wh\Phi_1|\Psi_1\ra \right)^* |\wh \Psi_2\ra .
\end{align}
 Then, making use of Eqs.(\ref{whwh-Psi}) and (\ref{wh-<wh|>-c-num}), one gets that
\begin{align}\label{wh-<wh|>-gen-1}
 \wh{\la \wh \Phi|\Psi\ra} =   (-1)^{N_f(\Phi_2)} \la \Phi_2| \left( \la \wh \Psi_1| \Phi_1\ra \right) |\wh \Psi_2\ra  .
\end{align}

\subsection{Adjoint operators for ACC-scalar products}\label{app-adjoint-operator}

 In this appendix, we discuss adjoint operators, which are defined with respect to 
 the ACC-scalar product. 
 In particular, we give detailed derivations for Eqs.(\ref{b-express}) and (\ref{b+-bdag}).

 For this purpose, it proves convenient to use the notation of $(|\Phi\ra ,|\Psi\ra )$ for ACC-scalar product
 of two arbitrary states $|\Psi\ra$ and $|\Phi\ra$, i.e., 
\begin{align}\label{SP(,)-ACC}
  (|\Phi\ra ,|\Psi\ra ) \equiv \la \wh{\Phi}|\Psi\ra.
\end{align}
 The adjoint operator of an operator $A$, indicated as $A^+$, is defined by the following relation, 
\begin{align}\label{A+-def}
 (|\Phi\ra , A^+|\Psi\ra ) \equiv (A|\Phi\ra ,|\Psi\ra ).
\end{align}

 We first derive the relation in Eq.(\ref{b-express}) for annihilation operators.
 Using Eq.(\ref{SP(,)-ACC}), one writes
\begin{align}\label{}
 (|\Phi\ra, b_M^{\alpha}(\bp)|\Psi\ra ) =  \la \wh \Phi| b_M^{\alpha}(\bp)|\Psi\ra.
\end{align}
 Meanwhile, making use of the definition of $b_M^{\alpha} (\bp)  := \left[ b_M^{\alpha \dag}(\bp) \right]^+$
 and Eq.(\ref{b-dag-def}), one gets that
\begin{align}\label{}
 (|\Phi\ra, b_M^{\alpha}(\bp)|\Psi\ra ) &= (b_M^{\alpha \dag}(\bp)|\Phi\ra , |\Psi\ra)
  = \left( \la \wh\Phi | \la \wh M^{\alpha}_\bp| \right) |\Psi\ra.
\end{align}
 Then, one gets Eq.(\ref{b-express}), i.e., 
\begin{align}\label{b-express-app}
  \la \wh \Phi| b_M^{\alpha}(\bp)|\Psi\ra 
  =  \la \wh\Phi | \la \wh M^{\alpha}_{\bp} | |\Psi\ra.
\end{align}

 Next, we derive Eq.(\ref{b+-bdag}).
 To this end, let us consider a term $(|\Phi\ra, \left( b_M^{\alpha}(\bp) \right)^+|\Psi\ra ) $.
 By definition, $(|\Phi\ra, \left( b_M^{\alpha}(\bp) \right)^+|\Psi\ra ) 
 =(b_M^{\alpha }(\bp)|\Phi\ra , |\Psi\ra)$.
 This relation, together with the equivalence of $b_M^{\alpha}(\bp)$ and $ \la \wh M^{\alpha}_{\bp} |$ 
 as shown in Eq.(\ref{b-express-app}), gives that
\begin{align}\label{SP-1-app}
 (|\Phi\ra, \left( b_M^{\alpha}(\bp) \right)^+|\Psi\ra ) 
  =    \left( \la \wh M^{\alpha}_{\bp}| \Phi\ra , |\Psi\ra \right).
\end{align}
 The rhs of Eq.(\ref{SP-1-app}) may be written in the form of $\la \Omega |\Psi\ra $, 
 where $\la \Omega|$ is obtained from $\wh{ \la \wh M^{\alpha}_{\bp}| \Phi\ra}$
 by changing its pure ket part to a bra, with the $c$-number part unchanged.

 To get $\la \Omega|$, we note that, when being used in the computation of physical quantities, 
 the final place where the term $(|\Phi\ra, \left( b_M^{\alpha}(\bp) \right)^+|\Psi\ra ) $ 
 appears should be in the computation of some $c$-number quantity.
 We assume that the term already has this $c$-number feature with respect to the mode $M$.
 This requires that, for such a term not definitely vanishing, 
 the state $|\Phi\ra$ should contain at least one state of the mode $M$. 
 With this property, making use of Eq.(\ref{wh-<wh|>-gen-1}), 
 one finds that $\la \Omega| = \la \wh\Phi| M^{\alpha }_{\bp}\ra$.
 This gives that
\begin{align}\label{}
 (|\Phi\ra, \left( b_M^{\alpha}(\bp) \right)^+|\Psi\ra )  
 = \la \wh\Phi| b_M^{\alpha \dag} (\bp) |\Psi\ra ,
\end{align}
 where the definition of creation operator in Eq.(\ref{b-dag-def}) has been used. 
 This justifies Eq.(\ref{b+-bdag}), i.e., $b_M^{\alpha \dag} (\bp) = \left[b_M^{\alpha } (\bp) \right]^+$.

 Moreover, for an arbitrary operator $A$, it is easy to check that
\begin{align}\label{A++=A}
 \left( A^+ \right)^+ = A.
\end{align}
 In fact, 
\begin{gather*}
 \la \wh \Phi|[ (A^+)^+ |\Psi\ra] =  ( |\Phi\ra , (A^+)^+ |\Psi\ra )
 \\  =  ( A^+|\Phi\ra ,  |\Psi\ra )  =  ( |\Psi\ra , A^+ |\Phi\ra )^*
 \\ =   ( A |\Psi\ra , |\Phi\ra )^*  = (|\Phi\ra , A |\Psi\ra ).
\end{gather*}
 Furthermore, for an operator $ A = c|\psi\ra \la \wh\phi|$, making use of Eq.(\ref{wh-<wh|>-gen-1}),
 one gets that
\begin{gather*}
 (|\Phi\ra , A^+|\Psi\ra ) = c^*  \left(  |\psi\ra \la \wh\phi|\Phi_1\ra |\Phi_2\ra , |\Psi\ra \right)
 \\ = c^* \la \wh\Phi| \left( |\phi \ra \la \wh\psi| \right) |\Psi\ra,
\end{gather*}
 which implies that
\begin{align}\label{A+-|><|}
 A^+ = c^*|\phi \ra \la \wh\psi|.
\end{align}

\section{Properties of z-mode and s-mode}\label{app-derive-properties}

 In Appendix \ref{app-Jz}, we discuss a method of computing the angular-momentum operator $\bJ_\VV$
 of $z$-mode and, as an application, derive Eq.(\ref{J3-matrix}) for its third component.
 In Appendix \ref{app-scalar-mode}, we show that the spinor spaces of $s$-mode and $\ov s$-mode
 are one-dimensional scalar spaces.

\subsection{Angular momentum of z-mode}\label{app-Jz}

 Let us consider an operator $X^k$ of $k=1,2,3$, defined by
\begin{align}\label{Xk}
 X^k: = \sigma s_b^k I_{\ov\WW} \sigma^T, 
\end{align}
 where $s_b^k$ are components of the angular-momentum operator $\bs_b$
 of $b$-mode and $I_{\ov \WW}$ is the identity operator on the space $\ov \WW$.
 Using the relation of $\ov \sigma = \sigma$ [Eq.(\ref{ov-s=s})], one finds that the complex conjugate
 of $X^k$ is written as
\begin{align}\label{}
 \ov X^k = \sigma \ov s_b^k I_{ \WW} \sigma^T.
\end{align}
 Noting that $\bs_{\ov b} = - {\ov \bs}_b$ [Eq.(\ref{sk=-sk'})], 
 one sees that components of $ \bJ_\VV $ in Eq.(\ref{J-VV}) are written in the following form,
\begin{align}\label{Jk-Xk}
  J^k_{\VV} = X^k - \ov X^k.
\end{align}
 Substituting Eq.(\ref{sij-app-1}) and the identity operator $I_{\ov \WW} = |\ov S^{C'}\ra \la \ov S_{C'}|$ [Eq.(\ref{I-ovW})]
 into Eq.(\ref{Xk}), one gets that
\begin{align}\label{X-1}
  X^k =  \sigma  (s_b^k)^{AB}|S_A\ra \la\la S_B|(|\ov S^{C'}\ra \la\la \ov S_{C'}|) \sigma^T.
\end{align}
 Then, substituting Eqs.(\ref{sigma}) and (\ref{sigma-T}) into Eq.(\ref{X-1}), one finds that
\begin{gather*}
X^k = \sigma^{\mu EF'} |T_\mu\ra \la\la S_{F'E}| (s_b^k)^{AB}|S_A\ra |\ov S^{C'}\ra   \la\la \ov S_{C'}|
 \\ \times \la\la S_B| \sigma^{\nu GH'}|S_{GH'}\ra \la\la T_\nu|
 \\ = \sigma^{\mu EF'}  (s_b^k)^{AB}  \sigma^{\nu GH'} |T_\mu\ra  \la\la T_\nu|
 \la\la S_{F'E}| |S_A\ra |\ov S^{C'}\ra   \la\la \ov S_{C'}|
 \\ \times \la\la S_B| |S_{GH'}\ra
 \\ = \sigma^{\mu EF'}  (s_b^k)^{AB}  \sigma^{\nu GH'} |T_\mu\ra  \la\la T_\nu|
 \epsilon_{EA} \epsilon_{F'}^{\ \ C'} \epsilon_{BG} \epsilon_{C'H'}
 \\ = \sigma^{\mu EC'}  (s_b^k)^{A}_{\ \ G}  \sigma^{\nu GH'} |T_\mu\ra  \la\la T_\nu|
 \epsilon_{EA}  \epsilon_{C'H'}
 \\ = \sigma^{\mu}_{ AH'}  (s_b^k)^{A}_{\ \ G}  \sigma^{\nu GH'} |T_\mu\ra  \la\la T_\nu|,
\end{gather*}
 where Eq.(\ref{eps-delta}) has been used. 
 Then, writing $X^k$ in the form of $X^k \equiv (X^k)^{\mu\nu} |T_\mu\ra  \la\la T_\nu|$, its components are written as
\begin{align}\label{}
 (X^k)^{\mu\nu} = \sigma^{\mu}_{ AH'}  (s_b^k)^{A}_{\ G}  \sigma^{\nu GH'}.
\end{align}

 To go further, we make use of the explicit expression of $ (s_b^3)^{A}_{\ B}$ in Eq.(\ref{sk-explicit-3})
 and find that
\begin{align}\label{}
 (X^3)^{\mu\nu} = \frac 12 \sigma^{\mu}_{ 0H'}   \sigma^{\nu 0H'} - \frac 12 \sigma^{\mu}_{ 1H'}   \sigma^{\nu 1H'}.
\end{align}
 Making use of the explicit expressions of the EW-symbols in Eqs.(\ref{sigma^AB}) and (\ref{sig-lower-AB}),
 direct computation shows that
\begin{gather}
  \sigma^{\mu}_{ 0H'}   \sigma^{\nu 0H'} = \frac 12 \left( \begin{array}{cccc} 1 & 0 & 0 & 1 \\ 0 & -1 & i & 0
  \\ 0 & -i & -1 & 0 \\ -1 & 0 & 0 & -1  \end{array} \right),
\\ \sigma^{\mu}_{ 1H'}   \sigma^{\nu 1H'} = \frac 12 \left( \begin{array}{cccc} 1 & 0 & 0 & -1 \\ 0 & -1 & -i & 0
  \\ 0 & i & -1 & 0 \\ 1 & 0 & 0 & -1  \end{array} \right).
\end{gather}
 This gives that
\begin{gather}\label{Xmn-final}
 (X^3)^{\mu\nu} = \frac 12 \left( \begin{array}{cccc} 0 & 0 & 0 & 1 \\ 0 & 0 & i & 0
  \\ 0 & -i & 0 & 0 \\ -1 & 0 & 0 & 0  \end{array} \right).
\end{gather}
 Finally, from Eqs.(\ref{Jk-Xk}) and (\ref{Xmn-final}), one gets  Eq.(\ref{J3-matrix}).

\subsection{$s$-mode as a scalar mode}\label{app-scalar-mode}

 In this section, we  show that the spinor space $\SP$ for an $s$-mode is a one-dimensional $SL(2,C)$-scalar space.
 Let us consider the product state $|\kappa\ra |\chi\ra$, 
 for two arbitrary spinors $|\kappa\ra = -\kappa^A|S_A\ra $ and $|\chi\ra = -\chi^A |S_A\ra$ in $\WW$.
 From Eq.(\ref{SAB-commu}), one sees that 
\begin{subequations}\label{SASA=0}
\begin{align}
 |S^0\ra |S^0\ra =0, \  |S^1\ra |S^1\ra =0, 
 \\  |S^0\ra |S^1\ra = - |S^1\ra |S^0\ra.
\end{align}
\end{subequations}
 Then, simple derivation gives
\begin{gather}\label{|s>-|kappa-chi>}
 |\kappa\ra |\chi\ra = (\kappa^0\chi^1 - \kappa^1\chi^0 )|S_{0}\ra |S_{1}\ra
 = \left| \begin{array}{cc} \kappa^0 & \kappa^1 \\ \chi^0 & \chi^1 \end{array} \right| |S_{01}\ra.
\end{gather}
 Similarly, the transformed-spinor product is written as
 $|\ww\kappa\ra |\ww\chi\ra = (\ww\kappa^0\ww\chi^1 - \ww\kappa^1\ww\chi^0 )|S_{01}\ra$.

 A generic $SL(2,C)$ transformation is written as $h^A_{\ \ B}$ [see Eq.(\ref{h-AB-app})], 
\begin{equation}\label{h-AB}
  h^{A}_{\ \ B} = \left( \begin{array}{cc} a & b \\ c & d \end{array} \right)
  \quad \text{with} \ ad-bc=1.
\end{equation}
 It transforms the components $\kappa^A$ to 
\begin{gather}
 \ww\kappa^A =h^A_{\ \ B} \kappa^B 
 = \left( \begin{array}{c} a\kappa^0 + b\kappa^1 \\ c\kappa^0 + d\kappa^1 \end{array} \right).
\end{gather}
 It is ready to compute that
\begin{align}\label{}\notag
  \ww\kappa^0\ww\chi^1- \ww\kappa^1\ww\chi^0 = (ad-bc)\kappa^0 \chi^1 + (bc-ad)\kappa^1\chi^0 
 \\ = \kappa^0 \chi^1 -\kappa^1\chi^0 .
\end{align}
 Then, one sees that
\begin{gather}\label{}
 |\ww\kappa\ra |\ww\chi\ra = |\kappa\ra |\chi\ra.
\end{gather}
 Hence, the product state $|\kappa\ra |\chi\ra$ is a scalar under $SL(2,C)$ transformations.
 One gets the same result for the spinor space $\ov \SP$.

\begin{widetext}

\section{Derivations for H-operators}\label{app-derive-relations}

 In this appendix, we give detailed derivations for some expressions of $H$-operators given in the main text. 

\subsection{H-operator of free process} \label{app-derive-eqs-free-process}

\noindent (i) Derivation of Eq.(\ref{H0M}):
 \\ 
 Substituting Eqs.(\ref{IM}) and (\ref{C-momentum-for-H0}) and that $\C^{M\to M}_{\rm sp}=1$
 into Eq.(\ref{H-gfp-origin}), and making use of Eq.(\ref{Mstate-IP}), one gets that
\begin{gather*}\label{}
 H_M^0 = \left( \int d\ww p  |M^\alpha_{\bp }\ra \la \wh{M}_{\bp \alpha}| \right)
  \times \left( \int d\ww p' |\bp'\ra   |\bp'|  \la \bp'| \right)
  \times \left( \int d\ww p''  |M^{\alpha''}_{\bp'' }\ra \la \wh{M}_{\bp'' \alpha''}| \right)
 \\ =  \int d\ww p d\ww p'  d\ww p''  |M^\alpha_{\bp }\ra \delta_{\alpha}^{\alpha''} |\bp| \delta^3(\bp-\bp')
 |\bp'| |\bp''| \delta^3(\bp''-\bp') \la \wh{M}_{\bp'' \alpha''}|
 = \int d\ww p  |M^\alpha_{\bp }\ra  |\bp| \la \wh{M}_{\bp \alpha}|.
\end{gather*}

\vspace{0.2cm}

\noindent (ii) Derivation of Eq.(\ref{p0-M-1}):
\\ Firstly, we compute that
 \begin{gather*}
 \la \wh M^\beta_\bq|H^0_M|M^\alpha_\bp\ra =
 \la \wh M^\beta_\bq|\int d\ww p' \   |M^{\alpha'}_{\bp' }\ra  |\bp'| \la  \wh M_{\bp' \alpha'}||M^\alpha_\bp\ra
 \\  = \la \wh M^\beta_\bq| \int d\ww p'  |M^{\alpha'}_{\bp' }\ra  |\bp'|
 |\bp|\delta^3(\bp - \bp') \delta_{\alpha'}^{ \alpha} 
   = \Upsilon^{\beta \alpha}  |\bq|\delta^3(\bq - \bp) |\bp|.
\end{gather*}
 Then, substituting this result into Eq.(\ref{p0-M}), one gets Eq.(\ref{p0-M-1}), i.e., 
\begin{align}\label{}
  p^0 = \int d\ww q \Upsilon(\alpha, \alpha)  |\bq|\delta^3(\bq - \bp) |\bp| =  \Upsilon(\alpha, \alpha) |\bp|.
\end{align}

\subsection{Dirac-spinor expression of FIP amplitude}\label{app-FIP-amplitudes}

 In this section, we derive the FIP amplitude given in 
 Eq.(\ref{Cfovf'B-Dirac}), which are written in terms of  Dirac spinors. 
 To this end,  one may insert the expression of the operator $\sigma$ in Eq.(\ref{sigma})
 into Eq.(\ref{Cfovf'B-expre-cd}).
 Making use of Eq.(\ref{gamma-mu}) for $\gamma^\mu$,  one gets that
\begin{gather}\notag
 C^{f \ov f'\to B}_{\alpha \alpha' \lambda } = \frac{1}{n_f\sqrt{n_V}} \la\la 
    \ov{\varepsilon}_{\lambda}(\bk)|\sigma^{\mu AB'} |T_\mu\ra \la S_{B'A}|
  \Big( c |u^{f}_\alpha(\bp)\ra |\ov v^{\ov f'}_{{\alpha'}}(\bp')\ra
   + d  |\ov v^{f}_{\alpha}(\bp)\ra |u^{\ov f'}_{\alpha'}(\bp')\ra \Big)
 \\ = \frac{1}{n_f\sqrt{n_V}} \sigma^{\mu AB'} {\ov\varepsilon}_{\lambda \mu}(\bk)
  \Big( c u^{f}_{\alpha A}(\bp) \ov v^{\ov f'}_{\alpha' B'}(\bp')
   - d  \ov v^{f}_{\alpha B'}(\bp) u^{\ov f'}_{\alpha' A}(\bp') \Big).
\end{gather}
 This amplitude may be expressed in terms of the Dirac spinors of $U^{\alpha }(\bp)$ and $V^{\alpha }(\bp)$ 
 defined in Eq.(\ref{UV-def}).
 Indeed, it is direct to check that
\begin{gather}\notag
  C^{f \ov f'\to B}_{\alpha {\alpha'} \lambda }
 =  \frac{1}{n_f\sqrt{n_V}}  {\ov\varepsilon}_{\lambda \mu}(\bk)
 \left( \begin{array}{cc}  \ov v^{\ov f'}_{\alpha' B'}(\bp'), - u^{\ov f'}_{\alpha' A}(\bp')\end{array}\right)
 \left( \begin{array}{cc} c \sigma^{\mu AB'} & 0 \\ 0 & d  \sigma^{\mu AB'} \end{array}\right)
 \left( \begin{array}{cc} u^{f}_{\alpha A}(\bp) \\ \ov v^{f}_{\alpha B'}(\bp) \end{array}\right)
 \\ = \frac{1}{n_f\sqrt{n_V}}  {\ov\varepsilon}_{\lambda \mu}(\bk)
 \left( \begin{array}{cc}  \ov v^{\ov f' B'}_{\alpha'}(\bp'), - u^{\ov f'}_{\alpha' A}(\bp')\end{array}\right)
 \left( \begin{array}{cc} c \ov\sigma^{\mu B'A} & 0 \\ 0 & d  \sigma^{\mu AB'} \end{array}\right)
 \left( \begin{array}{cc} u^{fA}_{\alpha }(\bp) \\ \ov v^{f}_{\alpha B'}(\bp) \end{array}\right),
 \label{Cfovf'B-exp1-app}
\end{gather}
 where Eqs.(\ref{f-AB}) and (\ref{sig-c}) have been used.
 The mode-species labels in the Weyl spinors used above, 
 e.g., the superscript $f$ in $u^{fA}_{\alpha }(\bp)$, indicate where the Weyl spinors come from.
 But, properties of these Weyl spinors are in fact independent of the mode species.
 Hence, it is unnecessary to use these mode-species labels in the final expression of 
 the interaction amplitude, as already done in Eq.(\ref{Cfovf'B-Dirac}). 
 Finally, noting the explicit expression of the $\gamma^\mu$ matrices in Eq.(\ref{gamma-mu}) 
 and the following relation,
\begin{align}\label{csig=Gam-gam-mu}
 \left( \begin{array}{cc} c \ov\sigma^{\mu}_{B'A}  & 0 \\ 0 & d\sigma^{\mu AB'} \end{array} \right)
 = \left( \begin{array}{cc} 0 & c \\ d  & 0 \end{array} \right)
 \left( \begin{array}{cc}  0 & \sigma^{\mu AB'} \\ \ov\sigma^{\mu}_{B'A} &0  \end{array} \right)
 = \frac{1}{\sqrt 2} \Gamma^{f \ov f'\to B} \gamma^\mu,
\end{align}
 one is ready to get Eq.(\ref{Cfovf'B-Dirac}).

\subsection{Derivations of H-operators of BPIPs} \label{app-derive-eqs-BPIP}

 In this section, we give detailed derivations for expressions of $H$-operator of BPIPs given in the main text,
 particularly, for Eqs.(\ref{H-BPIP-1-4}), (\ref{H-BPIP-1-4-UV}), and (\ref{HffB-BPIP-fields-1}).

\vspace{0.2cm}
\noindent (i) Derivation of  Eq.(\ref{H-BPIP-1-4}).
 \\ 
 (a) For $H^{f f' \to B}_{{\rm BPIP},1}$ in Eq.(\ref{H1}), one has $\varrho(\alpha)=\varrho(\alpha')=+$.
 Hence, this operator is directly obtained by replacing the ket and bras in Eq.(\ref{H-FIP1})
 by creation operator and annihilation operators, respectively. 
\\ (b) To compute $H^{f f' \to B}_{{\rm BPIP},2}$ Eq.(\ref{H2}), 
one substitutes Eqs.(\ref{H-FIP1}) and (\ref{VK1})  into Eq.(\ref{H-type1-BPIP2}).
 For $H^{f f' \to B}_{{\rm BPIP},2}$, $\varrho(\alpha)=+$ and $\varrho(\alpha')=-$, hence, 
 the VF pair should be $(f',\ov f')$, with the state of $\ov f'$
 appearing in a scalar product. 
 Thus, we have
\begin{gather}\label{}\notag
 H^{f \ov f' \to B}_{{\rm BPIP},2} =  \int d\ww q |{f'}_{\ov \bq \ov \beta} \ra
 \left( H_{\rm FIP}^{f \ov f'\to B} \right)^{\curvearrowleft} |{\ov f'}_{\bq \beta} \ra
  =  \int d\ww q |{f'}_{\ov \bq \ov \beta} \ra  \left( \int d\Omega \
 |B^\lambda_\bk\ra \delta^3_P C^{f \ov f'\to B}_{\alpha \alpha' \lambda }
 \la \wh{\ov f'}_{\bp'}^{ \alpha'}| \la   \wh{f}_{\bp}^{ \alpha}| \right)^{\curvearrowleft} |{\ov f'}_{\bq \beta} \ra
 \\ = \int d\Omega \ |B^\lambda_\bk\ra \delta^3_P C^{f \ov f'\to B}_{\alpha \alpha' \lambda }
  \int d\ww q |{f'}_{\ov \bq \ov \beta} \ra
 \la \wh{\ov f'}_{\bp'}^{ \alpha'}|{\ov f'}_{\bq \beta} \ra \la   \wh{f}_{\bp}^{ \alpha}|
 = \int d\Omega \ |B^\lambda_\bk\ra \delta^3_P C^{f \ov f'\to B}_{\alpha \alpha' \lambda }
  |{f'}_{\ov \bp'  \ov \alpha'} \ra   \la \wh{f}_{\bp}^{ \alpha}|.
\end{gather}
 For $\varrho(\alpha')=-$, one has $\varrho(\ov \alpha')=+$ and,
 as a result, $|{f'}_{\ov \bp'  \ov \alpha'} \ra = |{f'}_{\ov \bp'}^{ \ov \alpha'} \ra$.
 Then, it is ready to get Eq.(\ref{H2}). 
 \\ (c) Similarly, for $H^{f \ov f' \to B}_{{\rm BPIP},3}$ with $\varrho(\alpha) =-$ and $\varrho(\alpha')=+$, one has
\begin{gather}\label{}\notag
 H^{f \ov f' \to B}_{{\rm BPIP},3} = \int d\ww q |\ov{f}_{\ov \bq \ov \beta} \ra
 \left( H_{\rm FIP}^{f \ov f'\to B} \right)^{\curvearrowleft} |f_{\bq \beta} \ra
  = \int d\ww q |\ov{f}_{\ov \bq \ov \beta} \ra
 \left( \int d\Omega \
 |B^\lambda_\bk\ra \delta^3_P C^{f \ov f'\to B}_{\alpha \alpha' \lambda }
 \la \wh{\ov f'}_{\bp'}^{ \alpha'}| \la   \wh{f}_{\bp}^{ \alpha}| \right)^{\curvearrowleft} |f_{\bq  \beta} \ra
 \\ =  \int d\Omega \
 |B^\lambda_\bk\ra \delta^3_P C^{f \ov f'\to B}_{\alpha \alpha' \lambda }
 \int d\ww q |\ov{f}_{\ov \bq \ov \beta} \ra
 \la \wh{\ov f'}_{\bp'}^{ \alpha'}| \la   \wh{f}_{\bp}^{ \alpha} |f_{\bq  \beta} \ra
 =  \int d\Omega \
 |B^\lambda_\bk\ra \delta^3_P C^{f \ov f'\to B}_{\alpha \alpha' \lambda }
 |\ov{f}_{\ov \bp}^{\ov \alpha} \ra \varrho(\ov\alpha)  \la \wh{\ov f'}_{\bp'}^{ \alpha'}|.
\end{gather}
 This gives Eqs.(\ref{H3}) with $\varrho( \ov \alpha) =+$. 
\\ (d) For $H^{f \ov f' \to B}_{{\rm BPIP},4}$ with $\varrho(\alpha) =\varrho(\alpha') =-$, one has
\begin{gather}\label{}
 H^{f \ov f' \to B}_{{\rm BPIP},4}  = 
 \int d\ww q d\ww q' |\ov{f}_{\ov \bq \ov \beta} \ra \ \Big[ \ |{f'}_{\ov \bq' \ov \beta'} \ra
  \left( H_{\rm FIP}^{f \ov f'\to B} \right)^{\curvearrowleft} |{\ov f'}_{\bq'  \beta'} \ra 
  \Big]^{\curvearrowleft} |f_{\bq  \beta} \ra   \notag
 \\ = \int d\ww q d\ww q' |\ov{f}_{\ov \bq \ov \beta} \ra \ \Big[ \ |{f'}_{\ov \bq' \ov \beta'} \ra
  \left(  \int d\Omega \
 |B^\lambda_\bk\ra \delta^3_P C^{f \ov f'\to B}_{\alpha \alpha' \lambda }
 \la \wh{\ov f'}_{\bp'}^{ \alpha'}| \la   \wh{f}_{\bp}^{ \alpha}| \right)^{\curvearrowleft} |{\ov f'}_{\bq'  \beta'} \ra 
  \Big]^{\curvearrowleft} |f_{\bq \beta} \ra   \notag
 \\ =  \int d\Omega \ |B^\lambda_\bk\ra \delta^3_P C^{f \ov f'\to B}_{\alpha \alpha' \lambda }
 \int d\ww q d\ww q' |\ov{f}_{\ov \bq \ov \beta} \ra  \ |{f'}_{\ov \bq' \ov \beta'} \ra
 \la \wh{\ov f'}_{\bp'}^{ \alpha'}|{\ov f'}_{\bq' \beta'} \ra   \la   \wh{f}_{\bp}^{ \alpha}| f_{\bq \beta} \ra   \notag
 \\ =  \int d\Omega \ |B^\lambda_\bk\ra \delta^3_P C^{f \ov f'\to B}_{\alpha \alpha' \lambda }
  |\ov{f}_{\ov \bp}^{\ov \alpha} \ra  \ |{f'}_{\ov \bp'}^{\ov \alpha'} \ra \varrho(\ov\alpha') \varrho(\ov\alpha).
\end{gather}
 This gives Eq.(\ref{H4}) with $\varrho(\ov\alpha') = \varrho( \ov \alpha) =+$.

\vspace{0.2cm}
\noindent (ii) Derivation of Eq.(\ref{H-BPIP-1-4-UV}).
 \\ (a)  Equation (\ref{H1-UV})
 can be obtained directly by substituting Eq.(\ref{Cfovf'B-Dirac}) into Eq.(\ref{H1})
 with $(\varrho(\alpha),\varrho(\alpha'))= (+,+)$.
 To be consistent with expressions to be derived below for other $H$-operators, we reindex $(\bp',\alpha')$ as $(\bq,\beta)$.
\\ (b)  To prove Eq.(\ref{H2-UV}) from Eq.(\ref{H2}) with $(\varrho(\alpha),\varrho(\alpha'))= (+,-)$,
 we note that Eq.(\ref{uv-al-p=ovq}) implies that 
\begin{gather*}
 U^{\alpha }(\ov\bp) = \left( \begin{array}{c}   u^{\alpha  A}(\ov\bp) \\  \ov v_{B'}^{\alpha}(\ov\bp)   \end{array} \right)
  = \left( \begin{array}{c}   -v^{\alpha  A}(\bp) \\  \ov u_{B'}^{\alpha}(\bp)   \end{array} \right) = - V^{\alpha }(\bp), \quad
  V^{\alpha }(\ov\bp) = \left( \begin{array}{c}   v^{\alpha  B}(\ov\bp) \\  - \ov u_{A'}^{\alpha}(\ov\bp)  \end{array} \right)
  = \left( \begin{array}{c}   u^{\alpha  B}(\bp) \\ \ov v_{A'}^{\alpha}(\bp)  \end{array} \right) = U^{\alpha }(\bp),
\end{gather*}
 as a result, 
\begin{gather}\label{UV-p-ovp}
 V^{\alpha }(\bp) = - U^{\alpha }(\ov\bp), \quad U^{\alpha }(\bp) =  V^{\alpha }(\ov\bp).
\end{gather}
 Substituting Eq.(\ref{Cfovf'B-Dirac}) into Eq.(\ref{H2}) and using the first equality in Eq.(\ref{UV-p-ovp}), we get that
\begin{gather*}
   H^{f f' \to B}_{{\rm BPIP},2} = \int d\Omega b^{\lambda \dag}_{B}(\bk) \delta^3_P  
    {\ov\varepsilon}_{\lambda \mu}(\bk) V_{\alpha'}^{ \dag}(\bp')  \Gamma^{f \ov f'\to B} \gamma^\mu U_{\alpha }(\bp)
  b^{\ov \alpha' \dag }_{f'}(\ov \bp') b^{ \alpha}_{f}(\bp)
 \\ = -\int d\Omega b^{\lambda \dag}_{B}(\bk) \delta^3(\bk-\bp + \ov\bp')
    {\ov\varepsilon}_{\lambda \mu}(\bk) U_{\alpha'}^{ \dag}(\ov\bp')  \Gamma^{f \ov f'\to B} \gamma^\mu U_{\alpha }(\bp)
  b^{\ov \alpha' \dag }_{f'}(\ov \bp') b^{ \alpha}_{f}(\bp).
\end{gather*}
 Noting that $- U_{\alpha'}^{ \dag}(\ov\bp') = \varrho(\alpha') U_{\alpha'}^{ \dag}(\ov\bp')$
 and making use of Eqs.(\ref{Upsilon-ab}) and (\ref{alpha-raise}),
 we find that $  \varrho(\alpha') U_{\alpha'}^{ \dag}(\ov\bp') = U^{\alpha'  \dag}(\ov\bp')$. 
 Moreover, according to the definition in Eq.(\ref{uv-alpha}), 
 an index $\alpha'$ in the upper position is independent of the sign $\varrho$ in it;
 this gives that $U^{\alpha'  \dag}(\ov\bp') = U^{\ov\alpha'  \dag}(\ov\bp')$.
 Furthermore, since $\varrho(\ov\alpha') = +$, 
 we have $U^{\ov\alpha'  \dag}(\ov\bp') = \varrho(\ov\alpha') U_{\ov\alpha' }^{ \dag}(\ov\bp')
 = U_{\ov\alpha' }^{ \dag}(\ov\bp')$.
 Making use of the above properties, we get that
\begin{gather}
   H^{f f' \to B}_{{\rm BPIP},2} = \int d\ww p d\ww {p'} d\ww k b^{\lambda \dag}_{B}(\bk) \delta^3(\bk-\bp + \ov\bp')
    {\ov\varepsilon}_{\lambda \mu}(\bk) U_{\ov\alpha'}^{ \dag}(\ov\bp')  \Gamma^{f \ov f'\to B} \gamma^\mu U_{\alpha }(\bp)
  b^{\ov \alpha' \dag }_{f'}(\ov \bp') b^{ \alpha}_{f}(\bp).
\end{gather}
 To simplify the notation, we introduce a index $\beta$ with $\beta = \ov\alpha' = (r',+)$
 and a momentum index $\bq$ with $\bq = \ov\bp'$.
 Finally, changing the integration over $\bp'$ to that over $\bq$, we get Eq.(\ref{H2-UV}). 
 \\ (c) Equation (\ref{H3-UV}) can be proved in a similar way from Eq.(\ref{H3}) with $(\varrho(\alpha),\varrho(\alpha'))= (-,+)$.
 In fact, substituting Eq.(\ref{Cfovf'B-Dirac}) into Eq.(\ref{H3})  and using the second equality in Eq.(\ref{UV-p-ovp}), 
 we get that
\begin{gather} \notag
 H^{f f' \to B}_{{\rm BPIP},3}  =  \int d\Omega b^{\lambda \dag}_{B}(\bk)  \delta^3(\bk-\bp-\bp')
 {\ov\varepsilon}_{\lambda \mu}(\bk) V_{\alpha'}^{ \dag}(\bp')  \Gamma^{f \ov f'\to B} \gamma^\mu U_{\alpha }(\bp)
   b^{\ov \alpha \dag }_{\ov f}(\ov \bp) b^{ \alpha'}_{\ov f'}(\bp')
 \\ =\int d\Omega b^{\lambda \dag}_{B}(\bk)  \delta^3(\bk+ \ov\bp-\bp')
 {\ov\varepsilon}_{\lambda \mu}(\bk) V_{\alpha'}^{ \dag}(\bp')  \Gamma^{f \ov f'\to B} \gamma^\mu V_{\alpha }(\ov\bp)
   b^{\ov \alpha \dag }_{\ov f}(\ov \bp) b^{ \alpha'}_{\ov f'}(\bp') \notag
 \\ = - \int d\Omega b^{\lambda \dag}_{B}(\bk)  \delta^3(\bk+ \ov\bp-\bp')
 {\ov\varepsilon}_{\lambda \mu}(\bk) V_{\alpha'}^{ \dag}(\bp')  \Gamma^{f \ov f'\to B} \gamma^\mu V_{\ov\alpha }(\ov\bp)
   b^{\ov \alpha \dag }_{\ov f}(\ov \bp) b^{ \alpha'}_{\ov f'}(\bp'), \label{app-H3-derive-a'}
\end{gather}
 where in the derivation of the last equality we have used that 
 $V_{\alpha }(\ov\bp) = - V^{\alpha }(\ov\bp) =  - V^{\ov\alpha }(\ov\bp) =  - V_{\ov\alpha }(\ov\bp)$.
 Finally, one gets Eq.(\ref{H3-UV}) by changing indices in Eq.(\ref{app-H3-derive-a'})
 as $(\ov\bp,\ov\alpha) \to (\bp,\alpha)$ and  $(\bp',\alpha') \to (\bq,\beta)$.
 \\ (d) To prove Eq.(\ref{H4-UV}) from Eq.(\ref{H4}) with $(\varrho(\alpha),\varrho(\alpha'))= (-,-)$, 
 one may follow a procedure similar to that used above. 
 Let us substitute Eq.(\ref{Cfovf'B-Dirac}) into Eq.(\ref{H4}), getting that
\begin{gather*}
   H^{f f' \to B}_{{\rm BPIP},4}  =  \int d\Omega b^{\lambda \dag}_{B}(\bk) \delta^3_P 
{\ov\varepsilon}_{\lambda \mu}(\bk) V_{\alpha'}^{ \dag}(\bp')  \Gamma^{f \ov f'\to B} \gamma^\mu U_{\alpha }(\bp)
  b^{\ov \alpha \dag }_{\ov f}(\ov \bp) b^{\ov \alpha' \dag }_{ f'}(\ov \bp')
 \\  =  - \int d\Omega b^{\lambda \dag}_{B}(\bk) \delta^3_P 
{\ov\varepsilon}_{\lambda \mu}(\bk) U_{\alpha'}^{ \dag}(\ov\bp')  \Gamma^{f \ov f'\to B} \gamma^\mu V_{\alpha }(\ov\bp)
  b^{\ov \alpha \dag }_{\ov f}(\ov \bp) b^{\ov \alpha' \dag }_{ f'}(\ov \bp')
 \\  =  - \int d\Omega b^{\lambda \dag}_{B}(\bk) \delta^3(\bk-\bp-\bp')
{\ov\varepsilon}_{\lambda \mu}(\bk) U_{\ov\alpha'}^{ \dag}(\ov\bp')  \Gamma^{f \ov f'\to B} 
\gamma^\mu V_{\ov\alpha }(\ov\bp)
  b^{\ov \alpha \dag }_{\ov f}(\ov \bp) b^{\ov \alpha' \dag }_{ f'}(\ov \bp').
\end{gather*}
 Then, reindexing as $(\ov\bp,\ov\alpha) \to (\bp,\alpha)$ and $(\ov\bp',\ov\alpha') \to (\bq,\beta)$, 
 we get Eq.(\ref{H4-UV}).

\vspace{0.2cm}
\noindent (v) Derivation of Eq.(\ref{HffB-BPIP-fields-1}). 
 \\ Making use of Eqs.(\ref{psif-phiB-expan}), one computes that
\begin{gather*}
 \int d^3\bx \ \psi^\dag_{f'} \Gamma^{f \ov f'\to B} \gamma^\mu \psi_f \phi_{B\mu}^\dag
  = (2\pi)^3 \int d\ww q  d\ww p d\ww k  \delta^3(\bk + \bq -\bp) 
  {\ov\varepsilon}_{\lambda \mu}(\bk)   b^{\lambda \dag}_{B}(\bk)  U_{\beta}^{ \dag}(\bq)   
  \Gamma^{f \ov f'\to B} \gamma^\mu
  U_{\alpha }(\bp)  b^{ \beta \dag}_{f'}(\bq)  b^{ \alpha}_{f}(\bp)  
 \\ + \delta^3(\bk + \bq + \bp)  {\ov\varepsilon}_{\lambda \mu}(\bk)  b^{\lambda \dag}_{B}(\bk) U_{\beta}^{ \dag}(\bq) 
 \Gamma^{f \ov f'\to B} \gamma^\mu
  V_{\alpha }(\bp)  b^{ \beta \dag}_{f'}(\bq)  b^{\alpha \dag }_{\ov f}(\bp) 
 \\ + \delta^3(\bk - \bq - \bp)   {\ov\varepsilon}_{\lambda \mu}(\bk)  b^{\lambda \dag}_{B}(\bk)
 V_{\beta}^{ \dag}(\bq)  \Gamma^{f \ov f'\to B} \gamma^\mu
   U_{\alpha }(\bp) b^{\beta }_{\ov f'}(\bq)  b^{ \alpha}_{f}(\bp) 
 \\ + \delta^3(\bk - \bq + \bp)   {\ov\varepsilon}_{\lambda \mu}(\bk)  b^{\lambda \dag}_{B}(\bk)
 V_{\beta}^{ \dag}(\bq)  \Gamma^{f \ov f'\to B} \gamma^\mu
 V_{\alpha }(\bp) b^{\beta }_{\ov f'}(\bq)  b^{\alpha \dag }_{\ov f}(\bp).
\end{gather*}
 Taking the normal product, this gives Eq.(\ref{HffB-BPIP-fields-1}). 

\end{widetext}

\section{Relations between  helicity states of b-modes with opposite momenta}\label{app-spinor-ov-p}

 In this appendix, we derive relations between basic-mode helicity states
 $|u^r(\bp)\ra$ and those for  the opposite momentum $\ov\bp \equiv -\bp$.
 For the simplicity in discussion, we set the direction of $\bp$ along the $z$-axis.
 (Generalization to a generic direction is straightforward.)

 As is known, corresponding to a rotation performed in the ordinary three-dimensional space 
 by an angle $\psi$ around the $(\theta,\phi)$ direction,
 one has the following rotation in the spinor space $\WW$ \cite{Kim-group}, i.e.,
\begin{gather}
 R_\WW(\theta,\phi,\psi) =  e^{-i\frac{\psi}{2} {\bf n} \cdot {\bbs \sigma}}
  =  \left( I \cos \frac{\psi}{2} - i {\bf n} \cdot {\bbs \sigma} \sin \frac{\psi}{2} \right), \label{R-gen}
\end{gather}
 where $I$ indicates the unit matrix, ${\bbs \sigma} = (\sigma_x,\sigma_y,\sigma_z)$
 represents the vector of Pauli matrices, and ${\bf n}$ is the unit vector along the $(\theta,\phi)$ direction.
 Or, in the notation of Eq.(\ref{R-ang}), the Weyl-spinor rotation is written as
\begin{gather}\label{R-ang-b-mode}
 R_\WW(\theta,\phi,\psi) =  \exp\left( -i\theta_k s_b^k\right),
\end{gather}
 where $s_b^k$ are components of the angular momentum of basic mode,
 which are given in Eq.(\ref{sk-explicit}).

 Clearly, a rotation in the ordinary $(x,y,z)$ space around the $y$-axis by an angle $\psi=\pi$
 transforms $\bp$ to $\ov\bp$.
 This rotation corresponds to $R_\WW(\frac{\pi}{2},\frac{\pi}{2},\pi)$ in the space $\WW$
 and gives $[ w^{rA}(\ov\bp)]$, the matrix form of $w^{rA}(\ov\bp)$, i.e.,
\begin{gather}\label{}
 [ w^{rA}(\ov\bp)] =  R(\frac{\pi}{2},\frac{\pi}{2},\pi)  [ w^{rA}(\bp)].
\end{gather}
 From Eqs.(\ref{R-gen})-(\ref{R-ang-b-mode}), one gets that
\begin{align}\label{}
\notag
  R_\WW(\frac{\pi}{2},\frac{\pi}{2},\pi)
 & =  ( I \cos \frac{\psi}{2} - i \sigma_y \sin \frac{\psi}{2}),
 \\ & =  \left( \begin{array}{cc} \cos \frac{\psi}{2} & - \sin \frac{\psi}{2}
               \\ \sin \frac{\psi}{2} & \cos \frac{\psi}{2} \end{array} \right)
  =  \left( \begin{array}{cc} 0 & - 1 \\ 1 & 0 \end{array} \right).
\end{align}
 Then, with $[ w^{rA}(\bp)]$ in Eq.(\ref{urASM-RP0}), this gives that
\begin{gather}\label{wr-ovp-wrp-1}
 [{w}^{r=0,A}(\ov \bp)]  =  \left( \begin{array}{c} 0  \\ 1 \end{array} \right), \quad
 [{w}^{r=1,A}(\ov \bp)] =  - \left( \begin{array}{c} 1  \\ 0 \end{array} \right).
\end{gather}

 Comparing Eq.(\ref{urASM-RP0}) and Eq.(\ref{wr-ovp-wrp-1}), one sees that
\begin{gather}\label{ur=ur+1}
 |{w}^{r}(\ov \bp)\ra = (-1)^{r} |{w}^{r+1}(\bp)\ra .
\end{gather}
 Then, usig Eq.(\ref{w^r-w_s}), one gets that
\begin{gather}\label{ur=ur-ud}
 |{w}^{r}(\ov \bp)\ra = - |{w}_{r}(\bp)\ra .
\end{gather}
 To find an expression of $|{w}_{r}(\ov \bp)\ra$, one may substitute Eq.(\ref{ur=ur-ud})
 into the relation of $|{w}_{r}(\ov \bp)\ra =  |{w}^{s}(\ov \bp)\ra \epsilon_{sr}$, getting that
\begin{align}\label{}
 |{w}_{r}(\ov \bp)\ra =  - |{w}_{s}(\bp)\ra \epsilon_{sr} =  - |{w}^{t}(\bp)\ra \epsilon_{ts} \epsilon_{sr}.
\end{align}
 Then, using Eq.(\ref{epep-delta}), one gets that
\begin{gather}\label{w_rq=-w^rp}
 |{w}_{r}(\ov \bp)\ra =  |{w}^{r}(\bp)\ra.
\end{gather}
 \footnote{
 One should note that, due to the well-known difference between rotations in the real three-dimensional 
 space and those
 in the Weyl spinor space, the relation between $|{w}_{r}(\ov \bp)\ra$ and $|{w}^{r}(\bp)\ra$ can not be obtained
 by directly exchanging $\bp$ and $\ov \bp$ in Eq.(\ref{ur=ur-ud}),
 which would give a wrong relation of $|{w}_{r}(\ov \bp)\ra = - |{w}^{r}(\bp)\ra$.
 In fact, this latter relation corresponds to the rotation of
\begin{gather}\label{}
  R(\frac{\pi}{2},\frac{\pi}{2},3\pi)
  =  \left( \begin{array}{cc} 0 & 1 \\ -1 & 0 \end{array} \right).
\end{gather}
 It is easy to check that, under this rotation, the rhs of Eq.(\ref{ur=ur+1})
 should be multiplied by a factor $(-1)$ and, as a consequence, so do Eqs.(\ref{ur=ur-ud})-(\ref{uv-al-p=ovq}).
 }
 Finally, for the spinors of $u$ and $v$ defined in Eq.(\ref{uv-alpha}), 
 one gets that
\begin{subequations}\label{uv-al-p=ovq}
\begin{align}
 |{u}^{\alpha}(\ov\bp)\ra & = -  |{v}^{\alpha}(\bp)\ra, \label{uv-al-p=ovq1}
 \\ |{v}^{\alpha}(\ov\bp)\ra & =  |{u}^{\alpha}(\bp)\ra .\label{uv-al-p=ovq2}
\end{align}
\end{subequations}

\section{The matrices $\Gamma^{f \ov f' \to B}$}\label{app-G-matrix}

 In this appendix, we give detailed derivations for matrices $\Gamma^{f \ov f' \to B}$
 given in the main text, as well as for related quantities. 

\subsection{Expressions of c and d}\label{app-compute-cd}

 In this section, we derive expressions of the two quantities $c$ and $d$ [see Eq.(\ref{cd})], 
 which are useful when computing the matrices $\Gamma^{f \ov f' \to B}$ for $T_1$ and $T_2$.

 It proves convenient to write the layer phase $\vartheta^M_m$ of a mode $M$ as a function of the global phase 
 $\vartheta^M_1$ and the phase difference $\delta^M_\theta$.
 For two-layer and three-layer modes, the assumption ${\rm MA}^{\rm layer}_{\rm phase}$ gives that
\begin{align}\label{theta-Mm-deltatheta}
 \vartheta^M_m = \vartheta^M_1 + (m-1) \delta^M_\theta.
\end{align}
 For $u$-modes and $d$-modes, within each sections the layer phases satisfy Eq.(\ref{theta-Mm-deltatheta})
 for a corresponding $t$-mode. 
 Written explicitly, 
\begin{align}\label{theta-u-deltatheta}
 \vartheta^{u_i}_m = \left\{                    \begin{array}{ll}
                     \vartheta^{t_i}_1 + (m-1) \delta^{t_i}_\theta, & \hbox{for $m=1,2,3$;} \\
                     \vartheta^{t_i}_1 + (m-4) \delta^{t_i}_\theta , & \hbox{for $m=4,5,6$;}
                   \end{array}                 \right.
\end{align}
\begin{align}\label{theta-d-deltatheta}
 \vartheta^{d_i}_m = \left\{                    \begin{array}{ll}
                     \vartheta^{t_i}_1 + (m-1) \delta^{t_i}_\theta, & \hbox{for $m=1,2,3$;} \\
                     \vartheta^{t_i}_1 + (m-4) \delta^{t_i}_\theta + \frac{\pi}{2}, & \hbox{for $m=4,5,6$;}
                   \end{array}                 \right.
\end{align}
 where the phase $\pi/2$ for $d$-modes' layers with $m=4,5,6$ comes from the factor $i$ in
 states of their second sections [see Eq.(\ref{Sdi})].

 For $T_1$ in Eq.(\ref{T1}) with $n_f=2,3$,  making use of Eq.(\ref{theta-Mm-deltatheta})
 and noting the term $\delta_{nm}$, from Eq.(\ref{cd}) one gets that
\begin{subequations}\label{cd-T1}
\begin{align}\label{}
 c = e^{i(\vartheta^f_1+\vartheta^{f'}_1 )} 
 \sum_{m\in \M_z}  e^{i(m-1)(\delta^f_\theta + \delta^{f'}_\theta - \delta^B_\theta)},
 \\ d = e^{i(\vartheta^f_1+\vartheta^{f'}_1 )} 
 \sum_{m\in \M_{\ov z}}  e^{i(m-1)(\delta^f_\theta + \delta^{f'}_\theta - \delta^B_\theta)},
\end{align}
\end{subequations}
 where the fact of $\vartheta^{B}_1 =0$ has been used.

 The matrix $T_2$ in Eq.(\ref{T-zm}) is for two cases.
 In the first case, $f=f'=u_i$.
 From Eq.(\ref{Sui}), one sees that the two sections of a $u_i$-mode are identical.
 Then, noting that $\vartheta^B_1=0$ and $\vartheta^B_2 = \delta^B_\theta$, one writes Eq.(\ref{cd}) as
\begin{subequations}\label{cd-T2-u}
\begin{align}\label{}
 & c = 2\exp( i2\vartheta^{t_i}_1 )
   \sum_{m\in \M_z ( m\le 3)} e^{i2(m-1) \delta^t_\theta },
 \\ & d = 2\exp( i 2\vartheta^{t_i}_1 -i\delta^B_\theta) 
 \sum_{m\in \M_{\ov z} ( m\le 3)}  e^{i2(m-1) \delta^t_\theta }.
\end{align}
\end{subequations}
 In the second case, $f=f'=d_i$.
 In this case, the two sections of a $d_i$-mode 
 are antimodes of each other with an additional section phase of $\pi/2$
 and, as a result, expressions of $c$ and $d$ like those in Eq.(\ref{cd-T2-u}) are quite complicated. 
 Hence, we simply write them in terms of $\vartheta^{d_i}_m$, i.e., 
\begin{subequations}\label{cd-T2-d}
\begin{align}\label{}
 & c = \sum_{m\in \M_z} \exp (i 2 \vartheta^{d_i}_m),
 \\ & d = \exp(  -i\delta^B_\theta)  \sum_{m\in \M_{\ov z}} \exp (i 2 \vartheta^{d_i}_m).
\end{align}
\end{subequations}

\subsection{Derivation of some equations}\label{app-derive-Gamma}

 In this section, we give detailed derivations for Eqs.(\ref{Gamma-enu}), (\ref{Gamma-cd-ud-AZ}),  (\ref{Gamma-qijg-all}),
 and (\ref{Gam-gam-exc-1}).

\subsubsection{ Derivation of Eq.(\ref{Gamma-enu}) }\label{app-derive-G-enu}

 This equation is for two-layer modes with $T_1$. 
 As discussed in the main text, for two-layer modes, one has
\begin{align}\label{}
 &  \vartheta^e_1=\vartheta^\nu_1=0,
 \\ & \delta^\nu_\theta = \delta_\theta^{Z} = 0, \ \  \delta^e_\theta =\pi/2, \ \  \delta_\theta^{A} =\delta_\theta^W = \pi.
\end{align}
 Below, we discuss the FIPs separately. 
 \\ (a) For the pair of $e\ov e$. 
 For an FIP of $f\ov f' \to B$ with $f=f'=e$, from Eq.(\ref{S-eove}) one gets that
 $|\LL^{e}_1\ra |\LL^{\ov e}_1\ra =|u^e_\alpha(\bp)\ra |\ov v^{\ov e}_{{\alpha'}}(\bp')\ra$ 
 and $ |\LL^{e}_2\ra |\LL^{\ov e}_2\ra =|\ov v^e_{\alpha}(\bp)\ra |u^{\ov e}_{\alpha'}(\bp')\ra$.
 This gives that $\M_z =\{ 1 \}$ and $ \M_{\ov z} = \{ 2 \}$ for $\M_z$ and $ \M_{\ov z}$ defined in Eq.(\ref{Mz-Movz}).
 Then, from Eq.(\ref{cd-T1}), one gets that
\begin{align}\label{}
 & \text{$e \ov e  \to A$: \ } c=e^0=1,  \quad \&  \quad  d=e^{0} =1;
 \\ & \text{$e \ov e  \to Z$: \ } c=e^0=1,  \quad \&  \quad  d=e^{i\pi} =-1.
\end{align}
 \\ (b) For the pair of $\nu \ov \nu $, using Eq.(\ref{S-novn}), one finds that 
 $\M_z =\{ 1,2 \}$ and $ \M_{\ov z} = \emptyset$.
 This gives that
\begin{align}\label{}
 & \text{$\nu \ov \nu  \to A$: \ } c=1 + e^{-i\pi}=0,  \quad \&  \quad  d=0;
 \\ & \text{$\nu \ov \nu  \to Z$: \ } c=1+ e^0=2,  \quad \&  \quad  d=0.
\end{align}
 \\ (c) For both pairs of $\nu \ov e$ and $e \ov \nu$, one finds that $\M_z =\{ 1 \}$ and $ \M_{\ov z} = \emptyset$
 and this gives that
\begin{align}\label{}
 & \text{$\nu \ov e  \to W$: \ } c=e^0=1,  \quad \&  \quad  d=0;
 \\ & \text{$e \ov \nu  \to \ov W$: \ } c=e^0 =1,  \quad \&  \quad  d=0.
\end{align}
 Finally, substituting the above values of $c$ and $d$ into Eq.(\ref{Gamma1}),
 and noting that $n_f=2$, $n_V=2$ for $B=A$ or $ Z$, and $n_V = 1$ for $B=W$ or $ \ov W$, one gets Eq.(\ref{Gamma-enu}).

\subsubsection{ Derivation of Eq.(\ref{Gamma-cd-ud-AZ})} \label{proof-G-cd-ud-AZ}

 Equation (\ref{Gamma-cd-ud-AZ}) is for six-layer modes with $T_2$. 
 Let us first discuss $u$-mode, whose two sections are identical [see Eq.(\ref{Sui})]. 
 To compute the values of $c$ and $d$, one needs to find out the two sets of $\M_z $ and $\M_{\ov z}$.
 Making use of the explicit expressions of the spinor states of $u$-modes, one finds that
\begin{subequations}\label{M-uovu}
\begin{align} 
 u_1 \ov u_1 :  \qquad & \ \M_z =\{ 1,4 \}, \ \M_{\ov z} = \{ 2,3,5,6 \};
 \\ {u_2 \ov u_2:}   \qquad & \M_z =\{ 1,2,4,5 \}, \ \M_{\ov z} = \{ 3,6 \};
  \\ {u_3 \ov u_3: }   \qquad & \M_z =\{ 1,3,4,6 \}, \ \M_{\ov z} = \{ 2,5 \}.
\end{align}
\end{subequations}
 Then, it is ready to compute $c$ and $d$ for $u_i \ov u_i \to A,Z$ by making use of Eq.(\ref{cd-T2-u}), getting 
 the following results. 
\begin{subequations}\label{cd-uovu1}
\begin{align} \notag
 u_1 & \ov u_1  \to A : 
 \\ \notag & c = 2\exp( i2\vartheta^{t_1}_1 ),
 \\ & d = 2e^{i(2\vartheta^{t_1}_1 -\delta_\theta^{A}) } (e^{i2\delta_\theta^t} + e^{i4\delta_\theta^t})
  = 2e^{i2\vartheta^{t_1}_1 };
 \\ u_1 & \ov u_1  \to Z : \notag
 \\ \notag & c = 2\exp( i2\vartheta^{t_1}_1 ),
 \\ & d = 2e^{i(2\vartheta^{t_1}_1 -\delta_\theta^{Z}) } (e^{i2\delta_\theta^t} + e^{i4\delta_\theta^t})
  = -2e^{i2\vartheta^{t_1}_1 }.
\end{align}
\end{subequations}
\begin{subequations}\label{cd-uovu2}
\begin{align} 
 \notag u_2 & \ov u_2  \to A : 
 \\ \notag & c = 2\exp( i2\vartheta^{t_2}_1 ) (1+ e^{i2\delta_\theta^t}) = 2\exp( i2\vartheta^{t_2}_1 ) e^{i\pi/3},
 \\ & d = 2e^{i(2\vartheta^{t_2}_1 -\delta_\theta^{A}) } e^{i4\delta_\theta^t}
  = 2e^{i2\vartheta^{t_2}_1 } e^{i\delta_\theta^t};
 \\ \notag u_2 & \ov u_2  \to Z : 
 \\ \notag & c = 2\exp( i2\vartheta^{t_2}_1 ) (1+ e^{i2\delta_\theta^t}) = 2\exp( i2\vartheta^{t_2}_1 ) e^{i\pi/3},
 \\ & d = 2e^{i(2\vartheta^{t_2}_1 -\delta_\theta^{Z}) } e^{i4\delta_\theta^t}
  = -2e^{i2\vartheta^{t_2}_1 } e^{i\delta_\theta^t}.
\end{align}
\end{subequations}
\begin{subequations}\label{cd-uovu3}
\begin{align} 
 \notag u_3 & \ov u_3  \to A : 
 \\ \notag & c = 2\exp( i2\vartheta^{t_3}_1 ) (1+ e^{i4\delta_\theta^t}) = 2\exp( i2\vartheta^{t_3}_1 ) e^{-i\pi/3},
 \\ & d = 2e^{i(2\vartheta^{t_3}_1 -\delta_\theta^{A}) } e^{i2\delta_\theta^t}
  = 2e^{i2\vartheta^{t_3}_1 } e^{-i\pi/3};
 \\ \notag u_3 & \ov u_3  \to Z : 
 \\ \notag & c = 2\exp( i2\vartheta^{t_3}_1 ) (1+ e^{i4\delta_\theta^t}) = 2\exp( i2\vartheta^{t_3}_1 ) e^{-i\pi/3},
 \\ & d = 2e^{i(2\vartheta^{t_3}_1 -\delta_\theta^{Z}) } e^{i2\delta_\theta^t}
  = -2e^{i2\vartheta^{t_3}_1 } e^{-i\pi/3}.
\end{align}
\end{subequations}

 For $d$-modes,  making use of the explicit expressions of their spinor states, one finds that
\begin{subequations}\label{M-dovd}
\begin{align} 
 d_1 \ov d_1 :  \qquad & \ \M_z =\{ 1,5,6 \}, \ \M_{\ov z} = \{ 2,3,4 \};
 \\ {d_2 \ov d_2:}   \qquad & \M_z =\{ 1,2,6 \},  \M_{\ov z} = \{ 3,4,5 \};
  \\ {d_3 \ov d_3: }   \qquad & \M_z =\{ 1,3,5 \},  \M_{\ov z} = \{ 2,4,6 \}.
\end{align}
\end{subequations}
 Then, substituting Eq.(\ref{theta-d-deltatheta}) into Eq.(\ref{cd-T2-d}), one finds that
\begin{subequations}\label{cd-dovd1}
\begin{align} 
 \notag d_1 & \ov d_1  \to A: 
 \\ \notag & c = \exp( i2\vartheta^{t_1}_1 ) (1+ e^{ i2\delta_\theta^t +i\pi} +  e^{ i4\delta_\theta^t +i\pi})
 = 2\exp( i2\vartheta^{t_1}_1 ),
 \\ & d = \exp( i2\vartheta^{t_1}_1 -i\delta_\theta^{A}) (e^{ i2\delta_\theta^t } + e^{ i4\delta_\theta^t } + e^{i\pi})
 = 2\exp( i2\vartheta^{t_1}_1 );
\\ \notag d_1 & \ov d_1  \to Z: 
 \\ \notag & c = \exp( i2\vartheta^{t_1}_1 ) (1+ e^{ i2\delta_\theta^t +i\pi} +  e^{ i4\delta_\theta^t +i\pi})
 = 2\exp( i2\vartheta^{t_1}_1 ),
 \\ & d = \exp( i2\vartheta^{t_1}_1 -i\delta_\theta^{Z}) (e^{ i2\delta_\theta^t } + e^{ i4\delta_\theta^t } + e^{i\pi})
 = -2\exp( i2\vartheta^{t_1}_1 ).
\end{align}
\end{subequations}
\begin{subequations}\label{cd-dovd2}
\begin{align} 
 \notag d_2 & \ov d_2  \to A: 
 \\ \notag & c = \exp( i2\vartheta^{t_2}_1 ) (1+ e^{ i2\delta_\theta^t}  + e^{ i4\delta_\theta^t +i\pi})
 = 2\exp( i2\vartheta^{t_2}_1 + i\delta_\theta^t),
 \\ \notag & d = \exp( i2\vartheta^{t_2}_1 -i\delta_\theta^{A}) ( e^{ i4\delta_\theta^t } + e^{i\pi} + e^{ i2\delta_\theta^t +i\pi} )
 \\ & = 2\exp( i2\vartheta^{t_2}_1 + i\delta_\theta^t);
 \\ \notag d_2 & \ov d_2  \to Z: 
 \\ \notag & c = \exp( i2\vartheta^{t_2}_1 ) (1+ e^{ i2\delta_\theta^t}  + e^{ i4\delta_\theta^t +i\pi})
 = 2\exp( i2\vartheta^{t_2}_1 + i\delta_\theta^t),
 \\ \notag & d = \exp( i2\vartheta^{t_2}_1 -i\delta_\theta^{Z}) ( e^{ i4\delta_\theta^t } + e^{i\pi} + e^{ i2\delta_\theta^t +i\pi} )
 \\ & = -2\exp( i2\vartheta^{t_2}_1 + i\delta_\theta^t);
\end{align}
\end{subequations}
\begin{subequations}\label{cd-dovd3}
\begin{align} 
 \notag d_3 & \ov d_3  \to A: 
 \\ \notag & c = \exp( i2\vartheta^{t_3}_1 ) (1+ e^{ i4\delta_\theta^t}  + e^{ i2\delta_\theta^t +i\pi})
 = 2\exp( i2\vartheta^{t_3}_1 - i\delta_\theta^t),
 \\ \notag & d = \exp( i2\vartheta^{t_3}_1 -i\delta_\theta^{A}) ( e^{ i2\delta_\theta^t } + e^{i\pi} + e^{ i4\delta_\theta^t +i\pi} )
 \\ & = 2\exp( i2\vartheta^{t_3}_1 - i\delta_\theta^t);
 \\ \notag d_3 & \ov d_3  \to Z: 
 \\ \notag & c = \exp( i2\vartheta^{t_3}_1 ) (1+ e^{ i4\delta_\theta^t}  + e^{ i2\delta_\theta^t +i\pi})
 = 2\exp( i2\vartheta^{t_3}_1 - i\delta_\theta^t),
 \\ \notag & d = \exp( i2\vartheta^{t_3}_1 -i\delta_\theta^{Z}) ( e^{ i2\delta_\theta^t } + e^{i\pi} + e^{ i4\delta_\theta^t +i\pi} )
 \\ & = -2\exp( i2\vartheta^{t_3}_1 - i\delta_\theta^t);
\end{align}
\end{subequations}

\subsubsection{ Derivation of Eq.(\ref{Gamma-qijg-all})}\label{proof-G-qijg-all}

 Equation (\ref{Gamma-qijg-all}) is for three-layer modes with $T_1$. 
  For three-layer modes, we have
\begin{align}\label{theta-t123-app}
 & \vartheta^{t_1}_1 =0 , \quad \vartheta^{t_2}_1 = - \delta_\theta^t/2,
 \quad \vartheta^{t_3}_1 =  \delta_\theta^t/2,
 \\ & \delta_\theta^t = \pi/3, \quad \delta_\theta^{g} = 2\delta_\theta^{t}.
\end{align}
 For the $t$-modes and $g$-modes, one has
 $\delta^{t}_\theta + \delta^{t}_\theta - \delta^g_\theta=0$ and, as a result, Eq.(\ref{cd-T1}) becomes that
\begin{gather}\label{}
  c = e^{i(\vartheta^f_1+\vartheta^{f'}_1 )}  \sum_{m\in \M_z} 1,
 \\  d = e^{i(\vartheta^f_1+\vartheta^{f'}_1 )}  \sum_{m\in \M_{\ov z}} 1.
\end{gather}
 Using these expressions of $c$ and $d$, one finds that
\begin{subequations}\label{M-q-modes}
\begin{align} \notag
 t_1 \ov t_1 \to g_{1\ov 1}: & \ \M_z =\{ 1 \}  \ \& \ \M_{\ov z} = \{ 2,3 \},
 \\ & \to c=1,  \ \&  \  d=2;
 \\ \notag
  {t_2 \ov t_2\to g_{2\ov 2}:}  &  \ \M_z =\{ 1,2 \}  \ \& \ \M_{\ov z} = \{ 3 \}, 
 \\ &  \to c=2e^{-i \delta^t_\theta } \ \&  \  d=e^{-i \delta^t_\theta };  \label{M-t22}
 \\ \notag
  {t_3 \ov t_3\to g_{3\ov 3}:}  &  \ \M_z =\{ 1,3 \}  \ \& \ \M_{\ov z} = \{ 2 \}, 
 \\ &  \to c=2e^{i \delta^t_\theta } \ \&  \  d=e^{i \delta^t_\theta };  \label{M-t33}
 \\ \notag
  {t_1 \ov t_2\to g_{1\ov 2}:}  &  \ \M_z =\{ 1 \}  \ \& \ \M_{\ov z} = \{ 3 \}, 
 \\ &  \to c=e^{-i \delta^t_\theta/2 } \ \&  \  d=e^{-i \delta^t_\theta/2 };  \label{M-t12}
 \\ \notag
  {t_1 \ov t_3\to g_{1\ov 3}:}  &  \ \M_z =\{ 1 \}  \ \& \ \M_{\ov z} = \{ 2 \}, 
 \\ &  \to c=e^{i \delta^t_\theta/2 } \ \&  \  d=e^{i \delta^t_\theta/2 };  \label{M-t13}
 \\ \notag
  {t_2 \ov t_3\to g_{2\ov 3}:}  &  \ \M_z =\{ 1 \},  \M_{\ov z} = \emptyset, 
 \\ &  \to c=1 \ \&  \  d=0;  \label{M-t23}
\end{align}
\end{subequations}
 Finally, substituting the above values of $c$ and $d$ into Eq.(\ref{Gamma1})
 and making use of Eq.(\ref{Gamf=Gam-ovf}) to be proved later, one gets Eq.(\ref{Gamma-qijg-all}).

\subsubsection{ Derivation of Eq.(\ref{Gam-gam-exc-1})}\label{app-derive-Gam-dag}

  From the expression of 
\begin{align}\label{}
  \gamma^\mu = \sqrt 2 \left( \begin{array}{cc} 0 & \sigma^{\mu AB'}
 \\ \ov\sigma^{\mu}_{A'B} & 0\end{array}\right),
\end{align}
 one sees that
\begin{align}\label{ga-dag-1}
 \gamma^{\mu \dag} = \sqrt 2 \left( \begin{array}{cc} 0 &  \sigma^{\mu}_{AB'}
 \\ \ov \sigma^{\mu A'B}  & 0\end{array}\right).
\end{align}
 Changing indices on the rhs of Eq.(\ref{ga-dag-1}) by $A \leftrightarrow  B$,  one gets that
\begin{align}\label{}
 \gamma^{\mu \dag} = \sqrt 2 \left( \begin{array}{cc} 0 &  \sigma^{\mu}_{BA'} \\ \ov \sigma^{\mu B'A}  & 0\end{array}\right)
 =  \sqrt 2 \left( \begin{array}{cc} 0 &  \ov\sigma^{\mu}_{A'B} \\ \sigma^{\mu A B'}  & 0\end{array}\right),
\end{align}
 where Eq.(\ref{sig-c}) has been used in the derivation of the second equality.
 Then,  one finds that
\begin{align}\label{} \notag
 & \left(\Gamma^{f \ov f'\to B } \gamma^{\mu} \right)^\dag = \left( \gamma^{\mu} \right)^\dag
 \left( \Gamma^{f \ov f'\to B } \right)^\dag
 \\ & \notag = \sqrt 2 \left( \begin{array}{cc} 0 &  \ov\sigma^{\mu}_{A'B} \\ \sigma^{\mu A B'}  & 0\end{array}\right)
 \left( \begin{array}{cc} 0 & d^* \\ c^*  & 0 \end{array} \right)
 \\ & \notag =  \sqrt 2 \left( \begin{array}{cc} c^* \ov\sigma^{\mu}_{A'B} & 0 \\ 0 & d^*\sigma^{\mu A B'} \end{array}\right)
 \\ & = \sqrt 2  \left( \begin{array}{cc} 0 & c^* \\ d^*  & 0 \end{array} \right)
  \left( \begin{array}{cc} 0 & \sigma^{\mu AB'} \\ \ov\sigma^{\mu}_{A'B} & 0\end{array}\right)
 \\ &  = \left( \Gamma^{f \ov f'\to B } \right)^* \gamma^\mu.
\end{align}
 This gives Eq.(\ref{Gam-gam-exc-1}).

\subsection{A relation between $\Gamma$-matrices of FIPs of antimodes}\label{app-relation-G-matrix}

 In this section, we prove the following relation used above, 
\begin{align}\label{Gamf=Gam-ovf}
 \Gamma^{f' \ov f \to \ov B} = \Gamma^{ f \ov f' \to  B}.
\end{align}
 To be specific in discussion, let us consider the following expressions of the
 spinor states of $f$ and $\ov f'$,  namely,
\begin{gather}\label{Sf-1}
 |\cs_{f}^{\alpha }(\bp)\ra
 = \frac{1}{\sqrt{n_f}} \left( \begin{array}{c}  e^{i\vartheta_1^f} |u^\alpha(\bp)\ra
 \\ e^{i\vartheta_2^f} |\ov v^\alpha(\bp)
 \\ \vdots    \\ e^{i\vartheta_{n}^f} |u^\alpha(\bp)\ra \end{array} \right)
\end{gather}
 and
\begin{gather}\label{Sovf'-1}
 |\cs_{\ov f'}^{\beta }(\bq)\ra
 = \frac{1}{\sqrt{n_f}} \left( \begin{array}{c}  e^{i\vartheta_1^{f'}} |\ov v^\beta(\bq)\ra
 \\ e^{i\vartheta_2^{f'}} |u^\beta(\bq)
 \\ \vdots    \\ e^{i\vartheta_{n}^{f'}} |u^\beta(\bq)\ra \end{array} \right).
\end{gather}
 (Other expressions may be treated in exactly the same way.)
 The corresponding spinor states of $\ov f$ and $f'$ are obtained by replacements of, say,
 $|u_\alpha(\bp)\ra \leftrightarrow |\ov v_{\alpha}(\bp)\ra$, for all the layer states, resulting in
\begin{gather} \label{Sovf-1}
|\cs_{\ov f}^{\alpha }(\bp)\ra
 = \frac{1}{\sqrt{n_f}} \left( \begin{array}{c}  e^{i\vartheta_1^{f} } |\ov v^\alpha(\bp)\ra
 \\ e^{i\vartheta_2^{}} |u^\alpha(\bp)\ra
 \\ \vdots  \\ e^{i\vartheta_{n}^f} |\ov v^\alpha(\bp)\ra  \end{array} \right),
\end{gather}
\begin{gather}\label{Sf'-1}
 |\cs_{f'}^{\beta }(\bq)\ra
 = \frac{1}{\sqrt{n_f}} \left( \begin{array}{c}  e^{i\vartheta_1^{f'}} |u^\beta(\bq)\ra
 \\ e^{i\vartheta_2^{f'}} |\ov v^\beta(\bq)
 \\ \vdots    \\ e^{i\vartheta_{n}^{f'}} |\ov v^\beta(\bq)\ra \end{array} \right).
\end{gather}

 For the sake of clearness in comparing $C^{f' \ov f\to \ov B}_{\alpha' \alpha \lambda }$ and
 $C^{f \ov f'\to B}_{\alpha \alpha' \lambda } $, let us relabel the spinor states of $f'$ and $\ov f$ as
 $|\cs_{f'}^{\alpha }(\bp)\ra$ and $|\cs_{\ov f}^{\beta }(\bq)\ra$, respectively, i.e.,
\begin{gather}\label{Sf'-2}
 |\cs_{f'}^{\alpha }(\bp)\ra
 = \frac{1}{\sqrt{n_f}} \left( \begin{array}{c}  e^{i\vartheta_1^{f'}} |u^\alpha(\bp)\ra
 \\ e^{i\vartheta_2^{f'}} |\ov v^\alpha(\bp)
 \\ \vdots    \\ e^{i\vartheta_{n}^{f'}} |\ov v^\alpha(\bp)\ra \end{array} \right).
\end{gather}
\begin{gather} \label{Sovf-1}
|\cs_{\ov f}^{\beta }(\bq)\ra
 = \frac{1}{\sqrt{n_f}} \left( \begin{array}{c}  e^{i\vartheta_1^{f} } |\ov v^\beta(\bq)\ra
 \\ e^{i\vartheta_2^{}} |u^\beta(\bq)\ra
 \\ \vdots  \\ e^{i\vartheta_{n}^f} |\ov v^\beta(\bq)\ra  \end{array} \right),
\end{gather}
 Then, it is easy to see that $\overrightarrow D  |\cs_{f}^{ \alpha}(\bp) \cs_{\ov f'}^{ \beta}(\bq) \ra$
 has the same vector-type layers as $\overrightarrow D  |\cs_{f'}^{ \alpha}(\bp) \cs_{\ov f}^{ \beta}(\bq) \ra$.

 Moreover, we note that, for the matrices $T_k$ used in the studied model of modes, 
 ${T_k}$ for ${f' \ov f\to \ov B}$ is the same as that for ${f \ov f'\to B}$. 
 Then, one sees that the values of $c$ and $d$ for $C^{f' \ov f\to \ov B}_{\alpha' \alpha \lambda }$
 are the same as those given on the rhs of Eq.(\ref{Cfovf'B-expre-cd}) for $C^{f \ov f'\to B}_{\alpha \alpha' \lambda } $.
 This proves Eq.(\ref{Gamf=Gam-ovf}).


\end{document}